\documentclass[11pt]{article}
\usepackage{graphicx}
\usepackage{amsmath}
\usepackage{amsthm}
\usepackage{amssymb}
\usepackage{amsfonts}
\usepackage{epsfig}
\usepackage{color}
\def\Im {\mathop{\rm Im}\nolimits}
\def\arg {\mathop{\rm arg}\nolimits}
\def\Re {\mathop{\rm Re}\nolimits}

\def\tr {{\rm tr\;}}

\newtheorem{pro}{Proposition}

\newtheorem{lem}{Lemma}
\newtheorem{thm}{Theorem}
\newtheorem{rmk}{Remark}
\numberwithin{equation}{section}

\title{{Critical edge behavior in the modified Jacobi ensemble and Painlev\'{e}  equations  }}

\author{Shuai-Xia Xu$^a$ and Yu-Qiu Zhao$^b$\footnote{Corresponding author (Yu-Qiu Zhao).
 {\it{E-mail
address:}} {stszyq@mail.sysu.edu.cn} }}
  \date{
 {\it{$^a$Institut Franco-Chinois de l'Energie Nucl\'{e}aire, Sun Yat-sen University, GuangZhou
510275,  China}}\\
 {\it{$^b$Department of Mathematics, Sun Yat-sen University, GuangZhou
510275, China}}
}

\begin{document}

\maketitle

\noindent \hrule width 6.27in\vskip .3cm

\noindent {\bf{Abstract.}}
We study  the Jacobi unitary ensemble perturbed by an algebraic singularity at $t>1$.  For fixed $t$, this  is the
modified  Jacobi  ensemble  studied by  Kuijlaars {\it{et  al.}} The main focus here, however,   is the case when
the algebraic  singularity   approaches the hard edge, namely $t\to 1^+$.

In the double scaling limit case when  $t- 1$ is of the order of magnitude of $1/n^2$, $n$ being the size of the matrix,  the eigenvalue correlation kernel is shown  to have a
 new limiting  kernel     at the hard edge $1$,  described by the $\psi$-functions for a certain second-order nonlinear equation.
  The equation is related  to the Painlev\'{e} III equation by
  a M\"{o}bius transformation. It   also furnishes a
    generalization of    the Painlev\'{e} V equation, and can be reduced to a particular Painlev\'{e} V equation via the B\"{a}cklund transformations in special cases.
The transitions   of the limiting  kernel to Bessel kernels are also investigated, with     $n^2(t-1)$  being   large or  small.

In the present paper, the approach is based on the Deift-Zhou nonlinear steepest descent
analysis for Riemann-Hilbert problems.


\vspace{.2in} \noindent {\it {Keywords: }}
Riemann-Hilbert approach; uniform asymptotic approximation;
random matrix; modified Jacobi unitary ensemble; Painlev\'{e} III equation; Painlev\'{e} V equation\vskip .3cm


\newpage

\section{Introduction and statement of results} \indent\setcounter{section} {1}
Random matrices were introduced in nuclear physics  by Wigner in the 1950s to describe the statistics of energy level of  quantum  systems.
In the 1960s, Dyson obtained the sine kernel  limit of the correlations between eigenvalues in the Gaussian unitary ensemble. He then predicted that the same limit kernel should appear in general random matrix models. This is now known as the famous  universality conjecture   in   random matrices theory.

A unitary ensemble  is determined by the probability distributions (cf. \cite{deift, mehta})
\begin{equation}\label{ensemble-1}
\frac{1}{Z_n}   e^{- \textrm{tr}\, V(M)}dM,\ \ \
dM=\prod_{i=1}^{n}dM_{ii}\prod_{i=1}^{n-1}\prod_{j=i+1}^{n}d\Re
M_{ij}d\Im M_{ij}
\end{equation}on the vector space of $n\times n$   Hermitian  matrices $M=(M_{ij})_{n\times n}$, where $V$ is a certain potential function,  and $Z_n=\int  e^{- \textrm{tr}\, w(M)}dM$  is the normalization constant. For $V(x)=x^2/2$ on $\mathbb{R}$, we have the classical Gaussian unitary ensemble. While the case  $V(x)=-\ln \left \{(1-x)^{\alpha}(1+x)^{\beta}\right \}$ for $x\in (-1,1)$ gives   the classical Jacobi unitary ensemble, where $\alpha>-1$ and $\beta>-1$.
It is shown by Dyson \cite{dy} that the eigenvalues form  a  determinantal process    with the correlation kernel
 \begin{equation}\label{kernel-formula}
K_n(x,y)=e^{-\frac {V(x)}2} e^{-\frac {V(y)}2} \sum_{k=0}^{n-1}p_k(x)p_k(y),
\end{equation}
where $p_k(x)$ denotes the $k$-th degree orthonormal polynomial with
respect to the weight
$e^{-V(x)}$;  see also \cite{dai,deift,kv,magnus,mehta}.
Using the  Christoffel-Darboux formula, \eqref{kernel-formula} can be put into the following closed form
\begin{equation} \label{kernel-CD-formula}
  K_n(x,y)=\gamma^2_{n-1} e^{-\frac {V(x)}2} e^{-\frac {V(y)}2}\,
\frac {\pi_n(x)\pi_{n-1}(y)-\pi_{n-1}(x)\pi_n(y)}{x-y},
\end{equation}
where $\gamma_k$ is the leading coefficient of $p_k(x)$, and $\pi_k(x)$ is the monic polynomial such that $p_k(x)=\gamma_k \pi_k(x)$.
Thus, to justify the    universality  conjecture, or, more general, to study the limiting behavior of the kernel $K_n$ as the size $n$ tends to infinity,   a major step is to obtain  the   asymptotics of the associated orthogonal polynomials.

There are several   results worth mentioning.
For the Jacobi unitary ensemble (JUE), we have the limiting mean eigenvalue density
\begin{equation} \label{eigenvalue-density}
    \lim_{n\to \infty} \frac 1 n  K_n( x, x) = \frac 1  {\pi\sqrt{ 1-x^2}}, ~~ x\in (-1,1).
\end{equation}
Moreover, it is well-known that in the bulk of the spectrum,  the limiting behavior of $K_n$
is given by the sine kernel
\begin{equation} \label{sine-kernel}
   \mathbb{S}(x,y):=\frac{\sin \pi (x-y)}{x-y}
\end{equation}
in the sense that  the limit is independent of the precise reference point in the bulk.
The bulk universality is rigorously  proved for general unitary ensembles with real analytic potentials $V$ in  \eqref{ensemble-1} by  Deift {\it{et al.}}\;\cite{dkmv1} and  for ensembles with continuous potentials by Lubinsky  \cite{lubinsky08,lubinsky}.

However, at the  hard edge of the spectrum, namely at $\pm 1$, the eigenvalue density \eqref{eigenvalue-density} has a square-root singularity. The  hard edge universality is investigated and is  described in this case  by the Bessel  kernel
\begin{equation} \label{bessel-kernel}
  \mathbb{ J}_{\alpha}(x,y):=\frac{J_\alpha(\sqrt{x}) \sqrt{y} J'_\alpha(\sqrt{y}) - J_\alpha(\sqrt{y}) \sqrt{x} J'_\alpha(\sqrt{x})}{2(x-y)};
\end{equation}see,  e.g., Mehta \cite{mehta}. The hard edge universality is proved  by Kuijlaars {\it{et al.}}\;\cite{kmvv,kv} in the modified  $\mathrm{JUE}$ with   potential  $$V(x)=-\ln\left \{\left (1-x\right )^{\alpha}\left (1+x\right )^{\beta}h(x)\right \},~~ x\in(-1,1), $$
 where $\alpha>-1$, $\beta>-1$ and $h$ is positive and analytic on $[-1,1]$. The reader is also referred to the work of Lubinsky \cite{lubinsky08-2,lubinsky 09} for a new approach to obtain the Bessel kernels near the hard edge $x=1$ for unitary ensemble associated with the Jacobi weight perturbed by a continuous function $h$ with $h(1)>0$.
The analysis and results in \cite{kmvv} may be applied to  other Szeg\"{o} class weights of Jacobi type. However, in cases when the  weights decay fast at the endpoints, other kernels, such as the Airy kernel, may be  needed to describe the  edge behavior; cf.    \cite{xzz}.

It is   worth   pointing  out that, in the double scaling limit, some kernels involving higher transcendental functions, such as the Painlev\'e functions,  have  appeared  in certain critical situations. For example, when the eigenvalue density function vanishes at an interior point of the   support, the Painelv\'{e} II kernel appears
 as an appropriate  double scaling limit of the correlation kernel; see, e.g.,\;\cite{bi,ckv}. Other limiting  kernels  involving  Painelv\'{e} I transcendents  and Painlev\'{e} III transcendents    also appear  in the sense of  double scaling limits in  critical situations, such as  where the eigenvalue density functions vanish  to higher order than square root \cite{cv}, and   the potential possesses  a simple pole \cite{xdz2014}, respectively.

To see the appearance of the Painelv\'{e} type kernel in the double scaling limit, we mention a recent work on $\alpha$-generalized Airy kernel by Its, Kuijlaars and \"{O}stensson  \cite{ik1}.
For the Gaussian unitary ensemble  $\mathrm{GUE(n)}$,  described by the Gaussian measure
$$\frac{1}{Z_{\mathrm{GUE(n)}}}e^{- 2n  \mathrm{tr} M^2}dM,$$
it is well-known that at  the soft edge,   namely,   the edge of the support of the equilibrium measure,  the scaling limit of the correlation kernel is the  Airy kernel; cf. \cite[(24.2.1)]{mehta}, see also \cite{deift}.
In   \cite{ik1}, Its, Kuijlaars and \"{O}stensson  have investigated  the Guassian unitary ensemble   perturbed by an algebraic  singularity at the soft edge, of the form
$$\frac{1}{Z_{n}}|\det(M-I)|^{2\alpha}e^{- 2N  \mathrm{tr} M^2}dM.$$The    kernel $K_n$,   given in \eqref{kernel-formula}, is associated with
 the perturbed Gaussian weight
$$
|x-1|^{2\alpha} e^{-2N x^2},~~  x\in \mathbb{R}.
$$
As $ N/n\rightarrow1$, the  algebraic singularity in the perturbed term coalesces  with the soft edge, which in this case is the edge of the support of the equilibrium measure,   with respect to  the external field $2\frac N n x^2$. Instead of the Airy kernel limit,  a so-called $\alpha$-generalized Airy kernel is involved in this case.  The generalized kernel  is described in terms of a certain solution   of a Painlev\'{e} XXXIV equation.

Similar Painlev\'{e} asymptotics  can also be derived if the
Gaussian weight is perturbed by a Heaviside step function,
$$
e^{-2N x^2}\left\{\begin{array}{c}
                                1,\quad x<1; \\
                                \omega,\quad x>1,
                              \end{array}\right.
$$ with $\omega$ being a nonnegative complex constant.
The jump discontinuity  in the perturbed term also approaches   the soft edge   as $ N/n\rightarrow 1$; cf. Xu and Zhao \cite{xz2011}.

In the present work, we consider the perturbed Jacobi  unitary random matrix ensemble (pJUE)
\begin{equation}\label{ensemble}
\frac{1}{Z_n}   e^{- \textrm{tr}\,  \ln w(M)}dM,
\end{equation} where the weight
\begin{equation}\label{p-jacobi weight} w(x;t)=\left (1-x^2\right )^{\beta}\left (t^2-x^2\right )^{\alpha}h(x),\quad x\in(-1,1),
  \end{equation}
  with $t\in(1,d]$, $d>1$, $\beta>-1$, $\alpha\in \mathbb{R}$  and $h(z)$  is analytic in a domain containing $[-1,1]$, such that $h(x)>0$ for $x\in[-1,1]$.

We note that, if   $t$ keeps a positive distance from $[-1, 1]$, the ensemble \eqref{ensemble} is reduced to the
unitary ensemble of   Jacobi type  associated with the weight
$$(1-x^2)^{\beta}h_1(x),~~ x\in(-1,1),$$
where $h_1$  is again analytic  and positive  on  $[-1,1]$, which furnishes a special case of the modified Jacobi weight considered in   Kuijlaars {\it{et  al.}} \cite{kmvv, kv}.  The scaling limit of the eigenvalue correlation kernel \eqref{kernel-formula} at the edge $x=1$ is the Bessel kernel $\mathbb{J}_{\beta}$ of order $\beta$; cf.   \eqref{bessel-kernel}. While for   $t=1$, the same happens, the  ensemble is again reduced to the modified Jacobi ensemble with weight
 $$(1-x^2)^{\alpha+\beta}h(x),~~ x\in(-1,1),$$ where a further restriction $\alpha+\beta>-1$ is brought in.
Hence this time the scaling  limit of the eigenvalue correlation kernel at   $x=1$ is   the  Bessel kernel $\mathbb{J}_{\alpha+\beta}$,   of  order $\alpha+\beta$; cf.
\eqref{bessel-kernel}.

In the present paper, however, the main focus will be on the double scaling limit of the correlation kernel in the situation when the algebraic  singularity approaches the hard edge, that is   $t\to 1$,  as $n\to\infty$.

As mentioned earlier, unitary ensemble of modified Jacobi type has been studied in, e.g., Kuijlaars {\it{et  al.}} \cite{kmvv, kv}; see \eqref{sine-kernel} and \eqref{bessel-kernel} for the limiting kernels.
The results have been extended in  a  paper \cite{mms2006}  of Mart\'{i}nez-Finkelshtein, McLaughlin and   Saff,  to a positive weight on the unit circle with  Fisher-Hartwig singularities, of the form
$$ w(z) \prod_{k=0}^n \left | z-e^{i\theta_k}\right |^{\alpha_k},$$
where $\theta_k$ are real, $\alpha_k>-1$, and $w(z)$ is strictly positive and holomorphic on the unit circle. The asymptotics at the  singular points    can be expressed in terms of the Bessel functions of the first kind.

An  example has been provided by
  Claeys,  Its  and   Krasovsky  \cite{cik}, with  coalescing singularities of algebraic nature   in the weight with jumps, of the form
 $$(z-e^t)^{\alpha+\beta} (z-e^{-t})^{\alpha-\beta}z^{-\alpha+\beta}e^{-\pi i(\alpha+\beta)}e^{V(z)}, $$
where $\alpha\pm \beta \not=-1,-2,\cdots$, $t>0$, and $V(z)$ is a specific analytic function on the unit circle. The interesting part in \cite{cik} is the transition between Szeg\"{o} weight  and Fisher-Hartwig weight, as $t\to 0$.  A particular solution to a Painlev\'{e}  V is used to describe the intermediate asymptotics.
More recently, Claeys and Krasovsky \cite{ck} have studied  the Toeplitz determinants with merging algebraic singularities   and jumps, with weight
 $$e^{V(z)}z^{\beta_1+\beta_2}\prod_{j=1}^2|z-z_j|^{\alpha_j} g_{z_j,\beta_j}\; z_j^{-\beta_j},~~ z=e^{i\theta},~~ \theta\in[0,2\pi), $$
 where $z_1=e^{it}$, $z_2=e^{i(2\pi-t)}$, and the step functions
$$g_{z_j,\beta_j} =\left\{\begin{array}{cc}
                                                                  e^{i\pi\beta_j} & 0\leq\arg z< \arg z_j \\
                                                                  e^{-i\pi\beta_j} & \arg z_j\leq\arg z<2\pi .
                                                                \end{array}\right.$$
Certain Painlev\'{e} V functions are also involved to describe the transition between the asymptotics of the Toeplitz determinants with different types of singularities.

 By a change of variables $x=\cos \theta$, the polynomials with respect  to the weight \eqref{p-jacobi weight} on the interval $[-1,1]$ are converted to polynomials orthogonal
on the unit circle with the weight
\begin{equation}\label{p-jacobi weight-circle-0} w(\cos \theta)|\sin \theta |=2^{-2\alpha-2\beta-1}e^{V(z)}|z^2-1|^{2\beta+1}
\prod_{j=1}^4|z-z_j|^{\alpha},~~ z=e^{i\theta},~~ \theta\in[0,2\pi);
 \end{equation} see \cite[Thm 11.5]{s},
where $V(z)=\ln h(\frac {z+1/z}{2})$ is analytic on the unit circle, $z_1=\varphi(t)$, $z_2=1/\varphi(t)$, $z_3=-\varphi(t)$, $z_4=-1/\varphi(t)$,  and $\varphi(t)=t+\sqrt{t^2-1}$ for $t>1$.
As $t\to 1$, the Fisher-Hartwig singularities  $z_j\to 1$ for  $j=1,2$ and $z_j\to -1$ for $j=3,4$. If the parameter $\beta=-\frac 12$, there are no singularities
in \eqref{p-jacobi weight-circle-0} at $z=\pm1$ for $t>1$, and the case was   solved in \cite{cik}.

In the particular  case $\beta=-\frac 12$,  according to  \cite{cik},  the Painlev\'{e} V asymptotics is expected for the double scaling limit of the eigenvalue correlation kernel near the hard edge
for the perturbed Jacobi weight \eqref{p-jacobi weight}, as $t\to 1$ and $n\to \infty$. In a  preceding  paper  \cite{xz2013-2}, we showed
this is true,  and  the Painlev\'{e} V kernel plays a role in describing  the transition of the limiting kernel from $\mathbb{J}_{-\frac 12}$ to $\mathbb{J}_{\alpha-\frac 12}$. However,  for
$\beta$ other than half integers, there are extra Fisher-Hartwig singularities  at $z=\pm1$ in \eqref{p-jacobi weight-circle-0}, and in general the Painlev\'{e} V asymptotics is no longer valid.

   An  alternative  way to convert the orthogonality to the  unit circle is to introduce a re-scaling in an earlier stage in the weight \eqref{p-jacobi weight}, so that
$$w(tx)=t^{2\alpha}h(tx)(1-t^2x^2)^{\beta}(1-x^2)^{\alpha},~~ x\in(-  1/t, 1/t),~~ t>1. $$
Then by the same change of variables $x=\cos  \theta$ and applying  \cite[Thm 11.5]{s}, we have the polynomials orthogonal
on the unit circle with the weight
\begin{equation}\label{p-jacobi weight-circle-1}e^{\hat{V}(z)}|z^2-1|^{2\alpha+1}
\prod_{j=1}^4|z-e^{i\theta_j}|^{\beta},~~ z=e^{i\theta},~~ \theta\in(\theta_1,\theta_2)\cup(\theta_3,\theta_4),
 \end{equation}
 where $\hat{V}(z)$ is analytic on the unit circle, $\theta_1=\arccos (1/t)$, $\theta_2=\pi-\arccos (1/t)$, $\theta_3=\pi+\arccos (1/t)$, and $\theta_4=2\pi-\arccos (1/t)$.
 For $t>1$, there are   gaps on the unit circle.
As $t\to 1$, the gaps around the singularities $z=\pm1$ disappear, and the other Fisher-Hartwig singularities at the ends of the gaps merge to $\pm1$.
The transition asymptotics of Toeplitz determinants in \cite{cik} and \cite{ck} is inspiring. It is  of particular interest  to study the asymptotics of the Hankel determiants
with the weight \eqref{p-jacobi weight}, as has been addressed  in a separated paper \cite{zxz}.

In the present paper, we will focus on  the correlation kernel with respect to \eqref{p-jacobi weight} in the general setting $\beta>-1$ and $\alpha\in \mathbb{R}$.
The main goal   is to study the  transition asymptotics of the eigenvalue correlation kernel for the perturbed Jacobi unitary ensemble \eqref{p-jacobi weight}, varying  from the  Bessel   kernel  $\mathbb{J}_{\alpha+\beta}$   to  $\mathbb{J}_{\beta}$ as the parameter $t$ varies from $t=1$ to a fixed $d>1$. It is interesting that   a new limiting  kernel is obtained, which  involves a particular Painlev\'{e} III transcendent and, alternatively,  a solution to  a generalized   Painlev\'{e} V equation. To obtain our main results, the nonlinear steepest descent method developed by Deift and Zhou is applied.

\subsection{Modified Painlev\'{e} equation }

To state our results, we   briefly discuss several equations of Painlev\'{e} type.
{\pro{The function $y(s)$ in the present paper satisfies the equation of the
 Painlev\'{e} type
 \begin{equation}\label{nonlinear diff order 2-introduction} \frac {d^2y}{ds^2}-\frac{2y}{y^2-1}\left(\frac {dy}{ds}\right)^2+\frac{1}{s} \frac {dy}{ds} +\frac {y(y^2+1)}{4(y^2-1)}+\frac {y}{2s}-\Theta\frac{y}{s}+\gamma\frac{y^2+1}{2s}=0,\end{equation} where $\gamma$ and $\Theta$ are constants.
The equation is converted  to a generalized Painlev\'{e} V equation by putting   $\omega=y^2$, so that
\begin{equation}\label{modified Painleve V-introduction}
 \frac {d^2\omega} {ds^2} - \left ( \frac 1 {\omega-1} +\frac 1 {2\omega} \right ) \left (\frac {d\omega} {ds} \right )^2+\frac 1 s\frac {d\omega} {ds}  -\frac {(2\Theta-1)\omega} s+\frac {\omega(\omega+1)}{2(\omega-1)}\pm \gamma\frac{\sqrt{\omega}}{s}(\omega+1)=0, \end{equation}
which is reduced to the classical Painlev\'{e} V equation for $\gamma=0$. Alternatively,
   a change of unknown functions  $v(s)=\frac{y(s)+1}{y(s)-1}$ transforms  the equation \eqref{nonlinear diff order 2-introduction}  into the    Painlev\'{e} III equation
\begin{equation}\label{Painleve III cononic-introduction}
\frac {d^2v}{ds^2}-\frac{1}{v}\left(\frac {dv}{ds}\right)^2+\frac{1}{s} \frac {dv}{ds}  +\frac {1}{s}\left ( \frac {\Theta-\gamma-\frac 12}2 v^2-\frac {\Theta+\gamma-\frac 12}2\right ) -\frac {v^3} {16}+\frac{1}{16v}=0.
\end{equation}
Moreover, the equation \eqref{nonlinear diff order 2-introduction} is the
 compatibility condition, namely  $\Psi_{\lambda s}=\Psi_{s\lambda}$, for the following Lax pair
\begin{equation}\label{Lax pair-1-introduction}
 \Psi_\lambda(\lambda, s)= \left (\frac{s\sigma_3} 2+\frac{A(s)}{\lambda-\frac 12} +\frac{B(s)}{\lambda+\frac 12}+\frac{\gamma\sigma_1} {\lambda}
  \right )\Psi(\lambda, s),                                                  \end{equation}
\begin{equation}\label{Lax pair-2-introduction}
 \Psi_s(\lambda, s)=\left( \frac{\lambda\sigma_3} 2+u(s)\sigma_1  \right )
 \Psi(\lambda, s) ,                                                 \end{equation}
 where
\begin{equation}\label{coefficient A-B-introduction}
A(s)=\sigma_1B(s)\sigma_1,~~\mbox{and}~~B(s)=\left(
              \begin{array}{cc}
                b(s)+\frac{\Theta}2 & -(b(s)+\Theta)y(s) \\[0.2cm]
                b(s)/y(s) & -b(s)-\frac{\Theta}2 \\
              \end{array}
            \right) , \end{equation}
in which  $\sigma_1$ and $\sigma_3$ are the Pauli matrices; see \eqref{Pauli-matrices} below, and  $y(s)$ is a specific  solution of \eqref{nonlinear diff order 2-introduction},  while  $b(s)$ and $u(s)$ are determined  by the equations
\begin{equation}\label{nonlinear equations 2-introduction}s\frac {dy}{ds} =-\frac{s y}{2}+\frac{b(y^2-1)^2}{y} +\Theta(y^2-1)y-\gamma(y^2-1) \end{equation}
and \begin{equation}\label{coefficient u-introduction}u(s)=\frac{ b(s)/y(s)-(b(s)+\Theta)y(s)}{s}+\frac{\gamma} {s}. \end{equation}
 }}\vskip .3cm


In Section \ref{sec:2.4} below, we will also show that for integer $\gamma$, \eqref{modified Painleve V-introduction} can be reduced to the classic Painlev\'{e} V via  B\"{a}cklund transformations.

\subsection{The $\psi$-functions and the $\Psi$-kernel}\label{sec:1.2}
We give a brief description  of a pair of functions $\psi_1$ and $\psi_2$, upon which the limiting kernel will be constructed. The functions are determined via   a model RH problem      related to a special solution  of \eqref{nonlinear diff order 2-introduction}; see \eqref{psi-0 jump}-\eqref{psi-0 at 1/4}.
 Detailed analysis will be carried out in Section \ref{sec:2.2}.

 The model RH problem for $\Psi_0(\zeta,s)$ ($\Psi_0(\zeta)$, for short) is as follows.
 \begin{figure}[t]
 \begin{center}
   \includegraphics[width=7.5 cm]{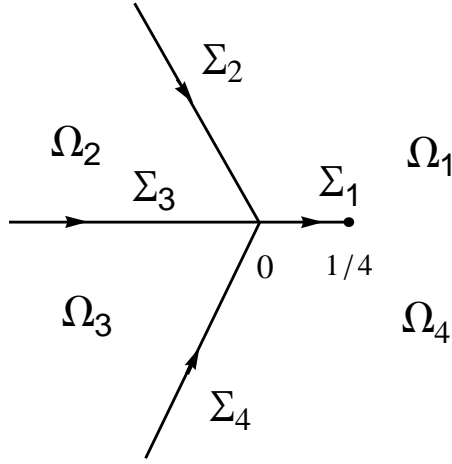} \end{center}
 \caption{\small{Contours in the $\zeta$-plane of the RH problem for $\Psi_0$.}}
 \label{figure 1}
 \end{figure}
\begin{description}
  \item(a)~~  $\Psi_0(\zeta)$ is analytic in
  $\zeta\in \mathbb{C}\backslash\cup^4_{j=1}{\Sigma}_j$, the contours  are  depicted in
Figure \ref{figure 1};

  \item(b)~~  $\Psi_0(\zeta)$  satisfies the jump condition
   \begin{equation}\label{psi-0 jump-int}
  \left(\Psi_0\right )_+(\zeta)=\left(\Psi_0\right )_-(\zeta)
      \left\{
         \begin{array}{ll}
            e^{-\pi i\Theta\sigma_3}, &  \zeta \in \Sigma_1, \\[.3cm]
            \left(
                               \begin{array}{cc}
                                 1 & 0 \\
                                 e^{\pi i(-\Theta+\gamma+\frac12)} & 1 \\
                                 \end{array}
                             \right), &   \zeta \in \Sigma_2,\\[.3cm]
         \left(
                               \begin{array}{cc}
                                 0 & 1 \\
                                 -1 & 0 \\
                                 \end{array}
                             \right),&\zeta \in \Sigma_3,\\[.3cm]
                  \left(
                               \begin{array}{cc}
                                 1 & 0 \\
                                 e^{-\pi i(-\Theta+\gamma+\frac12)} & 1 \\
                                 \end{array}
                             \right),&   \zeta \in \Sigma_4;
                             \end{array}\right .
       \end{equation}

\item(c)~~  The asymptotic behavior of $\Psi_0(\zeta)$  at infinity
  is
  \begin{equation}\label{psi-0 at infinity-int}\Psi_0(\zeta)=\zeta^{\frac{1}{4}\sigma_3}\frac{I-i\sigma_1}{\sqrt{2}}
    \left (I+\frac{\frac {\sigma(s)}s\sigma_3+iu(s)\sigma_1}{\sqrt{ \zeta}}+O\left (\frac 1{\zeta}\right )\right)
  e^{\frac {s\sqrt{\zeta}}2\sigma_3}\end{equation} for $\zeta\to \infty$, as $\arg\zeta\in (-\pi, \pi)$,
  where $s\in (0, \infty)$,   $\sigma=(b+\frac \Theta2)s-(su)^2$; see \eqref{coefficient u-introduction} for  $u(s)$;
\item(d)~~The behavior of $\Psi_0(\zeta)$  at the origin is
\begin{equation}\label{psi-0 at zero-int}
\Psi_0(\zeta) = O\left(1\right)\zeta^{(\frac 1 4+\frac \gamma 2) \sigma_3}
\left(
                                                                         \begin{array}{cc}
                                                                          O(1) &  O(1+c\ln \zeta) \\
                                                                           0 & O(1)\\
                                                                         \end{array}
                                                                       \right),
\end{equation}
for $\zeta\in \Omega_4$, $\zeta\to 0$,  the behavior in other sectors can be  determined by \eqref{psi-0 at zero-int} and the jump condition \eqref{psi-0 jump-int}, and $c=0$ for $\gamma-\frac 12\not\in \mathbb{N}$, $c=(-1)^{\gamma+\frac 12}$ for $\gamma-\frac 12 \in \mathbb{N}$;
\item(d)~~The behavior of $\Psi_0(\zeta)$  at $\zeta=\frac 14$ is
\begin{equation}\label{psi-0 at 1/4-int}
\Psi_0(\zeta)= \widehat{\Psi}^{(0)}(\zeta)(\zeta-  1/4)^{-\frac 1 2\Theta\sigma_3},~~\zeta\to   1 /4,~~\arg(\zeta- 1 /4) \in (-\pi, \pi),
\end{equation}where $\widehat{\Psi}^{(0)}(\zeta)$ is analytic at $\zeta=\frac 1 4$.
\end{description}

Using a  vanishing lemma argument; see Lemma \ref{vanishing-lemma} in Section \ref{sec:2.3} below, we have the following solvability result:
{\pro{\label{Existence-MRH-int}Assuming $\gamma>-\frac 32$ and $\Theta \in \mathbb{R}$,  for $s\in(0,\infty)$, there exists a unique  solution to the model RH problem for $\Psi_0(\zeta,s)$.
}}\vskip .5cm

Now we put parameters  $\gamma=\beta-\frac 12>-\frac 32$, $\Theta=-\alpha\in \mathbb{R}$ and the $\psi$-functions
\begin{equation}\label{}
 \left(
                                \begin{array}{c}
                                  \psi_1(x ,s) \\
                                  \psi_2(x,s) \\
                                \end{array}
                              \right)
 =\left (\Psi_0\right )_+ \left(x,s\right )         \left(
                                                           \begin{array}{cc}
                                                           e^{- \frac{\pi i}{2} (\alpha+\beta)}    & e^{  \frac{\pi i}{2} (\alpha+\beta)}  \\
                                                           \end{array}
                                                         \right)^T,~~x<0.\end{equation}
Accordingly we define the $\Psi$-kernel as
\begin{equation}\label{psi-kernel-int}K_{\Psi}(-u,-v;s)=\frac{  \psi_1(-u,s)\psi_2(-v,s)- \psi_1(-v,s)\psi_2(-u,s)   }{2\pi i(u-v)}  \end{equation}
for $u,v,s \in(0,\infty).$

Noting that for real $\Theta$, the complex conjugate    $\sigma_3\overline{\Psi_0(\overline{\zeta})}\sigma_3$ also satisfies the model RH problem for $\Psi_0$, then by Proposition \ref{Existence-MRH-int}, $\sigma_3\overline{\Psi_0(\overline{\zeta})}\sigma_3=\Psi_0(\zeta)$. In particular, the functions $u(s)$ and $\sigma(s)$   in \eqref{psi-0 at infinity-int},
being used to describe  the large-$\zeta$ behavior of $\Psi_0(\zeta, s)$,
 are real-valued for positive $s$.

Also, by applying  Proposition \ref{Existence-MRH-int}, and  in view of the  asymptotic conditions such as   \eqref{psi-0 at infinity-int}, we determine the coefficients of the Lax pair \eqref{coefficient A-B-introduction}
and  several quantities including   $u(s)$ and $\sigma(s)$ mentioned above. These are real functions analytic for $s\in (0, \infty)$.
Then from \eqref{coefficient u-introduction} and  $\sigma=(b+\frac \Theta2)s-(su)^2$, we see that the function $b(s)$ in  \eqref{coefficient A-B-introduction}  is real-valued and  analytic in $(0,+\infty)$,  and  $y(s)$, the special solution of \eqref{nonlinear diff order 2-introduction},     is  real and meromorphic in $(0,+\infty)$.

  As a corollary of Proposition \ref{Existence-MRH-int}, and taking into account
   the large-$s$  and small-$s$ asymptotics of the model RH problem provided in Sections \ref{sec:5} and \ref{sec:6}; see for example the  large-$\zeta$ asymptotic behavior  of $\Psi_0(\zeta, s)$   in \eqref{G-expand}   and \eqref{psi-0 at circle small s-2}, respectively for $s\to \infty$ and $s\to 0^{+}$, we have

{\pro{For $\gamma=\beta-\frac 12>-\frac 32$, $\Theta=-\alpha\in \mathbb{R}$, the functions $ \sigma(s)$ and $u(s)$, appearing  in  \eqref{psi-0 at infinity-int}, are real-valued, analytic in $(0,+\infty)$, with   boundary behavior
\begin{equation}\label{asymptotics of beta big s-intro} \sigma(s)=-\frac {\alpha} 2 s+O(1),~~ u(s)=O\left (\frac 1s\right )~~\mbox{as}~~s\to \infty,\end{equation}
 and
\begin{equation}\label{asymptotics of beta small s-intro}\sigma(s)=-(\alpha+\beta)^2-\frac 14+O\left(s^l\right ) ,~~ u(s)=-\frac{1}{2s}\left (1+O\left (s^l\right )\right)~~\mbox{as}~~s\to 0^+.
\end{equation}
Also, $b(s)$ is  analytic   and   $y(s)$  is  meromorphic in $s\in (0, \infty)$, taking  real values, with   boundary behavior
\begin{equation}\label{asymptotics of b-intro} b(s)=O\left (\frac 1s\right )~~\mbox{as} ~~s\to\infty ~~\mbox{and}~~b(s)= -\frac {(\alpha+\beta)^2} s+\frac \alpha 2+O\left (s^{l-1}\right )~~\mbox{as}~~s\to 0^+,\end{equation}
 and  \begin{equation}\label{asymptotics of beta big s-intro} y(s)=O(1)~~\mbox{as} ~~s\to\infty ~~\mbox{and}~~y(s)=1+O\left (s^l\right )~~\mbox{as}~~s\to 0^+.\end{equation}
In the small-$s$ behavior, the parameter $l=2\min\{1,1+\alpha+\beta\}$,  and there is an additional   restriction that  $\alpha+\beta>-1$.
}}\vskip .5cm

It is worth noting that by applying  the invertible piecewise transformations \eqref{hat psi} and  \eqref{psi-0}, we obtain the matrix solution  $\Psi\left (\lambda, s\right )$ of  the Lax pair \eqref{Lax pair-1-introduction} and \eqref{Lax pair-2-introduction}, of which \eqref{nonlinear diff order 2-introduction} is the compatibility condition.

\subsection{Main results }
Now we are ready to present our main results, including  a double scaling limit of the eigenvalue correlation kernel, in terms of a Painlev\'{e} type kernel,  when  parameter $s=4n \ln(t+\sqrt{t^2-1})$ is around  a finite positive   number, and the transition of the limiting  kernel to the Bessel kernels, respectively as $s\to 0^+$ and $s\to+\infty$.
\\

\noindent
\textbf{Limiting  kernel}\\

 The first main result  of the present paper is the $\Psi$-description of the limit of  the weighted polynomial kernel \eqref{kernel-formula}, associated with the weight \eqref{p-jacobi weight}.

{\thm{\label{thm-limit-kernel} Let   $\alpha\in \mathbb{R}$, $\beta>-1$, and  $K_n(x,y)$ be   the  weighted polynomial kernel \eqref{kernel-formula}   associated with the weight \eqref{p-jacobi weight}. Then   the following holds

\noindent
 \begin{description}

   \item[(i)] For $x\in (-1,1)$, we have the limiting eigenvalue density
\begin{equation}\label{limiting eigenvalue density introduction}
\frac 1 n K_n(x,x)  =\frac 1
{\pi \sqrt {1-x^2}} +
    O\left (\frac 1 n \right
) ,~~\mbox{as}~n\rightarrow \infty.
\end{equation}
The error term is uniform for $x$ in compact subsets of $(-1, 1)$.

   \item[(ii)] For fixed $x\in(-1,1)$, we have the following sine
 kernel limit uniformly for bounded real variables $u$ and $v$, as $n\rightarrow \infty$:
 \begin{equation}\label{sine-kernel-limit-introduction}  \frac{\pi\sqrt{1-x^2}}{n}K_n\left (x+\frac{\pi\sqrt{1-x^2}}{n}u,x+\frac{\pi\sqrt{1-x^2}}{n}v\right )=
 \frac{\sin\{\pi(u-v)\}}{\pi(u-v)}+O\left (\frac 1 n\right ). \end{equation}

  \item[(iii)]  At the edge of the spectrum,
   we have the double scaling limit as $n\rightarrow\infty$ and $t\to 1^+$ such that
  $$s=4n\ln(t+\sqrt{t^2-1})\to \tau, \quad \tau\in(0,\infty), $$
\begin{equation}\label{psi-kernel-double scaling}
\frac {s^2} {8n^2} K_n\left (1- \frac {s^2u} {8n^2}  , 1-\frac {s^2v} {8n^2};t \right )= K_{\Psi}(-u,-v;\tau)+O\left (n^{-2}\right)+O(s-\tau),
\end{equation}uniformly for $u,v,\tau$ in compact subsets of $(0,\infty)$,
where
   the $\Psi$-kernel $K_{\Psi}$ is  defined in \eqref{psi-kernel-int}.
\end{description}
 }}\vskip .5cm

The formulas in (i) and (ii) demonstrate the universality phenomenon. While the edge behavior is of special interest since its limiting kernel involves a Painlev\'{e} type equation.

 Generally, for  $1-r<x,\;  y< 1$ with positive $r$  and $t\in(1,d]$, we have the uniform estimate of the correlation kernel
$$
K_n(x,y)= \frac{
\left(
    - \psi_2\left (f_t(y)\right ), \psi_1\left (f_t(y)\right )\right )
\left (I+O(x-y)\right )
\left(
     \psi_1\left (f_t(x)\right ), \psi_2\left (f_t(x)\right)
\right)^T}{2\pi i(x-y)},
$$
where the $\psi$-function is defined in \eqref{psi-kernel-introduction} and the conformal mapping $f_t(x)=\frac 14\left(\frac {\ln\varphi(x)}{\ln\varphi(t)}\right)^2$, $\varphi(t)=t+\sqrt{t^2-1}$; see \eqref{kernel-psi-uniform}.
With the $\psi$-kernel as intermediate limiting kernel and in view of the   uniform estimate, we proceed to   the transition asymptotics between Bessel kernels.\\

\noindent
\textbf{Transition to Bessel kernel as $s\to \infty$}\\

In Theorem \ref{thm-limit-kernel}, we obtain the  $\Psi$-kernel of Painlev\'{e} type in the double scaling limit of
the correlation kernel near the hard edge.
In the $\Psi$-kernel $K_{\Psi}(-u,-v;s)$, the parameter  $s=4n\ln(t+\sqrt{t^2-1})\to \tau\in(0,\infty)$  describes  the gap between the hard edge $x=1$ and the
singularity $x=t$ of  the weight function \eqref{p-jacobi weight}.   It is also of interest to
consider the limit kernel as the parameter $s\to \infty$ or $s\to 0$, which reflects the separating and approaching of the hard edge   $x=1$ and the
singularity at $x=t$; cf. the weight \eqref{p-jacobi weight}.

It is worth noting that the double scaling limit case corresponding to $t-1=O(1/n^2)$.  As the distance between the hard edge $x=1$ and  the singularity $x=t$  becomes large  in the sense that  $  {t-1}\gg 1/n^2$, one has  $s=4n\ln(t+\sqrt{t^2-1})\to \infty$. It will be shown that  the limit kernel is reduced to  the Bessel kernel $\mathbb{J}_\beta$, just as in the case  when the hard edge $1$ is separated from the fixed singularity $t>1$,  previously considered in  Kuijlaars {\it{et  al.}} \cite{kmvv, kv}.

{\thm{\label{thm-limit-kernel-big s}{For $ \alpha\in \mathbb{R}$ and $\beta>-1$,} we have the Bessel  type approximation for large $s$.
\begin{description}
  \item[(i)] The  $\Psi$-kernel is approximated by the Bessel kernel as $s\to \infty$
 \begin{equation}\label{psi-kernel-introduction}
\frac {4}{s^2} K_{\Psi}\left (-\frac{4u}{s^2},-\frac{4v}{s^2};s\right )=\mathbb{ J}_{\beta}(u,v) \left(1+O\left(\frac 1s\right )\right ),
   \end{equation}
where the Bessel kernel $\mathbb{ J}_{\beta}(u,v)$ is given in \eqref{bessel-kernel}, and the error term is uniform for
 $u$ and $v$  in compact subsets of $(0, \infty)$. \\
  \item[(ii)]
If the parameter $t\in(1,d]$ and  $n\to \infty$  such  that
\begin{equation}\label{s-to-infty}
s=4n\ln(t+\sqrt{t^2-1})\to\infty.
\end{equation}
Then we have the Bessel kernel limit for $K_n(x,y)$:
\begin{equation}\label{Bessel-kernel-limit-big s-introduction}
 \frac{1}{2n^2}K_n\left (1-\frac{u}{2n^2},1-\frac{v}{2n^2};t\right )=\mathbb{ J}_{\beta}(u,v)  +O\left (\frac 1{n^2}\right )+O\left (\frac 1s\right ) ,
\end{equation}
where $\mathbb{ J}_{\beta}(u,v)$ is given in \eqref{bessel-kernel} and the error terms are uniform in compact subsets of  $u,v\in(0,+\infty)$.
\end{description}
}}
A proof of the theorem will be provided in Section \ref{sec:5}.
\\

\noindent\textbf{Transition to Bessel kernel as $s\to0$}\\

As the distance becomes small in the sense that  $t-1\ll 1/n^2$, then $s=4n\ln(t+\sqrt{t^2-1})\to 0^+$. We have the limiting  kernel $\mathbb{J}_{\alpha+\beta}$ in this case, just as the   case  when the singularity $x=t$ coincides with the hard edge $x=1$ in the perturbed weight \eqref{p-jacobi weight}, which again is the modified Jacobi weight investigated in  Kuijlaars {\it{et  al.}} \cite{kmvv, kv}.

{\thm{\label{thm-limit-kernel-small s} {For $\beta>-1$ and $\alpha+\beta>-1$, }we have the Bessel  type approximation for small $s$:
\begin{description}
  \item[(i)]  The  $\Psi$-kernel is approximated by the Bessel kernel as $s\to 0$
 \begin{equation}\label{psi-kernel-introduction}
\frac {4}{s^2} K_{\Psi}\left (-\frac{4u}{s^2},-\frac{4v}{s^2};s\right )=\mathbb{ J}_{\alpha+\beta}(u,v) \left (1+O\left (s^l\right )\right );
   \end{equation}
cf. \eqref{bessel-kernel} for  the Bessel kernel $\mathbb{ J}_{\alpha+\beta}(u,v)$, where the error term is uniform for  $u$ and $v$  in compact subsets of $(0, \infty)$, and  {$l=2\min\{1,\alpha+\beta+1\}$.}
  \item[(ii)]  If the parameter $t\to 1$ and  $n\to \infty$   such   that
\begin{equation}\label{s-to-0}
s=4n\ln(t+\sqrt{t^2-1})\to 0,
\end{equation}
then we have the Bessel kernel limit for $K_n(x,y)$:
\begin{equation}\label{Bessel-kernel-limit-small s-introduction}
 \frac{1}{2n^2}K_n\left (1-\frac{u}{2n^2},1-\frac{v}{2n^2};t\right )=\mathbb{ J}_{\alpha+\beta}(u,v)  +O\left (\frac 1{n^2}\right )+O\left (s^l\right ),
\end{equation}
where $\mathbb{ J}_{\alpha+\beta}(u,v)$ is given in \eqref{bessel-kernel}, the error terms are uniform for compact subsets of  $u,v\in(0,+\infty)$,   and {$l=2\min\{1,\alpha+\beta+1\}$.}
\end{description}
 }}

  \medskip

The rest of the paper is arranged as follows.
 In Section \ref{sec:2},
 we start by studying the Lax pair for the generalized  Painlev\'{e} V equation,
 and show that the compatibility of the Lax pair leads to the Painlev\'{e} III equation and the generalized Painlev\'{e} V equation. The RH problem for $\Psi(\zeta, s)$ associated with the Lax pair is transformed
  to the model RH problem for $\Psi_0(\zeta,s)$ with special  monodromy data. The solvability for
   $\Psi_0(\zeta, s)$ is then  proved for $s\in (0, \infty)$ by proving  Proposition \ref{Existence-MRH-int}.  Specific B\"{a}cklund transformations will also be established in this section, which implies that in the cases when $\gamma$ are integers (or, equivalently, when the parameter $\beta$ in \eqref{p-jacobi weight} are half-integers), the generalized Painlev\'{e} V can be reduced to the classic Painlev\'{e} V.
In Section \ref{sec:3}, we carry out, in full details,  the Riemann-Hilbert analysis  of the polynomials orthogonal with respect to  the weight functions    \eqref{p-jacobi weight}.
    Section \ref{sec:4} will be devoted to the proof of   Theorem \ref{thm-limit-kernel}, based  on the asymptotic results  obtained in the previous sections.
In   Section \ref{sec:5}, we investigate the transition from $\Psi$-kernel to the Bessel kernel $\mathbb{ J}_{\beta}$ as $s\to \infty$,  and prove  Theorem \ref{thm-limit-kernel-big s}.
In the last section, Section \ref{sec:6}, we consider the transition of  $\Psi$-kernel to the Bessel kernel $\mathbb{ J}_{\alpha+\beta}$ as $s\to 0^+$, and  prove Theorem \ref{thm-limit-kernel-small s}.  Thus we complete the   Bessel to Bessel transition
 as the parameter $t$ in \eqref{p-jacobi weight}varies from left to right in a finite interval $(1, d]$.

\section{Equations of  Painlev\'{e} type and a model Riemann-Hilbert problem}
 \indent\setcounter{section} {2}
\setcounter{equation} {0} \label{sec:2}
In the present section, we study a Lax pair system, of which the compatibility condition is a second-order nonlinear ordinary differential equation. The equation can be transformed    to a generalized version of the Painlev\'{e} V.  If  a certain parameter $\gamma=0$, the generalized Painlev\'{e} V is reduced to the classical  Painlev\'{e} V.
 Also, the second-order equation  can be transformed to the standard Painlev\'{e} III. Special cases are also investigated   when $\gamma$ are integers.   B\"{a}cklund transformations are determined,  so that the generalized Painlev\'{e} V equation is turned into a special  Painlev\'{e} V equation.

We also consider a model Riemann-Hilbert problem (RH problem) associated with the specific Lax pair. The solvability of the RH problem is justified. It is worth mentioning that the model problem will play a crucial role in the construction of parametrix in later sections, and in the description of the edge behavior transition.

\subsection{The Lax pair for the generalized  Painlev\'{e} V }
 \label{sec:2.1}
We consider the following Lax pair of
first-order systems
\begin{equation}\label{Lax pair-1}
 \Psi_\lambda= L\Psi,                                                  \end{equation}
\begin{equation}\label{Lax pair-2}
 \Psi_s=U\Psi ,                                                 \end{equation}
 where
 \begin{equation}\label{coefficient L}
 L(\lambda,s)=\frac{s\sigma_3} 2+\frac{A(s)}{\lambda-\frac 12} +\frac{B(s)}{\lambda+\frac 12}+\frac{\gamma\sigma_1} {\lambda},                                           \end{equation}
 \begin{equation}\label{coefficient U}
 U(\lambda,s)=\frac{\lambda\sigma_3} 2+u(s)\sigma_1,                                                  \end{equation}
 with coefficients
\begin{equation}\label{coefficient A-B}
A(s)=\sigma_1B(s)\sigma_1,~~
B(s)=\left(
              \begin{array}{cc}
                b(s)+\frac{\Theta}2 & -(b(s)+\Theta)y(s) \\[0.2cm]
                b(s)/y(s) & -b(s)-\frac{\Theta}2 \\
              \end{array}
            \right),\end{equation}and
\begin{equation}\label{coefficient u}u(s)=\frac{ b(s)/y(s)-(b(s)+\Theta)y(s)}{s}+\frac{\gamma} {s}, \end{equation}
where $\gamma$ and $\Theta$ are  constants, and the Pauli matrices are  defined as
\begin{equation}\label{Pauli-matrices}
\sigma_1=\left(
                   \begin{array}{cc}
                     0 &1 \\
                    1 & 0 \\
                   \end{array}
                 \right),   ~~\sigma_2=\left(
                   \begin{array}{cc}
                     0 & -i \\
                    i & 0 \\
                   \end{array}
                 \right),~~\mbox{and}~\sigma_3=\left(
                   \begin{array}{cc}
                     1 & 0 \\
                    0 & -1 \\
                   \end{array}
                 \right).
 \end{equation}

 For $\gamma=0$ the systems  \eqref{Lax pair-1}-\eqref{Lax pair-2} are reduced to the Lax pair for the Painlev\'{e} V equation of symmetry form, which differs from the ones in
\cite{fikn,fz}
 by a gauge transformation
$$\Psi(\lambda,s)= e^{-\frac {s\sigma_3} 4 }\Phi\left (\lambda+ 1/2,s\right ).$$

It is easily seen that $[L,U]=LU-UL$ is a
meromorphic function in $\lambda$ with only possible singularities at $\lambda=0$, $\pm1/2$ and $\infty$, while straightforward calculation gives
$$ [L,U]=O(1)~~\mbox{as}~~\lambda\to 0,~~\mbox{and}~~ [L,U]=O(1/\lambda)~~\mbox{as}~~\lambda\to \infty.$$
So, computing the singular parts of $[L,U]$ at $\lambda= \pm1/2$, we have
$$
[L,U] =\frac {[A,\frac 14\sigma_3+u\sigma_1]}{\lambda-\frac12}+\frac {[B,-\frac 14\sigma_3+u\sigma_1]}{\lambda+\frac12},$$
where
$$[B,-\frac 14\sigma_3+u\sigma_1]=-u\left\{ b /y +(b +\Theta)y \right\}\sigma_3+iu(2b+\Theta)\sigma_2-\frac 12\left(
                                                                                                                                \begin{array}{cc}
                                                                                                                                  0 & (b+\Theta)y \\
                                                                                                                                  b/y & 0 \\
                                                                                                                                \end{array}
                                                                                                                              \right),
$$
and
$$[A,\frac 14\sigma_3+u\sigma_1]=\sigma_1[B,-\frac 14\sigma_3+u\sigma_1]\sigma_1.
$$
Thus, the compatibility condition $L_s-U_\lambda+[L,U]=0$ is equivalent to
\begin{equation}\label{schlesinger equation} \left\{
                                                      \begin{array}{l}
                                                       \displaystyle{  \frac {d b}{ds}= u( {b}/{y}+y(b+\Theta))} \\[.2cm]
                                                       \displaystyle{ \frac {d }{ds}(b /y) =u(2b+\Theta)+\frac 12 (b/y)}\\[.2cm]
                                                        \displaystyle{ \frac {d}{ds}((b+\Theta)y) =u(2b+\Theta)-\frac 12(b+\Theta)y,}\\
                                                      \end{array}
                                                   \right. \end{equation}
where  $u(s)$ is given in   \eqref{coefficient u}.
It is readily verified that for $2b+\Theta\not\equiv 0$,  \eqref{schlesinger equation}  is in turn equivalent to the set of equations
  \begin{equation}\label{nonlinear equations}
  \left\{  \begin{array}{l}
             \displaystyle{s\frac {d b}{ds}=\frac{b^2}{y^2}-(b+\Theta)^2y^2+\gamma\left (\frac{b}{y}+y(b+\Theta)\right ) ,} \\[.2cm]
           \displaystyle{ s\frac {dy}{ds} =-\frac{sy}{2}+\frac{b(y^2-1)^2}{y} +\Theta(y^2-1)y-\gamma(y^2-1).}
           \end{array}
  \right .
     \end{equation}
From the second equation in \eqref{nonlinear equations} (cf. \eqref{nonlinear equations 2-introduction}),  $b(s)$ can be represented in terms of $y(s)$ and $y'(s)$. Substituting the representation into the first equation, we see that $y(s)$ solves   the following second-order nonlinear differential equation:
  \begin{equation}\label{nonlinear diff order 2} \frac {d^2y}{ds^2}-\frac{2y}{y^2-1}\left(\frac {dy}{ds}\right)^2+\frac{1}{s} \frac {dy}{ds}  +\frac {y(y^2+1)}{4(y^2-1)}+\frac {y}{2s}-\Theta\frac{y}{s}+\gamma\frac{y^2+1}{2s}=0.\end{equation}

 Let $\omega(s)=y^2(s)$, then we obtain the generalized Painlev\'{e} V equation

 \begin{equation}\label{modified Painleve V}  \frac {d^2\omega} {ds^2} - \left ( \frac 1 {\omega-1} +\frac 1 {2\omega} \right ) \left (\frac {d\omega} {ds} \right )^2+\frac 1 s\frac {d\omega} {ds}  -\frac {(2\Theta-1)\omega} s+\frac {\omega(\omega+1)}{2(\omega-1)}\pm \gamma\frac{\sqrt{\omega}}{s}(\omega+1)=0. \end{equation}
Note that for $\gamma=0$, the equation is reduced to a special Painlev\'{e} V equation; cf. \cite{fikn} and \cite{xz2013,xz2013-2}.

An interesting fact is that the equation   \eqref{nonlinear diff order 2} can be converted to a certain Painlev\'{e} III equation. Indeed, taking the following simple    M\"{o}bius transformation  of the unknown function
\begin{equation}\label{mobius-transform}v(s)=  \frac{y(s)+1}{y(s)-1},   \end{equation}
  we obtain the Painlev\'{e} III equation
\begin{equation}\label{Painleve III}\frac {d^2v}{ds^2}-\frac{1}{v}\left(\frac {dv}{ds}\right)^2+\frac{1}{s} \frac {dv}{ds}  +\frac {1}{s}\left ( \frac {\Theta-\gamma-\frac 12}2 v^2-\frac {\Theta+\gamma-\frac 12}2\right ) -\frac {v^3} {16}+\frac{1}{16v}=0; \end{equation}
cf. \cite{fikn,fz}.
All the quantities involved in the Lax pair \eqref{Lax pair-1}-\eqref{Lax pair-2} can now be determined by the solution $y$ to \eqref{nonlinear diff order 2}, or in turn by $\omega$ in \eqref{modified Painleve V} and $v$ in \eqref{Painleve III}.

To complete the subsection, we derive an equation for the function $u(s)$ given in \eqref{coefficient u}.
Indeed, a combination of the last two equations in  \eqref{schlesinger equation} yields
\begin{equation}\label{schlesinger equation-1}
                                                       \frac {d }{ds}\left (\frac b y-(b+\Theta)y\right ) =\frac 12\left (\frac{b}y+ (b+\Theta)y\right ).
                                                       \end{equation}
Then, in view of  \eqref{coefficient u},  \eqref{schlesinger equation} and \eqref{schlesinger equation-1}, we have
\begin{equation}\label{schlesinger equation-2} \left\{
                                                      \begin{array}{l}
                                                       \displaystyle{ \frac {d b}{ds}=2u\frac d{ds} (su) } \\[.2cm]
                                                       \displaystyle{ \frac {d }{ds}(su) =\frac 12\left (\frac{b}y+y(b+\Theta)\right )}\\[.2cm]
                                                       \displaystyle{ \frac {d^2 }{ds^2}(su) =u(2b+\Theta)+\frac14(su -\gamma).}
                                                      \end{array}
                                                   \right. \end{equation}
Representing $b(s)$ from the third equation of   \eqref{schlesinger equation-2} and substituting it to the  first equation, we find that
$$2u(su)'= \frac {d b}{ds}=\left(\frac {(su)''}{2u}-\frac {\Theta}2-\frac s8+\frac {\gamma }{8u}\right)',$$
from which we obtain a third order nonlinear differential equation for $u$
\begin{equation}\label{nonlinear equation-u}
su'''+u''\left (3-\frac {su'}u\right )-\frac{2 u'^2}u-4su'u^2-4u^3-\frac u4-\frac {\gamma u'}{4 u}=0. \end{equation}

For later use, we define an auxiliary  function
\begin{equation}\label{sigma-functon} \sigma(s)=(b(s)+  \Theta/2)s-(su)^2. \end{equation}
Then it is readily seen from \eqref{schlesinger equation-2} that
\begin{equation}\label{sigma-derivative} \sigma'(s)=b(s)+\frac \Theta2. \end{equation}

\subsection{A model Riemann-Hilbert problem  }\label{sec:2.2}

\begin{figure}[t]
 \begin{center}
   \includegraphics[width=7cm]{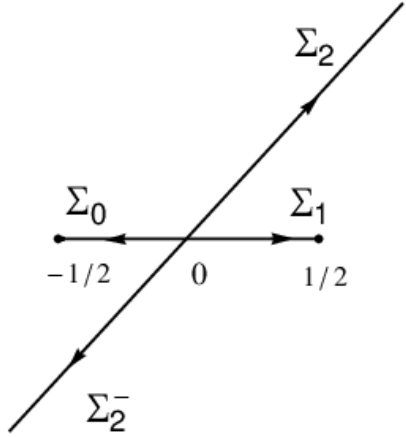} \end{center}
  \caption{\small{Contours and regions in the $\lambda$-plane:  The RH problem for the generalized Painlev\'{e} V.}}
 \label{figure 2}
    \end{figure}

In the present  subsection, we construct the two-by-two matrix-valued RH problem for $\Psi(\lambda)=\Psi(\lambda, s)$ in \eqref{Lax pair-1}-\eqref{Lax pair-2}. Notice that we have introduced in the Lax pair  an extra  regular singularity at $\lambda=0$, as compared with the Lax pair for the canonic  Painlev\'{e} V; cf. \cite[Prop. 5.9]{fikn} or \cite{fz}. Thus,   for  the RH problem for the Painlev\'{e} V, we have correspondingly an extra  singularity at $\lambda=0$; see \eqref{Psi at 0} below.
In the construction that follows, the symmetry relation   $\sigma_1\Psi(-\lambda)\sigma_1=\Psi(\lambda)$ will be used.
The regions and contours are illustrated in Figure \ref{figure 2}.

\begin{itemize}
  \item[(a)] $\Psi(\lambda)$ is analytic in
  $\mathbb{C}\backslash\{\Sigma_0\cup\Sigma_1\cup\Sigma_2\cup\Sigma^-_2 \}$;

  \item[(b)] $\Psi(\lambda)$  satisfies the jump condition
  \begin{equation}\label{Psi jumps}
  \Psi_+(\lambda)=\Psi_-(\lambda)\cdot
\left\{ \begin{array}{ll}
         \sigma_1J\sigma_1,  &  \lambda\in \Sigma_0, \\[0.2cm]
          J,  &  \lambda\in \Sigma_1, \\[0.2cm]

           S_1, & \lambda\in \Sigma_2,\\[0.2cm]

        S_2,   & \lambda\in \Sigma^-_2,
        \end{array}\right .\end{equation}
with
$$J=E_{1/2}^{-1}e^{-\pi i\Theta\sigma_3}E_{1/2},~~ S_{1}= \left(
                               \begin{array}{cc}
                                 1 & 0 \\
                                 s_0 & 1 \\
                                 \end{array}
                             \right) , ~~\mbox{and}~~ S_2=\sigma_1S_1\sigma_1= \left(
                               \begin{array}{cc}
                                 1 & s_0 \\
                                 0 & 1 \\
                                 \end{array}
                             \right) ,$$
where $s_0$ is  a complex constant, and the connection matrix $E_{1/2}$ is constant, such that $\det E_{1/2}=1$;
\item[(c)] The asymptotic behavior of $\Psi(\lambda)$  at infinity
  is
  \begin{equation}\label{Psi at infinity}\Psi(\lambda)=\left (I+\frac {c_1(s)\sigma_3+c_2(s)\sigma_2}{\lambda}+ O\left (\frac 1 {\lambda^2}\right )\right )e^{\frac 1 2 s\lambda \sigma_3}.\end{equation}
  Here use has been made of  the fact that  $\sigma_1\Psi(-\lambda)\sigma_1=\Psi(\lambda)$.  By substituting \eqref{Psi at infinity} into  \eqref{Lax pair-1}, and expanding both sides of \eqref{Lax pair-1} at infinity  in powers of $1/\lambda$,  we have  $$c_1(s)=\sigma(s)/s~~\mbox{and}~~c_2(s)=-iu(s),$$ where $u(s)$ and $\sigma(s)$ are introduced  respectively in \eqref{coefficient u} and \eqref{sigma-functon}.   In computing  the coefficients, we have also used   \eqref{coefficient L}, \eqref{coefficient A-B}, and the fact that by symmetry,  the $O(1/\lambda^2)$ term in \eqref{Psi at infinity} has a leading behavior  of the form $(\hat c_1(s) I +\hat c_2(s)\sigma_1)/\lambda^2$, with  scalar functions $\hat c_1(s)$ and
 { $$\hat c_2(s)=\frac 1s\left(u+u\sigma-\frac 12 \left (\frac by+(b+\Theta)y\right )\right);$$}

\item[(d)] The behavior of $\Psi(\lambda)$ at $\pm\frac 12$ are respectively
\begin{equation}\label{Psi at 1/2}
\Psi(\lambda)= \hat{\Psi}_{1/2}(\lambda)\left (\lambda- 1/2\right )^{-\frac12\Theta\sigma_3}E_{1/2}~~\mbox{as}~~\lambda \to   1/ 2,
\end{equation}and
\begin{equation}\label{Psi at -1/2}
\Psi(\lambda)= \hat{\Psi}_{-1/2}(\lambda)\left (\lambda+ 1/2\right )^{\frac12\Theta\sigma_3}E_{-1/2}~~\mbox{as}~~\lambda \to -  1/ 2,
\end{equation}
where the connection matrices $E_{-1/2}=\sigma_1E_{1/2}\sigma_1$,
the functions  $\hat{\Psi}_{\pm1/2}(\lambda)$ are  analytic  respectively at  $\lambda=\pm\frac 12$,
and the branch cut for $\lambda=1/2$ is $\Sigma_2^-$, joined by the line segment $[0,1/2]$, while the cut for $\lambda=-1/2$ is $[-1/2, 0]\cup \Sigma_2$;

\item[(e)]
The behavior of $\Psi(\lambda)$ at $\lambda=0$ can be described  by
\begin{equation}\label{Psi at origin}
\Psi_{Ori}(\lambda)=\hat{\Psi}_{0}(\lambda)\; \lambda^{\gamma \sigma_3}\left(
                                                                \begin{array}{cc}
                                                                  1 & c\ln \lambda \\
                                                                  0& 1 \\
                                                                \end{array}
                                                              \right),
                                                              \end{equation}
where $c=0$ for $\gamma-\frac 12 \not\in \mathbb{N}$, the branches are chosen such that $\arg \lambda\in (-3\pi/4, 5\pi/4)$,    and $\hat{\Psi}_{0}(\lambda)$ is analytic at $\lambda=0$, with Maclaurin  expansion
 $$\hat{\Psi}_{0}(\lambda)=\frac{1}{\sqrt{2}}(I-i\sigma_2)\left [I+\sum_{m=1}^{\infty}(a_{1m}I+a_{2m}\sigma_3)\lambda^{2m}+(b_{1m}\sigma_1+b_{2m}\sigma_2)\lambda^{2m-1}\right ],$$ with $b_{11}=\frac {4(b+\frac \Theta 2+\frac s8)+4\gamma((b+\Theta)y+\frac b y)}{4\gamma^2-1} $ and  $b_{21}=\frac {8\gamma i(b+\frac \Theta 2+\frac s8)+2i((b+\Theta)y+\frac b y)}{4\gamma^2-1}$.

The function $\Psi(\lambda)$, behaving as \eqref{Psi at infinity} at infinity  and fulfilling jump conditions \eqref{Psi jumps} on $\Sigma_2$ and $\Sigma_2^-$,  is related to this function via a connection matrix $E_0$, $\det E_0=1$,  such that
\begin{equation}\label{Psi at 0}
\Psi(\lambda)=\Psi_{Ori}(\lambda)E_0 \left\{
\begin{array}{ll}
E^{-1}_{1/2} e^{\pi i\Theta\sigma_3} E_{1/2},& \arg \lambda \in (-3\pi/4, 0),\\[.2cm]
I,       &  \arg \lambda\in (0, \pi/4), \\[.2cm]
S_1,      &   \arg \lambda\in (\pi/4, \pi), \\[.2cm]
S_1 E^{-1}_{-1/2} e^{-\pi i\Theta\sigma_3} E_{-1/2},& \arg \lambda \in (\pi, 5\pi/4).
                                 \end{array}
\right. \end{equation}
\end{itemize}
\vskip .5cm

\noindent
{\rmk{The monodromy  data $\{S_1,S_2,E_{0},E_{1/2},E_{-1/2}\}$ are constrained by the cyclic condition,
\begin{equation}\label{cyclic condition}
E_{1/2}^{-1}e^{\pi i\Theta\sigma_3}E_{1/2}E_0^{-1}e^{-2\pi i\gamma\sigma_3}\left(
                                                                             \begin{array}{cc}
                                                                               1 & -2c\pi i \\
                                                                            0 & 1 \\
                                                                             \end{array}
                                                                           \right)
E_0( E_{-1/2}S_2)^{-1}e^{-\pi i\Theta\sigma_3}( E_{-1/2}S_2)=S_1S_2,
\end{equation}
where $c=0$ for $\gamma-\frac 12 \not\in \mathbb{N}$.
Each of the matrices  $E_{0}$, $E_{1/2}$ and $E_{-1/2}$ is
determined by \eqref{cyclic condition}   up to a  left-multiplicative diagonal matrix  $ \mbox{diag}(d,d^{-1})$; cf. \cite{fz}, see also \cite[p.69, p.108]{fikn} for derivation of cyclic condition, and \cite[p.205]{fikn} for a similar description of the behavior at the origin.}}
\vskip 1 cm

\begin{figure}[t]
 \begin{center}
   \includegraphics[width=7 cm]{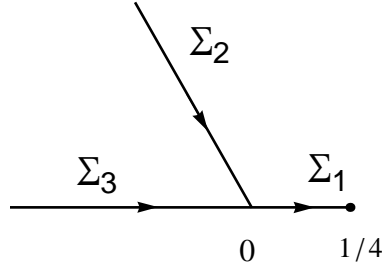} \end{center}
 \caption{\small{Contours in the $\zeta$-plane of the RH problem for the generalized Painlev\'{e} V
 transcendent. The ray $\Sigma_2$ is initially the positive imaginary axis, and can be rotated in the upper half plane.}}
 \label{figure 3}
 \end{figure}

In view of  the symmetry  $\sigma_1\Psi(-\lambda)\sigma_1=\Psi(\lambda)$, we
 expect a new RH problem on half of the $\lambda$-plane. To this end, we introduce a change of variables  $\lambda=\sqrt{\zeta}$, where  $\arg \zeta \in [-\pi, \pi]$, corresponding to $\arg \lambda \in [-\pi/2, \pi/2]
 $.
Now we take
\begin{equation}\label{hat psi}
\widehat{\Psi}(\zeta, s)=\zeta^{\frac{1}{4}\sigma_3}\frac{I+i\sigma_2}{\sqrt{2}}
\Psi\left (\sqrt{\zeta}, s\right ),~~   \arg \zeta \in [-\pi, \pi]         .
\end{equation}
Then, $\widehat{\Psi}(\zeta, s)$($\widehat{\Psi}(\zeta)$, for short)   solves the following RH
problem; see Figure \ref{figure 3}  for   the
contours, where $\Sigma_1$ and $\Sigma_2$ denote the images of the original ones, with direction being adjusted.

\begin{description}
  \item(a)~~  $\widehat{\Psi}(\zeta)$ is analytic in
  $\mathbb{C}\backslash {\Sigma}_j, j=1,2,3$;

  \item(b)~~  $\widehat{\Psi}(\zeta)$  satisfies the jump condition
   \begin{equation}\label{hat psi jumps}
  \widehat{\Psi}_+(\zeta)=\widehat{\Psi}_-(\zeta)\cdot
\left\{ \begin{array}{ll}
          E_{1/2}^{-1}e^{-\pi i\Theta\sigma_3}E_{1/2},,  &  \zeta\in  \Sigma_1, \\[0.2cm]

           S_1^{-1}, & \zeta\in  \Sigma_2,\\[0.2cm]

        \left(
                               \begin{array}{cc}
                                 0 & i \\
                                 i & 0 \\
                                 \end{array}
                             \right),   & \zeta\in  \Sigma_3;
        \end{array}\right .\end{equation}

\item(c)~~  The asymptotic behavior of $\widehat{\Psi}(\zeta)$  at infinity
  is
  \begin{equation}\label{hat psi infinity}\widehat{\Psi}(\zeta)=
  \zeta^{\frac{1}{4}\sigma_3}\frac{I+i\sigma_2}{\sqrt{2}}
  \left (I+ \frac {c_1(s)\sigma_3+c_2(s)\sigma_2}{\sqrt{\zeta}}+ O\left (1/{\zeta}\right )\right )e^{\frac 1 2 s\sqrt{\zeta}\; \sigma_3},\end{equation}
  where $\arg\zeta\in (-\pi, \pi)$, $sc_1(s)=\sigma(s)$ and $c_2(s)=-iu(s)$; see \eqref{coefficient u} and \eqref{sigma-functon} for definition of $u(s)$ and $\sigma(s)$;

\item(d)~~The behavior of $\widehat{\Psi}(\zeta)$  at $\zeta=\frac 14$ is
\begin{equation}\label{hat psi at 1/4}
\widehat{\Psi}(\zeta)=
\widehat{\Psi}^{(1)}(\zeta)(\zeta- 1/4)^{-\frac 1 2\Theta\sigma_3}
E_{1/2},
\end{equation}where $\widehat{\Psi}^{(1)}(\zeta)$ is analytic at $\zeta=\frac 1 4$;
\item(e)~~The behavior of $\widehat{\Psi}(\zeta)$  at $\zeta=0$ is
\begin{equation}\label{hat psi at 0}
\widehat{\Psi}(\zeta)=\widehat{\Psi}^{(0)}(\zeta)\zeta^{(\frac14+ \frac{\gamma}{2} )\sigma_3}\left(
                                                                                               \begin{array}{cc}
                                                                                                 1 & \frac c2 \ln \zeta \\
                                                                                               0 & 1 \\
                                                                                               \end{array}
                                                                                             \right)
\left\{\begin{array}{ll}E_0, &  \arg \zeta \in(-\pi,  \pi/2), \\[.2cm]
 E_0S_1,& \arg \zeta \in(\pi/2,\pi)
                                                                                        \end{array}\right.
\end{equation}
as $\zeta\to 0$, where $c=0$ for $\gamma-\frac 12\not \in \mathbb{N}$, and $\widehat{\Psi}^{(0)}(\zeta)$ is analytic at $\zeta=0$;
\end{description}
\vskip .5cm

Now we rewrite  the cyclic condition  \eqref{cyclic condition}   as
\begin{equation}\label{cyclic condition-2}(JS_1\sigma_1)^2=E_0^{-1}e^{-2\pi i\gamma\sigma_3}\left(
                                                                                                   \begin{array}{cc}
                                                                                                     1 & -2c\pi i \\
                                                                                                     0 & 1 \\
                                                                                                   \end{array}
                                                                                                 \right)
E_0,\end{equation}
where $c=0$ for $\gamma-\frac 12\not \in \mathbb{N}$,   and
$$J=E_{1/2}^{-1}e^{-\pi i\Theta\sigma_3}E_{1/2}= \left(
                                    \begin{array}{cc}
                                      e^{-\pi i \Theta} & 0 \\
                                     2i a \sin \pi\Theta  & e^{\pi i \Theta} \\
                                    \end{array}
                                  \right)      ~~\mbox{if}~~E_{1/2}:=\left(
                                    \begin{array}{cc}
                                      1 & 0 \\
                                     a & 1 \\
                                    \end{array}
                                  \right).
$$
We proceed to determine the  connection matrices, assuming $E_{1/2}$ takes the specific form.
Comparing the trace on both side of equation \eqref{cyclic condition-2} gives
\begin{equation}s_0 e^{\pi i \Theta}+2i a \sin \pi\Theta=\pm2i\sin (\pi  \gamma).\end{equation}
Taking the minus  sign in   the above equation, we see that the
factorization of $$JS_1\sigma_1=\left(
                                    \begin{array}{cc}
                                     0 &  e^{-\pi i \Theta} \\
                                     e^{\pi i \Theta}& -2i\sin (\pi  \gamma)\\
                                    \end{array}
                                  \right)$$
may take the following  form: For  $\gamma-\frac 12\not \in \mathbb{N}$,
$$JS_1\sigma_1=\left(
                                    \begin{array}{cc}
                                      e^{-\pi i \Theta} &  e^{-\pi i \Theta} \\
                                     e^{-\pi i \gamma}& -e^{\pi i \gamma}\\
                                    \end{array}
                                  \right)\left(
                                           \begin{array}{cc}
                                               e^{-\pi i\gamma} & 0 \\
                                             0 & -e^{\pi i\gamma} \\
                                           \end{array}
                                         \right)
                                    \left(
                                    \begin{array}{cc}
                                      e^{-\pi i \Theta} &  e^{-\pi i \Theta} \\
e^{-\pi i \gamma}&  -e^{\pi i \gamma}\\
                                    \end{array}
                                  \right)^{-1}, $$
and, for  $\gamma-\frac 12 \in \mathbb{N}$,
$$JS_1\sigma_1=\left(
                                    \begin{array}{cc}
                                      e^{-\pi i \Theta} &  0 \\
                                      e^{-\pi i \gamma}& 1\\
                                    \end{array}
                                  \right)\left(
                                           \begin{array}{cc}
                                               e^{-\pi i\gamma} & 1 \\
                                             0 & -e^{\pi i\gamma} \\
                                           \end{array}
                                         \right)
                                    \left(
                                    \begin{array}{cc}
                                      e^{-\pi i \Theta} & 0 \\
                                      e^{-\pi i \gamma}&  1\\
                                    \end{array}
                                  \right)^{-1}.$$
 Therefore, comparing these with \eqref{cyclic condition-2},  we   can determine
\begin{equation}\label{E matrix-1/2}\quad E_{1/2}=\left(
                                    \begin{array}{cc}
                                      1 & 0 \\
                                     \frac {- 2i\sin (\pi  \gamma)-s_0 e^{\pi i \Theta }}{2i  \sin \pi\Theta} & 1 \\
                                    \end{array}
                                  \right), \quad \Theta\not\in \mathbb{Z}
\end{equation}
and
\begin{equation}\label{E matrix-1}
E_0=\left\{\begin{array}{ll}
             \left(\frac{e^{\pi i \Theta}}{- 2\cos( \pi\gamma)}\right)^{\frac12}\left(
                                    \begin{array}{cc}
                                      - e^{\pi i \gamma} & - e^{-\pi i \Theta} \\
                                    - e^{-\pi i \gamma}& e^{-\pi i \Theta}\\
                                    \end{array}
                                  \right), & \gamma-\frac 12 \not\in \mathbb{N}, \\[.3cm]
            \left(
                \begin{array}{cc}
                   e^{\pi i\Theta/2} & 0 \\
                 - e^{\pi i(\Theta/2- \gamma)} & e^{-\pi i \Theta/2} \\
                \end{array}
              \right),
              & \gamma-\frac 12\in \mathbb{N},
           \end{array}
\right.\end{equation} each up to a  left-multiplicative diagonal matrix  $d^{\sigma_3}$. Accordingly, we have
\begin{equation}\label{c-monodromy}c=\left\{\begin{array}{ll}
                                              0, & \gamma-\frac 12\not\in \mathbb{N},\\
                                              \frac 1 \pi (-1)^{\gamma+\frac 1 2},    &\gamma-\frac 12\in \mathbb{N}.
                                            \end{array}
\right.
\end{equation}

For later use, we choose the specific Stokes multiplier
\begin{equation}\label{Stoke s}s_0=-2i\sin(\pi (\gamma-\Theta)).\end{equation}
Substituting \eqref{Stoke s}  in \eqref{E matrix-1/2}, we obtain
\begin{equation}\label{E matrix-2}
E_{1/2}=\left(
                                    \begin{array}{cc}
                                      1 & 0 \\
                                      -e^{\pi i (\Theta-\gamma) }  & 1 \\
                                    \end{array}
                                  \right).
\end{equation}
In later sections, we also need to compute
\begin{equation}\label{E0E1/2-}
E_0E_{1/2}^{-1}=\left\{
                         \begin{array}{ll}
                         \left( \frac{e^{\pi i \Theta}}{- 2\cos( \pi\gamma)}\right)^{1/2}\left(
                                                                         \begin{array}{cc}
                                                                           -2 \cos(\pi\gamma) & -e^{-\pi i\Theta} \\
                                                                           0  & e^{-\pi i\Theta} \\
                                                                         \end{array}
                                                                       \right), & \gamma-\frac 12\not\in \mathbb{N}, \\[.3cm]
                           \left(
                                    \begin{array}{cc}
                                     e^{\pi i\Theta/2} & 0 \\
                                      0 &   e^{-\pi i\Theta/2} \\
                                    \end{array}
                                  \right), & \gamma-\frac 12\in \mathbb{N}.  \\
                         \end{array}
                      \right.
\end{equation}

Now we are in a position to state the model RH problem, to be applied   later  to the Riemann-Hilbert analysis,  for the matrix function
\begin{equation}\label{psi-0}
\Psi_0(\zeta, s)=\left\{\begin{array}{ll}
                        e^{-\frac 14\pi i \sigma_3}\widehat{\Psi}(\zeta, s)E_{1/2}^{-1}e^{\frac 14\pi i \sigma_3}, \quad &\zeta\in \Omega_1\cup\Omega_4, \\[.1cm]
                        e^{-\frac 14\pi i \sigma_3} \widehat{\Psi}(\zeta,s)e^{\frac 14\pi i \sigma_3},  \quad &\zeta\in \Omega_2\cup \Omega_3.
                      \end{array}
\right.\end{equation}
The contours and regions are    illustrated in
Figure \ref{figure 1}.
\begin{description}
  \item(a)~~  $\Psi_0(\zeta,s)$ ($\Psi_0(\zeta)$, for short)  is analytic in
  $\zeta\in \mathbb{C}\backslash\cup^4_{j=1}{\Sigma}_j$;

  \item(b)~~  $\Psi_0(\zeta)$  satisfies the jump condition
   \begin{equation}\label{psi-0 jump}
  \left(\Psi_0\right )_+(\zeta)=\left(\Psi_0\right )_-(\zeta)
      \left\{
         \begin{array}{ll}
            e^{-\pi i\Theta\sigma_3}, &  \zeta \in \Sigma_1, \\[.3cm]
            \left(
                               \begin{array}{cc}
                                 1 & 0 \\
                                 e^{\pi i(-\Theta+\gamma+\frac12)} & 1 \\
                                 \end{array}
                             \right), &   \zeta \in \Sigma_2,\\[.3cm]
         \left(
                               \begin{array}{cc}
                                 0 & 1 \\
                                 -1 & 0 \\
                                 \end{array}
                             \right),&\zeta \in \Sigma_3,\\[.3cm]
                  \left(
                               \begin{array}{cc}
                                 1 & 0 \\
                                 e^{-\pi i(-\Theta+\gamma+\frac12)} & 1 \\
                                 \end{array}
                             \right),&   \zeta \in \Sigma_4;
                             \end{array}\right .
       \end{equation}

\item(c)~~  The asymptotic behavior of $\Psi_0(\zeta)$  at infinity
  is
  \begin{equation}\label{psi-0 at infinity}\Psi_0(\zeta)=\zeta^{\frac{1}{4}\sigma_3}\frac{I-i\sigma_1}{\sqrt{2}}
    \left (I+\frac{c_1(s)\sigma_3-c_2(s)\sigma_1}{\sqrt{ \zeta}}+O\left (\frac 1{\zeta}\right )\right)
  e^{\frac {s\sqrt{\zeta}}2\sigma_3}\end{equation} for $\zeta\to \infty$, as $\arg\zeta\in (-\pi, \pi)$,
  where $s\in (0, \infty)$,   $sc_1(s)=\sigma(s)$ and  $c_2(s)=-iu(s)$; see \eqref{coefficient u} and \eqref{sigma-functon} for  $u(s)$ and $\sigma(s)$, respectively;
\item(d)~~The behavior of $\Psi_0(\zeta)$  at the origin is
\begin{equation}\label{psi-0 at zero}
\Psi_0(\zeta) = O\left(1\right)\zeta^{(\frac 1 4+\frac \gamma 2) \sigma_3}
\left(
                                                                         \begin{array}{cc}
                                                                          O(1) &  O(1+c\ln \zeta) \\
                                                                           0 & O(1)\\
                                                                         \end{array}
                                                                       \right),
\end{equation}
for $\zeta\in \Omega_4$, $\zeta\to 0$, and the behavior in other sectors can be  determined by \eqref{psi-0 at zero} and the jump condition \eqref{psi-0 jump}. Here use has also been made of \eqref{E0E1/2-}, and $c$ is given in \eqref{c-monodromy};
\item(d)~~The behavior of $\Psi_0(\zeta)$  at $\zeta=\frac 14$ is
\begin{equation}\label{psi-0 at 1/4}
\Psi_0(\zeta)= \widehat{\Psi}^{(0)}(\zeta)(\zeta-  1/4)^{-\frac 1 2\Theta\sigma_3},~~\zeta\to   1 /4,~~\arg(\zeta- 1 /4) \in (-\pi, \pi),
\end{equation}where $\widehat{\Psi}^{(0)}(\zeta)$ is analytic at $\zeta=\frac 1 4$.
\end{description}

\vskip .5cm
\subsection{Solvability of the Riemann-Hilbert problem}\label{sec:2.3}

We turn to the solvability of the RH problem for $\Psi_0(\zeta, s)$ for $s\in (0, \infty)$. To this aim,
we put \eqref{psi-0 at infinity} in the form
\begin{equation}\label{psi-0 at infinity-1}\Psi_0(\zeta, s)=\left (I+(ic_1(s)-c_2(s))\sigma_{+}+O\left (\frac 1{ \zeta}\right )\right)\zeta^{\frac{1}{4}\sigma_3}\frac{I-i\sigma_1}{\sqrt{2}}
  e^{\frac {s\sqrt{\zeta}}2\sigma_3},~\arg\zeta\in (-\pi, \pi),\end{equation} as $\zeta\rightarrow\infty$,
where $\sigma_{+}=\left(
                      \begin{array}{cc}
                        0 & 1 \\
                        0 & 0 \\
                      \end{array}
                    \right)  $. Here use has been made of the fact that the leading $O(1/\zeta)$ term in \eqref{psi-0 at infinity} is of the form $(\tilde c_1(s) I+\tilde c_2(s)\sigma_2)/\zeta$, with scalar $\tilde c_1$ and $\tilde c_2$. Similar argument has previously been used  in the derivation of \eqref{Psi at infinity}.

 We obtain the solvability of the RHP for $\Psi_0(\zeta, s)$ by proving  a vanishing lemma. Similar arguments can be found in, e.g.,   \cite{fikn,fz,ik1,xz2011}. In particular, a  vanishing lemma for Painlev\'{e} $V$ with  special Stokes multipliers is given in \cite[Lemma 4.7]{cik}, and a proof of the lemma, very similar to the one we will carry out, has been given in \cite{xz2013-2}. The main difference in the present case lies in an extra singularity at the origin, indicated by the parameter $\gamma$.

{\lem{\label{vanishing-lemma}Assume that the homogeneous RH problem for
$\Psi_0^{(1)}(\zeta)=\Psi_0^{(1)}(\zeta, s)$ adapts  the same jump
\eqref{psi-0 jump} and the same boundary conditions
\eqref{psi-0 at zero}-\eqref{psi-0 at 1/4} as $\Psi_0(\zeta)$,
 with the behavior \eqref{psi-0 at infinity-1} at infinity
   being altered  to
 \begin{equation}\label{psi-01}\Psi_0^{(1)}(\zeta, s)=O\left (\frac 1 \zeta\right )
  \zeta^{\frac{1}{4}\sigma_3}\frac{I-i\sigma_1}{\sqrt{2}}
  e^{\frac {s\sqrt{\zeta}}2\sigma_3},~~\arg \zeta\in (-\pi, \pi),~~\zeta\rightarrow \infty.\end{equation}
  If $\gamma>-3/2$, and  the parameter $s\in(0,\infty) $,
   then   $\Psi_0^{(1)}(\zeta, s)\equiv 0$.
}}\vskip .2cm

\noindent {\sc{Proof}}. First, we remove the exponential factor  at
infinity and eliminate the jumps on $\Sigma_2$ and
$\Sigma_4$ by defining
\begin{equation}\label{psi-02}
\Psi_0^{(2)}(\zeta)=\left \{
\begin{array}{ll}
  \Psi_0^{(1)}(\zeta)e^{-\frac{s\sqrt{\zeta}}{2}\sigma_3}, & \mbox{for $\zeta\in\Omega_1\cup\Omega_4$ ,}
  \\[.3cm]
  \Psi_0^{(1)}(\zeta)e^{-\frac{s\sqrt{\zeta}}{2}\sigma_3} \left( \begin{array}{cc}
                                 1 & 0 \\
                                  e^{\pi i(-\Theta+\gamma+ 1/2)}e^{-s\sqrt{\zeta}} & 1 \\
                               \end{array}
                             \right) , & \mbox{for $\zeta\in\Omega_2$
                             ,}\\[.3cm]
\Psi_0^{(1)}(\zeta)e^{-\frac{s\sqrt{\zeta}}{2}\sigma_3} \left( \begin{array}{cc}
                                 1 & 0 \\
                                  -e^{-\pi i(-\Theta+\gamma+ 1/2)}e^{-s\sqrt{\zeta}} & 1 \\
                               \end{array}
                             \right) , & \mbox{for   $\zeta\in\Omega_3$; }
\end{array}\right.\end{equation}
cf.  Figure \ref{figure 1} for the  regions $\Omega_1$-$\Omega_4$, where $\arg\zeta\in (-\pi, \pi)$.
It is easily verified  that $\Psi_0^{(2)}(\zeta)$ solves the
following RH problem:
\begin{description}
  \item(a)~~  $\Psi_0^{(2)}(\zeta)$  is analytic in
  $\zeta\in \mathbb{C}\backslash{\Sigma}_1\cup {\Sigma}_3$ (see Figure \ref{figure 1});

  \item(b)~~  $\Psi_0^{(2)}(\zeta)$  satisfies the jump condition
   \begin{equation}\label{psi-02 jump}
  \left(\Psi_0^{(2)}\right )_+(\zeta)=\left(\Psi_0^{(2)}\right )_-(\zeta)
  \left\{
    \begin{array}{ll}
       e^{-\pi i\Theta\sigma_3}, & \zeta \in \Sigma_1,  \\[.3cm]
      \left(
                               \begin{array}{cc}
                                 e^{-\pi i \Theta_1}e^{-s\sqrt{\zeta}_{+}} & 1 \\
                                 0 & e^{ \pi i  \Theta_1}e^{s\sqrt{\zeta}_{+}} \\
                                 \end{array}
                             \right), &   \zeta \in \Sigma_3,
    \end{array}\right .
    \end{equation}
where $\Theta_1=\Theta-\gamma- 1/2$,    $\arg \zeta\in(-\pi,\pi)$, and $\sqrt{\zeta}_{+}=i\sqrt{|\zeta|}$ for $\zeta\in  \Sigma_3$;
\item(c)~~  The asymptotic behavior of $\Psi_0^{(2)}(\zeta)$  at infinity
  is
  \begin{equation}\label{psi-02 at infinity}\Psi_0^{(2)}(\zeta)=O\left(\zeta^{-\frac 34}\right),~~\zeta\rightarrow\infty
  ;\end{equation}
\item(d)~~The behavior of $\Psi_0^{(2)}(\zeta)$  at the origin is
\begin{equation}\label{psi-02 at 0}
\Psi_0^{(2)}(\zeta)= O\left(1
                             \right)\zeta^{(\frac 14+\frac{\gamma}{2})\sigma_3} \left(
                                                                                \begin{array}{cc}
                                                                                   O(1) &   O(1+c\ln \zeta) \\
                                                                                  0 &  O(1) \\
                                                                                \end{array}
                                                                              \right),
\end{equation}
where $c=0$ for $\gamma-\frac 12 \not\in \mathbb{N}$;
\item(e)~~The behavior of $\Psi_0^{(2)}(\zeta)$  at $\zeta=\frac 14$ is

\begin{equation}\label{psi-02 at 1/4}
\Psi_0^{(2)}(\zeta)=O(1)(\zeta-  1/4)^{-\frac 1 2\Theta\sigma_3}.
\end{equation}
\end{description}

We carry out yet another transformation to move the oscillating entries  in the jump matrices to off-diagonal, as follows:
\begin{equation}\label{psi-03}
\Psi_0^{(3)}(\zeta)=\left \{
\begin{array}{ll}
  \Psi_0^{(2)}(\zeta) \left( \begin{array}{cc}
                                 0 & -1 \\
                                  1 & 0 \\
                               \end{array}
                             \right) , & \mbox{for}~ \Im\zeta>0,
                              \\[.2cm]
\Psi_0^{(2)}(\zeta), & \mbox{for}~ \Im\zeta<0.
\end{array}\right .\end{equation}
Then $\Psi_0^{(3)}(\zeta)$ solves  a RH problem  with different jumps
\begin{equation}\label{psi-03-jump}
\left (\Psi_0^{(3)}\right )_+(\zeta)=\left (\Psi_0^{(3)}\right
)_-(\zeta)J^{(3)}(\zeta),\end{equation}
where
\begin{equation}\label{psi-03-jump-1}J^{(3)}(\zeta)=\left \{
\begin{array}{ll}
  \left(
                               \begin{array}{cc}
                                  1& - e^{-\pi i( \Theta-\gamma- 1/2)}e^{-s\sqrt{\zeta}_{+}}\\
                                 e^{ \pi i( \Theta-\gamma- 1/2)}e^{s\sqrt{\zeta}_{+}} & 0 \\
                                 \end{array}
                             \right), &    \zeta \in(-\infty,0) ,\\[.5cm]
\left(
                               \begin{array}{cc}
                                 0 & -e^{-\pi i\Theta} \\
                                e^{\pi i\Theta} & 0 \\
                                 \end{array}
                             \right),
&  \zeta \in(0,  1/4) ,\\ [.5cm]
 \left(
                               \begin{array}{cc}
                                 0 &-1 \\
                                1& 0 \\
                                 \end{array}
                             \right),
&   \zeta \in(  1/4,\infty).
 \end{array}\right. \end{equation}
At infinity, $\Psi_0^{(3)}$ behaves the same as $\Psi_0^{(2)}$ does; see \eqref{psi-02 at infinity}. While the  behavior  at $\zeta=0$ changes to
\begin{equation}\label{psi-03 at 0}
\Psi_0^{(3)}(\zeta)=O\left(1
                             \right)\zeta^{(\frac 14+\frac{\gamma}{2})\sigma_3} \left(
                                                                                \begin{array}{cc}
                                                                                   O(1) &   O(1+c\ln \zeta) \\
                                                                                  0 &  O(1) \\
                                                                                \end{array}
                                                                              \right)\left\{\begin{array}{ll}
                                                                                \sigma_2,& \arg \zeta\in(0,\pi),\\[.2cm]
                                                                                I,& \arg \zeta\in(-\pi,0)
                                                                              \end{array}
                                                                      \right.\end{equation}as $\zeta\to 0$,
where $c=0$ for $\gamma-\frac 12 \not\in \mathbb{N}$,
and the  condition at $\zeta= 1/4$ now takes the form
\begin{equation}\label{psi-03 at 1/4}
\Psi_0^{(3)}(\zeta)\sigma_2=O(1)\left (\zeta-    1 /4\right )^{-\frac 1 2\Theta\sigma_3}
\left\{\begin{array}{ll}
                                                                                \sigma_2,& \arg\zeta \in (0, \pi),\\[.2cm]
                                                                                I,& \arg \zeta\in(-\pi,0).
                                                                              \end{array}
                                                                      \right.\end{equation}

 It is readily seen that
\begin{equation}\label{non-negative}
(J^{(3)}(\zeta))^{*}+J^{(3)}(\zeta)=\left\{\begin{array}{ll}
                                          \left(
                                            \begin{array}{cc}
                                              0 & 0 \\
                                             0 & 0 \\
                                            \end{array}
                                          \right), & \zeta \in (0, +\infty), \\[.3cm]
                                           \left(
                                            \begin{array}{cc}
                                              2 & 0 \\
                                             0 & 0 \\
                                            \end{array}
                                          \right), & \zeta \in (-\infty, 0),
                                           \end{array}\right.
\end{equation}
where $X^*$ denotes the Hermitian conjugate of a matrix $X$.

 Next, we define an auxiliary matrix function
 \begin{equation}\label{h-0}
H(\zeta)=\Psi_{0}^{(3)}(\zeta) \left ( \Psi_{0}^{(3)}(\bar \zeta\
)\right )^*   \quad\mbox{for}~~ \zeta\not \in \mathbb{R}.
\end{equation}
Then $H(\zeta)$
 is analytic in $\mathbb{C}\backslash \mathbb{R}$. Since  $\Psi_0^{(3)}$
behaves  the same as $\Psi_0^{(2)}$  at infinity,
a combination  of \eqref{h-0} and \eqref{psi-02 at infinity}
yields
\begin{equation}\label{h-0 at infiniy}H(\zeta)=O\left (\zeta^{-  3/ 2}\right )~~ \mbox{as}~~\zeta\rightarrow\infty. \end{equation}
Similarly,   combining   \eqref{h-0} with \eqref{psi-03 at 0} and \eqref{psi-03 at 1/4} gives
\begin{equation}\label{h-0 at 1/4}H(\zeta)=O(1)~~\mbox{as}~~\zeta\rightarrow   1/4,
  \end{equation}
and
\begin{equation}\label{h-0 at 0}H(\zeta)= O\left(\zeta^{\frac 12 +\gamma}\ln \zeta\right)
~~ \mbox{as}~~ \zeta\rightarrow 0.
  \end{equation}
Here  use has been  made of the fact that $\tau^{\sigma_3}\sigma_2 \tau^{\sigma_3}= \sigma_2$ for non-vanishing scalar $\tau$, and that
$$
\left(
                                                                                \begin{array}{cc}
                                                                                   O(1) &   O(\ln \zeta) \\
                                                                                  0 &  O(1) \\
                                                                                \end{array}
                                                                              \right)
\sigma_2 \left(
                                                                                \begin{array}{cc}
                                                                                   O(1) &   O(\ln \zeta) \\
                                                                                  0 &  O(1) \\
                                                                                \end{array}
                                                                              \right)^*= \left(
                                                                                \begin{array}{cc}
                                                                                   O(\ln \zeta)   &  O(1)\\
                                                                                  O(1) &  0 \\
                                                                                \end{array}
                                                                              \right).
$$
Thus, for $\gamma>-3/2$, applying  Cauchy's integral theorem, we have
\begin{equation}\label{integral of H on R}\int_{\mathbb{R}}H_+(\zeta)d\zeta=0.
  \end{equation}
Now in view of \eqref{h-0}, and adding  to  \eqref{integral of H on R} its
Hermitian conjugate, we have
\begin{equation}\label{integral of H on R-}
2\int_{-\infty}^0\left (\Psi_0^{(3)}\right )_-(\zeta) \left(
                               \begin{array}{cc}
                                1 & 0 \\
                                 0 & 0 \\
                                 \end{array}
                             \right)
\left (\Psi_0^{(3)}\right )_-^*(\zeta) d\zeta= 0.
\end{equation}

A straightforward  consequence is that  the first column of $\left(\Psi_0^{(3)}\right)_-(\zeta)$ vanishes for $\zeta\in (-\infty,0)$.
 Furthermore, it follows from  \eqref{psi-03-jump-1} that the second column of $\left(\Psi_{0}^{(3)}\right)_{+}(\zeta)$ also vanishes  for  $\zeta\in (-\infty,0)$.

The jump $J^{(3)}(\zeta)$ in \eqref{psi-03-jump-1} admits an analytic continuation   in a neighborhood of $(-\infty,0)$. Accordingly, using  \eqref{psi-03-jump} we can extend
 $\Psi_0^{(3)}(\zeta)$ from $\arg\zeta\in (0, \pi)$  analytically to a larger sector $\arg \zeta \in (0, 2\pi)$, such that $\left (\Psi_0^{(3)}\right )_{12}(\zeta) = \left (\Psi_0^{(3)}\right )_{22}(\zeta)  = 0$ for $\arg \zeta=\pi$. Hence we  have
\begin{equation}\label{vanish upper plane}       \left (\Psi_0^{(3)}\right )_{12}(\zeta) = \left (\Psi_0^{(3)}\right )_{22}(\zeta)  = 0,~~\Im\zeta >0.\end{equation}
Similarly, by analytically  extending $\Psi_0^{(3)}(\zeta)$  to $\arg \zeta\in (-2\pi, 0)$, we have
\begin{equation}\label{vanish lower plane}      \left (\Psi_0^{(3)}\right )_{11}(\zeta) = \left (\Psi_0^{(3)}\right )_{21}(\zeta)  = 0,~~\Im\zeta <0.\end{equation}
The reader is referred to \cite{xz2013-2} for a similar argument.

Now we proceed to exam the other entries of
$\Psi_{0}^{(3)}(\zeta)$ by appealing to
 Carlson's
theorem (cf. \cite[p.236]{rs}).
To this aim, for $k=1,2$,  we define  scalar
functions
\begin{equation}\label{g-k}
g_k(\zeta)=\left \{
\begin{array}{ll}\left(
 \Psi_0^{(3)}(\zeta)\right)_{k1},
 ~\mbox{for}~0<\arg \zeta <\pi, \\[.2cm]
\left(
 \Psi_0^{(3)}(\zeta)\right)_{k2},
 ~\mbox{for}~-\pi<\arg\zeta<0.
\end{array}\right .\end{equation}
From \eqref{psi-03-jump-1} and \eqref{vanish upper plane}-\eqref{vanish lower plane},  we see  that
 each  $g_k(\zeta)$ is analytic in
  $\mathbb{C}\backslash (-\infty, 1/4]$,  and satisfies
  the jump conditions
  \begin{equation}\label{(g-k-jump-1}
  \left (g_k\right )_+(\zeta)=\left (g_k\right )_-(\zeta)e^{-\pi i(\gamma-\Theta+ 1/2)}e^{s\sqrt{\zeta}_{+}} , \quad \zeta \in (-\infty,0),
  \end{equation}where $\sqrt{\zeta}_{+}=i\sqrt{|\zeta|}$,
 and
 \begin{equation}\label{g-k-jump-2}
  \left (g_k\right )_+(\zeta)=\left (g_k\right )_-(\zeta)e^{\pi i\Theta} , \quad \zeta \in (0,  1/4).
  \end{equation}
The sector of
  analyticity can be extended as  follows:
\begin{equation}\label{g-k extend}
 \hat{g}_k(\zeta)=\left \{
\begin{array}{ll} g_k(e^{-2\pi i}\zeta ) e^{-\pi i(\gamma-\Theta+ 1/2)}e^{s\sqrt{\zeta}},
\quad &
 \mbox{for}~  \pi \leq\arg \zeta <2\pi, \\[.2cm]
g_k(e^{2\pi i}\zeta ) e^{\pi i(\gamma-\Theta+ 1/2)}e^{s\sqrt{\zeta}},\quad
&
 \mbox{for}~  -2\pi<\arg\zeta\leq-\pi.
\end{array}\right .
\end{equation}
Thus $\hat{g}_k(\zeta)$ is now analytic in a cut-sector $-2\pi<\arg \zeta<2\pi$ and $\zeta\not\in[0, 1/4]$.
It is worth noting that $\hat{g}_k(\zeta)$ can be further extended analytically   to  $-2\pi\leq \arg \zeta\leq 2\pi$ and $|\zeta|> 1/4$, and that    for $s\in (0,\infty)$, the exponential term $|e^{ s\sqrt{\zeta}}|\leq1$ for $\pi \leq\arg \zeta <2\pi$ and $-2\pi<\arg\zeta\leq-\pi$.

If we put
\begin{equation}\label{h-k}
 h_k(\zeta)=\hat{g}_k((\zeta+1)^4)  \quad \mbox{for} \ \arg\zeta\in [-\pi/2,
 \pi/2],
\end{equation}
then the above discussion implies that $h_k(\zeta)$ is analytic in
$\Re \zeta >0$, continuous and bounded in  $\Re\zeta\geq 0$, and
satisfies the  decay condition on the imaginary axis
\begin{equation}\label{decay of h-k}
 |h_k(\zeta)|=O\left ( e^{- |\zeta|^2}\right ), \quad \mbox{for} \ \Re \zeta=0~ \mbox{as}~ |\zeta|\rightarrow \infty.
\end{equation}  Hence   Carlson's
theorem applies, and we have $h_k(\zeta)\equiv 0$ for
$\Re \zeta>0$. Tracing back, we see that all entries  of $\Psi_{0}^{(3)}(\zeta)$
vanish for $\zeta\not\in \mathbb{R}$; cf. \eqref{g-k},\eqref{vanish upper plane}-\eqref{vanish lower plane}. Therefore,  $\Psi_{0}^{(3)}(\zeta)$ vanishes identically, which implies that  $\Psi_{0}^{(1)}(\zeta)$ vanishes identically. This completes the proof of the vanishing lemma.\hfill\qed\\

The solvability of  the RH problem for $\Psi_0$ follows from the   vanishing lemma. As briefly indicated in \cite[p.104]{fikn}, the RH problem is equivalent to a Cauchy-type
singular integral equations, the corresponding singular integral operator is a Fredholm operator of index zero. The vanishing
lemma states that the null space is trivial, which implies that the singular integral equation (and thus $\Psi_0$) is solvable as a result of the
Fredholm alternative theorem. More details can be found in \cite[Proposition 2.4]{ik1}; see also \cite{deift,dkmv1,fikn,fz} for  standard methods connecting RH problems with integral equations.

Now we have the   solvability result given  in Proposition \ref{Existence-MRH-int}, which states that for $\gamma>-3/2$, $\Theta\in \mathbb{ R}$, and $s\in(0,\infty)$, there exists a unique  solution $\Psi_0(\zeta,s)$   to the  RH problem
 \eqref{psi-0 jump}-\eqref{psi-0 at 1/4}.


\subsection{B\"{a}cklund transformation }\label{sec:2.4}
From \eqref{modified Painleve V}, we see that the generalized Painlev\'{e} V equation is reduced to the classical  Painlev\'{e} V equation as $\gamma=0$.
In this section, we study the B\"{a}cklund transformation of the generalized Painlev\'{e} V equation. We will show that, by applying a certain B\"{a}cklund transformation, \eqref{modified Painleve V} is turned into an equation of the same form, with only the parameter $\gamma$ being replaced by   $\tilde \gamma=-\gamma\pm 1$. In particular, when   $\gamma$ is an integer, making use of such  B\"{a}cklund transformations   $|\gamma|$ times, the equation   \eqref{modified Painleve V} can be   transformed
 to a  specific Painlev\'{e} V equation.

We seek a rational gauge transformation
\begin{equation}\label{Backlund transformation}\tilde{\Psi}(\lambda,s)=F(\lambda,s) \Psi(\lambda,s),\end{equation}
which preserves the canonical   asymptotic structure of $\Psi$ at infinity and at the regular singularity $\zeta=\pm 1/2$; cf. \eqref{Psi at infinity} and \eqref{Psi at 1/2}-\eqref{Psi at -1/2}, and shifts the formal   monodromy exponent at the origin.  In \eqref{Backlund transformation}, we take
\begin{equation}\label{Backlund transformation-F}F(\lambda,s)=I+\frac {F_1(s)}{\lambda};\end{equation} compare, e.g., Fokas {\it{et  al.}} \cite[(6.1.2)-(6.1.4)]{fikn}. We assume that the form of the  $\lambda$-equation \eqref{Lax pair-1} is preserved for $\tilde \Psi$, with $L$ being replaced  by a certain
$$\tilde L(\lambda, s)=\tilde C(s) +\frac {\tilde A(s)}{\lambda-\frac 1 2}+\frac {\tilde B(s)}{\lambda+\frac 1 2}+\frac {\tilde\gamma \sigma_1}\lambda,$$ with a shifted $\tilde \gamma$. Similar discussion can be found in \cite[Ch. 6]{fikn}. Substituting $\tilde \Psi$ into    \eqref{Lax pair-1}, we have
\begin{equation}\label{F-lambda-system}
 F_\lambda +F L=\tilde L F.
\end{equation}
The equation \eqref{F-lambda-system} splits into five equations
$$\begin{array}{rl}
   \lambda^{-2}:  & -F_1+\gamma F_1\sigma_1=\tilde \gamma \sigma_1 F_1, \\[.1cm]
    1: & \frac  s 2 \sigma_3=\tilde C,\\[.1cm]
    \left (\lambda-\frac 1 2\right )^{-1}: & \left ( I+2F_1\right ) A=\tilde A\left ( I+2F_1\right ),\\[.1cm]
    \left (\lambda+\frac 1 2\right )^{-1}: & \left ( I-2F_1\right ) B=\tilde B\left ( I-2F_1\right ),\\[.1cm]
  \lambda ^{-1}:  &  \gamma\sigma_1 +\frac s 2 F_1\sigma_3-2F_1 A+2F_1B= \tilde \gamma\sigma_1 +\tilde C F_1 -2 \tilde A F_1 +2\tilde B F_1;
      \end{array}
$$where $A$ and $B$ are the specified matrices given in \eqref{coefficient A-B}.

Assuming that $\det F(\lambda, s)\equiv 1$, or, equivalently, $\tr F_1(s)\equiv 0$ and $\det F_1(s)\equiv 0$, from the first equation  we have non-vanishing $F_1$ if only $\gamma+\tilde \gamma=\pm 1$. More precisely, we can write
\begin{equation}\label{F1-representation}
   F_1(s)=\kappa(s)\left ( \sigma_3\pm i\sigma_2\right )~~\mbox{for}~~\gamma+\tilde \gamma=\pm 1,
\end{equation}where $\kappa(s)$ is a scalar function to be determined.

From the second to the fourth equation, we obtain
\begin{equation}\label{A-B-C}
\tilde C=\frac  s 2 \sigma_3,~~ \tilde A=\left ( I+2F_1\right ) A \left ( I-2F_1\right ),~~\mbox{and}~~ \tilde B=\left ( I-2F_1\right ) B \left ( I+2F_1\right ).
\end{equation}
Substituting these into the fifth equation, we see that
 \begin{equation}\label{kappa-representation}
   \kappa(s)=\frac {\gamma-\tilde \gamma}{\pm 8  \left ( b+\frac \Theta 2\right )+\frac { 4b} y+4(b+\Theta) y\pm s }~~\mbox{for}~~\gamma+\tilde \gamma=\pm 1.
\end{equation}
The rational gauge transformation \eqref{Backlund transformation} is thus determined.

The B\"{a}cklund transformation can be deduced from \eqref{A-B-C}. Indeed, assuming that $\tilde B(s)$ takes the form of $B(s)$ as in \eqref{coefficient A-B},  for $\gamma + \tilde \gamma =1$,
we define a set of functions $\tilde y(s)$, $\tilde b(s)$ and $\tilde\Theta(s)$ as
\begin{equation}\label{b-y}
\left\{
\begin{array}{l}
  \tilde b+\frac {\tilde\Theta} 2= (1-8\kappa^2 )\left (b+\frac \Theta 2\right )-2\kappa(1+2\kappa) \frac b y +2\kappa(1-2\kappa) (b+\Theta) y, \\[.1cm]
-( \tilde b+\tilde\Theta )\tilde y=4\kappa (1-2\kappa)  \left (b+\frac \Theta 2\right )-4\kappa^2 \frac b y - (1-2\kappa)^2 (b+\Theta) y, \\[.1cm]
\frac {\tilde b}{\tilde y}=4\kappa (1+2\kappa)  \left (b+\frac \Theta 2\right )+(1+2\kappa)^2 \frac b y + 4\kappa^2 (b+\Theta) y.
\end{array}\right .
\end{equation}
It is readily seen that $\det B(s)=\det\tilde B(s)$, which implies  $\tilde \Theta^2=\Theta^2$.  Hence $\tilde \Theta$ is independent of $s$, and we may put  $\tilde \Theta =\Theta$. Straightforward verification then shows that
\begin{equation}\label{tilde schlesinger equation} \left\{
                                                      \begin{array}{l}
                                                       \displaystyle{  \frac {d \left (\tilde b+  \Theta/ 2\right )}{ds}= \tilde u\left (   {\tilde b} /{\tilde y}+\tilde y(\tilde b+\Theta)\right )} \\[.2cm]
                                                       \displaystyle{ \frac {d }{ds}(\tilde b /\tilde y) =2\tilde u(\tilde b+\Theta/2)+\frac 12 (\tilde b/\tilde y)}\\[.2cm]
                                                        \displaystyle{ \frac {d}{ds}((\tilde b+\Theta)\tilde y) =2 \tilde u(\tilde b+\Theta/2)-\frac 12(\tilde b+\Theta)\tilde y,}\\
                                                      \end{array}
                                                   \right. \end{equation}
where
\begin{equation}\label{coefficient tilde u}\tilde u(s)=\frac{\tilde b(s)/\tilde y(s)-(\tilde b(s)+\Theta)\tilde y(s)}{s}+\frac{\tilde \gamma} {s}, \end{equation}bearing in mind that $\gamma+\tilde\gamma=\pm 1$.
Comparing the equations \eqref{tilde schlesinger equation} with \eqref{schlesinger equation}, and the definition \eqref{coefficient tilde u} with \eqref{coefficient u}, we see a clear correspondence between the set of quantities such as $\tilde u$ and $u$, with $\tilde \gamma$ corresponding to $\gamma$. Hence we obtain a differential system  of the form \eqref{nonlinear equations}, and, eventually, wee see that $\tilde y$ solves the equation
 \begin{equation}\label{nonlinear diff order 2 tilde} \frac {d^2y}{ds^2}-\frac{2y}{y^2-1}\left(\frac {dy}{ds}\right)^2+\frac{1}{s} \frac {dy}{ds}  +\frac {y(y^2+1)}{4(y^2-1)}+\frac {y}{2s}-\Theta\frac{y}{s}+\tilde \gamma\frac{y^2+1}{2s}=0,\end{equation} which differs from \eqref{nonlinear diff order 2} with only the constant $\gamma$  being replaced with $\tilde\gamma=-\gamma \pm 1$. One more step further, we find that
  $\tilde \omega(s)=\tilde y^2(s)$ solves the generalized Painlev\'{e} V equation
 \begin{equation}\label{modified Painleve V tilde}  \frac {d^2\omega} {ds^2} - \left ( \frac 1 {w-1} +\frac 1 {2\omega} \right ) \left (\frac {d\omega} {ds} \right )^2+\frac 1 s\frac {d\omega} {ds}  -\frac {(2\Theta-1)\omega} s+\frac {\omega(\omega+1)}{2(w-1)}\pm \tilde \gamma\frac{\sqrt{\omega}}{s}(\omega+1)=0. \end{equation} Again, the equation differs from \eqref{modified Painleve V} in the parameter $\tilde \gamma=-\gamma \pm 1$.

We note that for $|\gamma| \geq 1$, we can always make $|\tilde \gamma|=|\gamma|-1$ by choosing the proper sign in $\pm$. Therefore, for integer $\gamma$, applying the gauge transformation \eqref{Backlund transformation} (and correspondingly, the B\"{a}cklund transformation \eqref{b-y}) $|\gamma|$ times, the constant $\tilde\gamma$ in \eqref{modified Painleve V tilde} is turned into $0$, and, as mentioned earlier,
  the equation is thus reduced to a special Painlev\'{e} V equation; cf. \cite{fikn} and \cite{xz2013,xz2013-2}.

\section{ Nonlinear steepest descent analysis }
\indent\setcounter{section} {3}
\setcounter{equation} {0} \label{sec:3}

\noindent
We begin with  a RH problem for $Y$, associated
with the orthogonal polynomials with respect to  the specific weight
 $w(x)$ given in \eqref{p-jacobi weight}. Such a  remarkable           connection between the orthogonal polynomials and
 RH problems is observed by    Fokas, Its and Kitaev \cite{fik}.
  Then, we apply the nonlinear
steepest descent analysis  developed by Deift and Zhou {\it{et al.}}
\cite{dkmv1,dkmv2} to the RH problem for $Y$; see also  Bleher and  Its \cite{bi}.
The idea is to obtain,
via a series of invertible transformations $Y
\rightarrow T \rightarrow S \rightarrow R$, eventually the  RH problem for $R$ with  jumps   in a sense   close to the identity
matrix.  Tracing back,   the uniform asymptotics of the orthogonal
polynomials in the complex plane is obtained for large degree  $n$.
A key step is  the construction of a certain local parametrix in the neighborhood of the
singular point $t$ and the hard edge. Constructing the parametrix  will be our main focus   in this section.

\subsection{  Riemann-Hilbert problem for orthogonal polynomials }
Initially,  the Riemann-Hilbert problem for orthogonal polynomials
is as follows (cf. \cite{fik}).
\begin{description}
  \item(Y1)~~  $Y(z)$ is analytic in
  $\mathbb{C}\backslash [-1,1]$;

  \item(Y2)~~  $Y(z)$  satisfies the jump condition
  \begin{equation}\label{}
  Y_+(x)=Y_-(x) \left(
                               \begin{array}{cc}
                                 1 & w(x) \\
                                 0 & 1 \\
                                 \end{array}
                             \right),
\qquad x\in (-1,1),\end{equation} where $w(x)=(1-x^2)^{\beta} (t^2-x^2)^\alpha h(x)$ is the
weight function defined  in
\eqref{p-jacobi weight};
  \item(Y3)~~  The asymptotic behavior of $Y(z)$  at infinity is
  \begin{equation}\label{}Y(z)=\left (I+O\left (  1 /z\right )\right )\left(
                               \begin{array}{cc}
                                 z^n & 0 \\
                                 0 & z^{-n} \\
                               \end{array}
                             \right),\quad \mbox{as}\quad z\rightarrow
                             \infty ;\end{equation}
\item(Y4)~~The asymptotic behavior of $Y(z)$   at the endpoints $z=\pm1$ are
 \begin{equation}\label{Y at 0}Y(z)=\left\{ \begin{array}{ll}
                                               \left(
                               \begin{array}{cc}
                                O( 1) & O\left ( (z\pm1)^{\beta}\right ) \\
                                O( 1) & O\left((z\pm1)^{\beta}\right )
                                  \\
                               \end{array}
                             \right),& \mbox{for}~-1<\beta<0,\\[0.4cm]
                                               \left(
                               \begin{array}{cc}
                                O( 1) & O\left ( \ln(z\pm1)\right ) \\
                                O( 1) & O\left(\ln(z\pm1)\right ) \\
                               \end{array}
                             \right), &\mbox{for}~\beta=0,\\[0.4cm]
                                               \left(
                               \begin{array}{cc}
                                O( 1) &  O( 1) \\
                                O( 1) &  O( 1)
                                  \\
                               \end{array}
                             \right), &\mbox{for}~ \beta>0.
                                            \end{array}\right.\end{equation}

\end{description}
\vskip .5cm

 By virtue of the    Sochocki-Plemelj formula   and Liouville's theorem, it is known that the above RH problem for $Y$
 has a unique solution $Y(z)=Y(z;n)$,
\begin{equation}\label{Y polynomial solution}
Y(z)= \left (\begin{array}{cc}
\pi_n(z)& \frac 1 {2\pi i}
\int_{-1} ^{1}\frac {\pi_n(s) w(s) }{s-z} ds\\[0.2cm]
-2\pi i \gamma_{n-1}^2 \;\pi_{n-1}(z)& -   \gamma_{n-1}^2\;
\int_{-1} ^{1}\frac {\pi_{n-1}(s) w(s) }{s-z} ds \end{array} \right ),
\end{equation}
where  $\pi_n(z)$ is the monic polynomial, and $p_n(z)=\gamma_{n}\pi_n(z)$
is the  orthonormal  polynomial with respect to the weight
$w(x)=w(x;t)$ in \eqref{p-jacobi weight};    cf., e.g.,  \cite{deift} and \cite{fik}.\\
\vskip .3cm

\noindent

\subsection{ The first transformation  $Y\rightarrow T$ }

The first transformation  $Y\rightarrow T$ is defined as
\begin{equation}\label{Y to T}T(z)= 2^{ n \sigma_3}Y(z) \varphi(z)^ {-n \sigma_3} \end{equation} for $z\in
\mathbb{C}\backslash[-1,1]$,  where  $\varphi(z)=z+\sqrt{z^2-1}$ is a conformal map from  $\mathbb{C}\backslash[-1,1]$ onto the exterior of the unit circle, with the branches specified as
 $\arg (z\pm1)\in(-\pi, \pi)$, such that $\varphi(z)\sim 2z$ as $z\rightarrow \infty$. The transformation
\eqref{Y to T} accomplishes a normalization of $Y(z)$ at infinity,
and $T$ solves the RH problem:
\begin{description}
\item(T1)~~ $T(z)$ is analytic in
$\mathbb{C}\backslash [-1,1]$;
\item(T2)~~  The jump condition is
\begin{equation}\label{T jump}T_+(x)=T_-(x)
\left(
       \begin{array}{cc}
        \varphi_+(x)^{-2n }& w(x)\\
         0 &\varphi_-(x)^{-2n } \\
       \end{array}
     \right) ,\quad x \in (-1, 1),
\end{equation}where  $\varphi_\pm(x)$ are the boundary values of $\varphi(z)$,  respectively from above $(-1, 1)$ and from below;
\item(T3)~~   The asymptotic behavior of $T(z)$  at infinity
\begin{equation}\label{}T(z)= I+O(1/z)\quad \mbox{as}\quad z\rightarrow \infty ;\end{equation}
\item(T4)~~ $T(z)$ behaves the same as $Y(z)$  at the end points  $\pm 1$, as described in \eqref{Y at 0}.
\end{description}

\subsection{ The second  transformation  $T\longrightarrow S$}
\begin{figure}[t]
 \begin{center}
   \includegraphics[width=4 in]{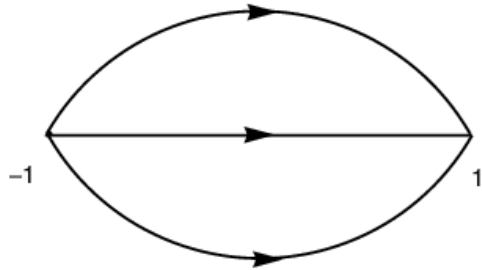} \end{center}
 \caption{ \small{Contours of  the RH problem for $S(z)$.}}
   \label{figure 5}
 \end{figure}

The Riemann-Hilbert problem for $T$ is oscillatory in the sense that
the jump matrix in \eqref{T jump} has  oscillating  diagonal
entries on the interval $(-1, 1)$. To remove the
oscillation, we introduce
 the second transformation $T\longrightarrow S$,  based on a factorization of
the oscillatory jump matrix
\begin{equation}\label{factory of T jump}
\left(
       \begin{array}{cc}
        \varphi_+^{-2n } & w \\
         0 &\varphi_-^{-2n }  \\
       \end{array}
     \right)=\left( \begin{array}{cc}
                                 1 & 0 \\
                                  \varphi_-^{-2n } w ^{-1} & 1 \\
                               \end{array}
                             \right)\left(
                              \begin{array}{cc}
                                0 & w  \\
                               -w ^{-1}& 0\\
                               \end{array}
                             \right)\left(
                              \begin{array}{cc}
                                1 & 0 \\
                                \varphi_+^{-2n } w ^{-1}  & 1 \\
                               \end{array}
                             \right),
\end{equation} where use has been made of the  fact that $\varphi_+(x)\varphi_-(x)=1$ for $x\in (-1, 1)$. Accordingly,
  we define a piecewise matrix-valued
function
\begin{equation}\label{T to S}
S(z)=\left \{
\begin{array}{ll}
  T(z), & \mbox{for $z$ outside the lens shaped region;}
  \\[0.4cm]
  T(z)\left(
                              \begin{array}{cc}
                                1 & 0 \\
                                -\varphi(z)^{-2n }w(z)^{-1}  & 1 \\
                               \end{array}
                             \right), & \mbox{for $z$ in the upper lens
                             region;}\\[0.4cm]
T(z) \left(
                              \begin{array}{cc}
                                1 & 0 \\
                                \varphi(z)^{-2n }w(z)^{-1}  & 1 \\
                               \end{array}
                             \right), & \mbox{for  $z$  in  the  lower
                             lens
                             region, }
\end{array}\right .\end{equation}where
the regions are depicted  in  Figure \ref{figure 5}, and
$$w(z)=(1-z^2)^{\beta} (t^2-z^2)^\alpha h(z),~~z\in \Omega \backslash \{  (-\infty, -1]\cup [1, \infty)\} $$  denotes the analytic continuation of $w(x)$,
with   $\arg (1\pm z)\in(-\pi, \pi)$ and $\arg (t\pm z)\in(-\pi, \pi)$, where $\Omega$ is the domain of analyticity of $h(z)$, such that $[-1, 1]\subset  \Omega$.

  Then  $S$ solves the Riemann-Hilbert problem:
\begin{description}
  \item(S1)~~ $S(z)$ is analytic  in  $\mathbb{C}\backslash\Sigma_S$, where $\Sigma_S$ are the deformed contours consisting of  $(-1, 1)$  and the upper and lower lens boundaries,  illustrated in Figure \ref{figure 5};
  \item(S2)~~The jump condition is
  \begin{equation}\label{S jump}S_+(x)=S_-(x)
  \left\{ \begin{array}{ll}
             \left(
                              \begin{array}{cc}
                                0 & w(x) \\
                               -w(x)^{-1}  & 0 \\
                               \end{array}
                             \right), &   \mbox{for}~x\in
  (-1,1),~\mbox{and}\\[.4cm]
           \left(
                              \begin{array}{cc}
                                1 & 0 \\
                                \varphi(z)^{-2n }w(z)^{-1}  & 1 \\
                               \end{array}
                             \right), &   \mbox{on~the~lens~boundaries};
          \end{array}\right .
 \end{equation}

  \item(S3)~~ The asymptotic behavior at infinity is
  \begin{equation}\label{}S(z)= I+O( 1/ z),~~~\mbox{as}~~z\rightarrow
                             \infty  ;\end{equation}
 \item(S4)~~  At the endpoints $\pm 1$, we have for $-1<\beta<0$
 \begin{equation}\label{S at 0 minus}S(z)=  \left(
                               \begin{array}{cc}
                                O( 1) & O\left ( (z\pm1)^{\beta}\right ) \\
                                O( 1) & O\left((z\pm1)^{\beta}\right )
                                  \\
                               \end{array}\right) \quad \mbox{as}\quad z\rightarrow\pm1,
\end{equation}
while for $\beta=0$,
\begin{equation}\label{S at 0-0}S(z)=  O(\ln(z\pm1) )   \quad \mbox{as}\quad z\rightarrow
                             \pm1,\end{equation}

and  for $\beta>0$,
 \begin{equation}\label{S at 0 plus}S(z)= \left\{\begin{array}{ll}
                                             \left(
                               \begin{array}{cc}
                             O\left ( (z\pm1)^{-\beta}\right ) & O( 1)\\
                                O\left((z\pm1)^{-\beta}\right )&O( 1)
                                  \\
                               \end{array}\right)& \mbox{as}~z\rightarrow\pm1, ~\mbox{inside of the lens,} \\[.4cm]
                                       O(1)
                                 & \mbox{as}~z\rightarrow
                             \pm1, ~\mbox{outside of the lens.}
                                          \end{array}\right.
\end{equation}

\end{description}

\subsection{ Global parametrix }\label{sec:3.4}

From \eqref{S jump}, we see  that the jump matrix for $S$ on the lens-shaped boundary  is of the
form $J_S(z)=I$,  plus an exponentially  small  term.
 The only jump of significance
is attached to $(-1, 1)$. We are now in a position to
solve the following  limiting Riemann-Hilbert problem for $N_t(z)$,
\begin{description}
\item(N1)~~  $N_t(z)$ is analytic  in  $\mathbb{C}\backslash
[-1,1]$;
\item(N2)~~ The jump condition is  \begin{equation}\label{N-t jump} \left (N_t\right )_+(x)=\left (N_t\right )_-(x)\left(
       \begin{array}{cc}
       0 & w(x) \\
       - {w(x)}^{-1} & 0 \\
       \end{array}
       \right)~~~\mbox{for}~~x\in  (-1,1);\end{equation}
\item(N3)~~ The asymptotic behavior at infinity is    \begin{equation}\label{} N_t(z)= I+O( 1/ z) ,~~~\mbox{as}~~z\rightarrow\infty .\end{equation}
    \end{description}

Since  $i\sigma_2=  M_1^{-1}  (-i \sigma_3)  M_1$, where $M_1=
(I+i\sigma_1) /{\sqrt{2}}$; see \eqref{Pauli-matrices} for the definition of the Pauli matrices,   a
solution to the above RH problem can be constructed explicitly as (cf. \cite{kv})
\begin{equation}\label{global parametrix}
N_t(z)=   D_t(\infty)^{\sigma_3}M_1^{-1}a (z)^{-\sigma_3}M_1 D_t(z)^{-\sigma_3}  ,\end{equation} where $a
(z)=\left(\frac{z-1} {z+1}\right)^{1/4}$ for $z\in \mathbb{C}\backslash [-1, 1]$, the
branches are  chosen such that $a (x) $ is  positive for $x>1$ and $a_+(x)/a_-(x)= i$ for $x\in (-1, 1)$, and the Szeg\"{o} function associated with  $w(x)$ takes the form
\begin{equation}\label{szego function}
D_t(z)=\left(\frac{z^2-1}{\varphi(z)^2}\right)^{\beta/2}\exp\left(\frac{\sqrt{z^2-1}}{2\pi}\int_{-1}^1\frac{\ln\left\{(t^2-x^2)^{\alpha}h(x)\right\}}{\sqrt{1-x^2}}\frac{dx}{z-x}\right), ~~z\in \mathbb{C}\backslash [-1, 1],
 \end{equation}
 which is a non-zero analytic function on   $\mathbb{C}\backslash [-1, 1]$ such that $\left ( D_t\right )_+(x) \left (D_t\right )_-(x)=w(x)$ for $x\in (-1, 1)$.    In   \eqref{szego function} the  principal branches  are taken, namely, $\arg (z\pm 1)\in (-\pi, \pi)$, and   $\varphi(z)$ for $z\in \mathbb{C}\backslash [-1, 1]$ is defined in   \eqref{Y to T}. It is readily seen that the limit at infinity is
  $$D_t(\infty)=2^{-\beta }  \exp \left(\frac 1{2\pi}\int_{-1}^1\frac{\ln\left\{(t^2-x^2)^\alpha h(x)\right\} \; dx }{\sqrt{1-x^2}} \right).$$

For each $t>1$,   the jump for $SN_t^{-1}$ is close to the unit matrix in the open curves $\Sigma_S\backslash\{\pm 1\}$, yet this is not true at    the endpoints:  $SN_t^{-1}$ is not even bounded near  $\pm1$.
Thus local parametrices have to be constructed  in neighborhoods of these endpoints.

\subsection{Local parametrix $P^{(1)}(z)$ }\label{sec:3.5}

In the present subsection, we focus on the construction of the
parametrix at the right endpoint $z=1$, or, more precisely,   in the neighborhood  $U(1,r)=\{z: ~|z-1|<r\} $,  $r$ being fixed and sufficiently small. The parametrix $P^{(1)}(z)$  should solve  the following  RH problem:
\begin{description}
  \item(a)~~ $P^{(1)}(z)$ is analytic in $U(1,r) \backslash  \Sigma_S$, where
  $\Sigma_S$ are the deformed contours depicted in  Figure \ref{figure 5};
  \item(b)~~ On $\Sigma_S \cap U(1,r)$, $P^{(1)}(z)$ satisfies the same jump conditions as $S(z)$ does, see \eqref{S jump};
  \item(c)~~  $P^{(1)}(z)$ fulfills the following  matching condition
   on  $\partial U(1,r)$:
\begin{equation}\label{matching condition}
P^{(1)}(z)N_t^{-1}(z)=I+ O\left (n^{-1}\right );
 \end{equation}
 \item(d)~~ The asymptotic behavior of  $P^{(1)}(z)$ at  the endpoint $z=1$ is as described in \eqref{S at 0 minus}-\eqref{S at 0 plus}.

\end{description}

To construct $P^{(1)}(z)$, we transform the RH problem for $P^{(1)}$ to a
new RH problem for $\hat{P}^{(1)}$, with constant jump matrices, as
\begin{equation}\label{P-1 to hat P-1}
\hat{P}^{(1)}(z)=P^{(1)}(z)\varphi(z)^{n\sigma_3}W(z)^{\frac 12\sigma_3},
 \end{equation}
in which
$$W(z)=(z^2-1)^{\beta}(z^2-t^2)^{\alpha }h(z),~~z\in \Omega\backslash (-\infty, t],$$ such that
 $\arg(z\pm1)\in(-\pi, \pi)$ and $\arg(z\pm t)\in(-\pi, \pi)$,   where $\Omega$ is the domain of analyticity of $h(z)$  such that $[-1, 1]\subset  \Omega$.
  We note that $W(z)$ is related to, but different from,  the function $w(z)$ introduced in \eqref{T to S}.
Then $\hat{P}^{(1)}$ solves the following RH problem:

\begin{description}
  \item(a)~~  $\hat{P}^{(1)}(z)$ is analytic in
 $U(1,r) \backslash  \Sigma_S$
(see Figure \ref{figure 5});
  \item(b)~~  $\hat{P}^{(1)}(z)$  possesses  the following constant  jumps
  \begin{equation}\label{hat p-1 jump}
  \hat{P}^{(1)}_{+}(z)=\hat{P}^{(1)}_{-}(z)
  \left\{\begin{array}{ll}
           \left(
                               \begin{array}{cc}
                                0 & 1 \\
                               -1 & 0\\
                                 \end{array}
                             \right), &  z \in (1-r, 1), \\[.4cm]
           \left(
                               \begin{array}{cc}
                                 1 & 0 \\
                               e^{(\alpha+\beta)\pi i} & 1 \\
                                 \end{array}
                             \right), &   \mbox{on~the~upper~lens~boundary}, \\[.4cm]
         \left(
                               \begin{array}{cc}
                                 1 & 0 \\
                               e^{-(\alpha+\beta)\pi i} & 1 \\
                                 \end{array}
                             \right),  &  \mbox{on~the~lower~lens~boundary}, \\[.4cm]
                                                          e^{\alpha\pi i\sigma_3},&
   z \in (1,t);
                                \end{array}\right .
   \end{equation}

 \item(c)~~The behavior of $\hat{P}^{(1)}(z)$  at $z=t$ is
 \begin{equation}\label{hat-P-1 at t} \hat{P}^{(1)}(z)=O(1)\; (z- t)^{\frac{\alpha}{2}\sigma_3}
                               , \quad\mbox{as} \quad   z\rightarrow t; \end{equation}
 \item(d)~~The behavior of $\hat{P}^{(1)}(z)$  at  $z=1$ is,   for $-1<\beta<0$,
\begin{equation}\label{hat P-1 at 1 minus}\hat{P}^{(1)}(z)=    O\left((z-1)^{\beta/2}\right )\quad\mbox{as}\quad z\rightarrow1,
\end{equation}while for $\beta=0$,
\begin{equation}\label{hat P-1 at 1-0}\hat{P}^{(1)}(z)=
                                O(\ln(z-1) )\quad \mbox{as}\quad z\rightarrow
                             1 ,\end{equation}

 and for $\beta>0$,
 \begin{equation}\label{hat P-1 at 1 plus}\hat{P}^{(1)}(z)= \left\{\begin{array}{ll}
                                               O\left ( (z-1)^{-\beta/2}\right ) & \mbox{as}\quad z\rightarrow 1,~\mbox{inside of the lens,} \\[.1cm]
                                       O\left ( 1 \right ) (z-1)^{\frac \beta 2\sigma_3}  & \mbox{as}\quad z\rightarrow
                             1,~\mbox{outside of the lens}.
                                          \end{array}\right.
\end{equation}

\end{description}
\vskip .5cm

The RH problem for $\hat{P}^{(1)}$ shares exactly  the same jumps
as those for $\Psi_0$ in \eqref{psi-0 jump}, with the parameter $\Theta=-\alpha$, $\gamma=\beta-\frac12$.
The behavior of  $\hat{P}^{(1)}$ at $z=t$ in \eqref{hat-P-1 at t} is the same as that of  $\Psi_0$ at $\zeta=\frac 14$ in \eqref{psi-0 at 1/4}.
We proceed to construct  $\hat{P}^{(1)}(z)$
out of   $\Psi_0(\zeta, s)$, bearing in mind  the matching condition
\eqref{matching condition}.

We define a  conformal mapping in   a $z$-neighborhood $U(1,r)$  of $z=1$ and $z=t$ as follows
\begin{equation}\label{conformal mapping}
f_t(z)=\frac{\left (\ln \varphi(z)\right )^2}{\rho_t}=\frac{2(z-1)}{\rho_t}(1+O(z-1)),\quad z\in U(1,r)
\end{equation}
with $f_t(1)=0$ and $f_t(t)=\frac14$, where  $\rho_t= 4 \left ( \ln \varphi(t)\right )^2=8(t-1)+O\left ( \left (t-1)^2\right )\right )$ as $t\rightarrow 1$.
Making use of the conformal mapping, we seek  a $\hat{P}^{(1)}$ of the form
\begin{equation}\label{hat-P-psi}
\hat{P}^{(1)}(z)=E(z)\Psi_0\left ( f_t(z),2n\sqrt{\rho_t }\right ),~~z\in U(1,r),
 \end{equation}     accordingly,
\begin{equation}\label{P-1-psi}
P^{(1)}(z)=E(z)\Psi_0\left (f_t(z),2n\sqrt{\rho_t }\right )\varphi(z)^{-n\sigma_3}W(z)^{-\frac 12\sigma_3} ,\end{equation} where $\Psi_0(\zeta)=\Psi_0(\zeta, s)$ is the solution to  the RH problem
\eqref{psi-0 jump}-\eqref{psi-0 at 1/4}, and $E(z)$
is an analytic matrix-valued  function in the neighborhood $U(1,r)$, to be determined by the matching condition \eqref{matching condition}.

First, we  introduce
\begin{equation}\label{E}
 E(z)=N_t(z) W(z)^{\frac 12\sigma_3}\left\{G(f_t(z))\right\}^{-1},\end{equation}
where $G(\zeta)$ is a specific matrix function  defined as
\begin{equation}\label{G}G(\zeta)=\zeta^{\frac 14 \sigma_3}\frac {I-i\sigma_1}{\sqrt{2}}\exp\left\{ \left ( \frac {\alpha\sqrt\zeta} 2 \int^{\frac 1 4}_0 \frac 1 {\sqrt \tau } \frac {d\tau}{\tau-\zeta}\right )\sigma_3\right\},~\zeta\in \mathbb{C}\backslash (-\infty, 1/4], \end{equation}with $\arg\zeta\in (-\pi, \pi)$,
and  satisfying  the jump conditions
\begin{equation}\label{G-jumps}
 G_+(x)=G_-(x)(i\sigma_2)~\mbox{for}~ x\in(-\infty,0),~~\mbox{and}~~  G_+(x)=G_-(x)e^{\pi i\alpha\sigma_3}~\mbox{for}~x\in(0,  1/4).\end{equation}

In view of \eqref{N-t jump} and  \eqref{G-jumps},
 it is readily verified from \eqref{E} that
\begin{equation*}\label{}
 E_+(x)
=E_-(x)~~\mbox{for}~~x\in U(1, r)\cap \mathbb{R},
\end{equation*}
where $r>t$. Hence $E(z)$ is analytic in  $U(1, r)\setminus\{1, t\}$.

Next, we show that $E(z)$ is also analytic at $z=1$ and $z=t$. Indeed,
it follows from  \eqref{global parametrix} and  \eqref{szego function} that
$$N_t(z) W(z)^{\frac 12\sigma_3}=O\left ( (z-1)^{-\frac 1 4}\right )~\mbox{as}~z\rightarrow 1, ~\mbox{and}~N_t(z) W(z)^{\frac 12\sigma_3}=O(1) (z-t)^{\frac \alpha2\sigma_3}~\mbox{as}~z\rightarrow t.$$

 Also, from  \eqref{conformal mapping} and   \eqref{G} we see that $G(f_t(z))=\left ( (z-1)^{-1/4}\right )$ as $z\rightarrow 1$,
 and the integral in \eqref{G} implies that $G(f_t(z))=O(1) (z-t)^{\frac \alpha2\sigma_3}$ as $z\rightarrow t$.
Substituting  these estimates into
\eqref{E} gives
\begin{equation*}\label{} E(z)=O\left ( (z-1)^{-1/2}\right )~~\mbox{as}~z\rightarrow 1,\end{equation*}
and
\begin{equation*}\label{} E(z)=O(1)~~\mbox{as}~z\rightarrow t,\end{equation*}
which means that $E(z)$ has at most isolated  weak singularities  at $z=1,t$, and hence the singularity is removable. Thus $E(z)$ is analytic in the neighborhood $U(1,r)$.

Now what remains is to show that the matching condition \eqref{matching condition} is fulfilled. To this end, we note that, from Section \ref{sec:5.1} below, $G(\zeta)$ solves a limiting RH problem  with jumps \eqref{G-jumps}, such that  $\Psi_0(\zeta, s)$ is approximated by $G(\zeta)$ for $s$ large, and $\zeta$ being kept away from the origin;
cf. \eqref{psi-G} below. Hence we have
\begin{equation}\label{psi-0 at infinity big t}
\Psi_0\left ( f_t(z),2n\sqrt{\rho_t }\right )\varphi(z)^{-n\sigma_3} = G (f_t(z))\left(I +O(\frac 1{n\sqrt{\rho_t }})\right),\end{equation}
uniformly for $z\in\partial U(1,r)$ as $n\sqrt{\rho_t }\to \infty$.
For $z\in\partial U(1,r)$,  we have
\begin{equation}\label{W-N  at circle}W^{\sigma_3}(z),~N_t^{-1}(z),~a^{\sigma_3}(z),~ D_t^{\sigma_3}(z)=O(1);\end{equation}
cf. Section \ref{sec:3.4} for the definitions of these quantities.
Thus by combining \eqref{P-1-psi} with \eqref{psi-0 at infinity big t} and \eqref{W-N  at circle}, we see that
\begin{equation}\label{matching condition estimate-big t}
P^{(1)} N_t^{-1}
  =  N_t W ^{\frac 12\sigma_3}\left (I+O\left(    \frac 1{n\sqrt{\rho_t }} \right)\right )
W ^{-\frac 12\sigma_3}N_t^{-1}
 =  I+O\left(   \frac 1 {n\sqrt{\rho_t }} \right),
\end{equation} for $|z-1|=r$, where
  $t$ is taken so that  $n\sqrt{\rho_t }\to \infty$. Thus the matching condition \eqref{matching condition} is fulfilled with obvious modification if $n\sqrt{\rho_t }$ is unbounded.

If  $n\sqrt{\rho_t }\approx 2\sqrt 2\;  n\sqrt{t-1}\in (0,\delta)$ bounded,  then the conformal mapping \eqref{conformal mapping} satisfies
\begin{equation}\label{conformal mapping big}
1/f_t(z) =O(1/n^2)
\end{equation}
for $|z-1|=r$.
Then, for $\zeta=f_t(z)\gg 1/4$,  the matrix function $G(\zeta)$ in  \eqref{G} is approximated as
\begin{equation}\label{G-small t}G(f_t(z))=(f_t(z))^{\frac 14 \sigma_3}\frac {I-i\sigma_1}{\sqrt{2}}\left (I+O\left(\frac 1 n\right )\right ).\end{equation}
Thus, combining  \eqref{G-small t} with   the expansion \eqref{psi-0 at infinity} of $\Psi_0$ at infinity and \eqref{psi-0 at circle small s}, we have
\begin{equation}\label{matching condition estimate-t}
P^{(1)} N_t^{-1}
  =  N_t W ^{\frac 12\sigma_3}\left (I+O\left(    \frac 1n \right)\right )
W ^{-\frac 12\sigma_3}N_t^{-1}
 =  I+O\left(   \frac 1n \right)
\end{equation} for $|z-1|=r$. Thus the matching condition \eqref{matching condition} is also fulfilled for bounded $n\sqrt{\rho_t }$.

We have completed the construction of the local parametrix    $P^{(1)}(z)$  at
the right edge $z=1$, in which  a generalized fifth Painlev\'{e} transcendent  is involved. Similarly, we can state and construct the parametrix $P^{(-1)}(z)$  at
the left edge $z=-1$.

\subsection{The final transformation $S\rightarrow R$}\label{sec:3.6}

Now we bring in the final transformation by defining
\begin{equation}\label{S to R}
R(z)=\left\{ \begin{array}{ll}
                S(z)N_t^{-1}(z), & z\in \mathbb{C}\backslash \left \{ U(-1,r)\cup U(1,r)\cup \Sigma_S \right \};\\[.1cm]
               S(z) (P^{(-1)})^{-1}(z), & z\in   U(-1,r)\backslash \Sigma_{P^{(-1)}} ;  \\[.1cm]
               S(z)  (P^{(1)})^{-1}(z), & z\in   U(1,r)\backslash
               \Sigma_{P^{(1)}} ;
             \end{array}\right.
\end{equation}
comparing Figure \ref{figure 6} for the regions involved, where $U(\pm1,r)$
are the   disks of radius $r$, centered
respectively at $\pm1$. So defined, the matrix-valued function
$R(z)$ satisfies a Riemann-Hilbert problem on the remaining contours
$\Sigma_R$ illustrated in Figure \ref{figure 6}, as follows:
 \begin{figure}[t]
 \begin{center}
   \includegraphics[width=9cm]{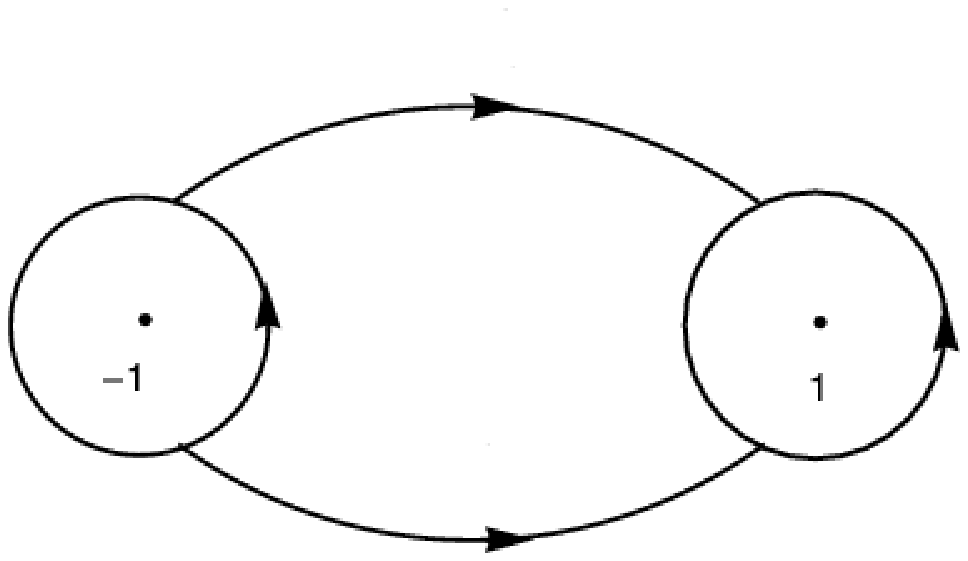} \end{center}
 \caption{\small{The remaining contours $\Sigma_R$: contours of the RH problem for $R(z)$.}}
 \label{figure 6}
 \end{figure}
\begin{description}
  \item(R1)~~ $R(z)$ is analytic in $\mathbb{C}  \backslash \Sigma_R$ (see Figure \ref{figure 6});
  \item(R2)~~ $R(z)$ satisfies the  jump conditions
  \begin{equation}\label{R jump}
R_+(z)=R_-(z)J_{R}(z), ~~z\in\Sigma_R,\end{equation}where
$$J_R(z)=\left\{ \begin{array}{ll}
                   N_t(z) \left(P^{(-1)}\right)^{-1}(z), ~ &   z\in\partial U(-1,r),\\[.1cm]
                      N_t(z) \left(P^{(1)}\right)^{-1}(z),&  z\in\partial
                    U(1,r),\\[.1cm]
N_t(z)J_{S}(z)N_t^{-1}(z), ~& \mbox{otherwise},
                 \end{array}\right .$$
                 where $J_S(z)$ is the jump for $S$, given in \eqref{S jump};
  \item(R3)~~  $R(z)$ demonstrates the following behavior at infinity:
  \begin{equation}\label{R at infinity}
  R(z)= I+ O\left({1}/{z} \right),~~\mbox{as}~ z\rightarrow\infty .
  \end{equation}
\end{description}

We note that  $R(z)$ has    removable singularities  at $z=\pm1$. Indeed, since $S(z)$ and   $P^{(1)}(z)$ share the same jump with in $U(1,r)$,
  we see that  $z=1$  is at most an isolated singularity for  $R(z)$. Again, $S(z)$ and   $P^{(1)}(z)$ satisfy the same behavior \eqref{S at 0 minus}-\eqref{S at 0 plus}, hence we have
  $$R(z)=O ((z-1)^\beta )~\mbox{as}~z\to 1,~-1<\beta<0~~\mbox{and}~~R(z)=O\left ((\ln(z))^2\right )~\mbox{as}~z\to 1,~\beta=0.$$
While for  the case $\beta>0$, we have
 \begin{equation*}
  R(z)=\left\{
  \begin{array}{ll}
    O\left ((z-1)^{-\beta}\right ), &\mbox{as}~z\to 1~\mbox{from inside of the lens}, \\
    O(1), &\mbox{as}~z\to 1~\mbox{from outside of the lens},
  \end{array}\right .\end{equation*}
Thus in each case, $R(z)$ has removable singularity at $z=1$. Similar argument applies to $z=-1$.

It follows from the matching condition \eqref{matching condition} of the local parametrices, the definition of $\varphi$, and the definition of $N_t$ in
\eqref{N-t jump} that
\begin{equation}\label{R-jump-error}J_R(z)=\left\{ \begin{array}{ll}
                     I+O(n^{-1}),&  z\in\partial
                    U(\pm1,r),\\[.1cm]
I+O(e^{-cn}), ~& z\in  \Sigma_R\backslash \partial
                    U(\pm1,r),
                 \end{array}\right .
\end{equation}
where $c$ is a positive constant, and the error term is uniform
for $z$ on the corresponding contours. Hence we have
\begin{equation}\label{R-jump estimates}\|J_R(z)-I\|_{L^2\cap L^{\infty}(\Sigma_R)}=O(n^{-1}).
\end{equation}
Then, applying  the now standard  procedure of norm
estimation  of Cauchy operator and using the technique of deformation of contours (cf. \cite{deift,dkmv2}), it
follows from  \eqref{R-jump estimates} that
 \begin{equation}\label{R estimate}R(z)=I+O(n^{-1}),\end{equation} uniformly for $z$ in the whole complex plane.

This completes the nonlinear steepest descent analysis. In the next section. we will show that the  orthogonal polynomial  kernel  (the Christoffel-Darboux kernel) can be represented  in terms of
the solution to
 the RH problem for $Y$, formulated at
 the very
beginning of this  section. The large-$n$ asymptotic
behavior of the kernel can then be obtained.

\section{  Proof of Theorem \ref{thm-limit-kernel}} \indent\setcounter{section} {4}
\setcounter{equation} {0} \label{sec:4}
The orthonormal polynomials $p_n(z)=\gamma_n\pi_n(z)$ satisfy  the
three-term recurrence relation
\begin{equation}\label{recurrence relation}
B_{n}p_{n+1}(z)+(A_{n}-z)p_{n}(z)+B_{n-1}p_{n-1}(z)=0,~~n=0,1,\cdots,
\end{equation}where $B_{-1}=0$, and $B_{n}=\gamma_{n}/\gamma_{n+1}$ for $n=0,1,2, \cdots$.
  From the three-term recurrence relation, it is
readily seen that  the following Christoffel-Darboux formula holds (see, e.g., \cite{s}):
\begin{equation}\label{c-d formula}
\sum_{k=0}^{n-1}p_k(x)p_k(y)=\gamma^2_{n-1}\frac{\pi_n(x)\pi_{n-1}(y)-\pi_n(y)\pi_{n-1}(x)}{x-y}.
\end{equation}
Hence,  in terms of the matrix-valued function $Y(z)$ defined in \eqref{Y polynomial solution},  the kernel $K_n(x,y)$ in
\eqref{kernel-formula} can be written as
\begin{equation}\label{kernel in terms of Y}
K_n(x,y)=\frac{\sqrt{w(x)w(y)}}{2\pi i(x-y)}\left\{
Y_+^{-1}(y)Y_+(x)\right \}_{21},~~~x,y\in(-1,1).
\end{equation}

\subsection{Proof of Theorem \ref{thm-limit-kernel} (ii): The sine kernel limit}
Assume that $x, y\in I_\delta=[-1+\delta, 1-\delta]$, with $\delta >0$ fixed, such that $0<r<\delta$, where $r$ is  the radius of  $U(\pm 1, r)$. Substituting the transformations \eqref{Y to T} and \eqref{T to S} to \eqref{kernel in terms of Y}, we have
\begin{equation}\label{kernel in S}
K_n(x,y)=\frac{\sqrt{w(x)w(y)}}{2\pi i(x-y)}\left\{
\left(
  \begin{array}{cc}
    1 & 0 \\
    \frac {-1} {w(y)} & 1 \\
  \end{array}
\right)\varphi_+^{-n\sigma_3}(y)
S_+^{-1}(y)S_+(x)
\varphi_+^{n\sigma_3}(x)
\left(
  \begin{array}{cc}
    1 & 0 \\
     \frac 1 {w(x)} & 1 \\
  \end{array}
\right)
\right \}_{21}.
\end{equation}
On the other hand, in view of  the jump condition \eqref{R jump} and the uniform estimate \eqref{R-jump estimates}, we have
$$
R(z)=I +\frac 1{2\pi i}\int_{\Sigma_R}\frac {R_-(\zeta) \left ( J_R(\zeta) -I\right ) d\zeta} {\zeta-z},~~z\not\in \Sigma_R,
$$ from which we conclude that
$$
R(x)=I +O\left (\frac 1 n\right ),~~\left . \frac {d R}{dz} \right |_{z=x} =O\left (\frac 1 n\right ),
$$ as long as  $I_\delta$ keeps a constant distance from $\Sigma_R$.
Hence we see that
$$
R^{-1}(y)R(x)=I +O\left ( (x-y)/n \right ),
$$ uniformly for $x, y \in I_\delta$. Observing that both  $\left (N_t\right )_+$ and   $\frac {d}{dx}\left (N_t\right )_+$  are uniformly of $O(1)$ on $I_\delta$, and accordingly
$\left ( N_t\right )_+^{-1}(y) \left ( N_t\right )_+(x)= I+O(x-y)$, uniformly again for  $x, y\in I_\delta$, and combining  these with \eqref{S to R},
we have
\begin{equation}\label{S-1-S}
S_+^{-1}(y)S_+(x)
=\left ( N_t\right )_+^{-1}(y)R^{-1}(y)R(x)\left ( N_t\right )_+(x)=I+O(x-y).
\end{equation}
Substituting \eqref{S-1-S} in \eqref{kernel in S}   then yields
\begin{equation}\label{kernel in phi}
K_n(x,y)=\frac{1}{2\pi i(x-y)}\left [
\sqrt{\frac {w(y)}{w(x)}} \left (\frac  {\varphi_+(y)}    {\varphi_+(x)} \right )^n -  \sqrt{\frac {w(x)}{w(y)}} \left (\frac  {\varphi_+(x)}    {\varphi_+(y)} \right )^n
\right ]+O(1),
\end{equation} uniformly for $x,y\in I_\delta$. In deriving  \eqref{kernel in phi}, use has been made of the fact that $\varphi_+(x)=e^{i\arccos x}$, so that $| \varphi_+(x)|=1$. Noting that $\sqrt{w(x)/w(y)},~\sqrt{w(y)/w(x)}=1+O(x-y)$, from  \eqref{kernel in phi}  we
further obtain
\begin{equation}\label{kernel in sine}
K_n(x,y)=\frac{\sin\left [ n(\arccos y-\arccos x)\right ]}{ \pi  (x-y)}+O(1).
\end{equation}
It then readily follows that
\begin{equation}\label{kernel-in-sine-final}
\frac {\pi \sqrt{1-x^2}} n \; K_n\left (x+\frac{\pi \sqrt {1-x^2} \; u} n ,x+\frac{\pi \sqrt {1-x^2} \; v} n\right )=\frac{\sin\{\pi (u-v)\}  }{ \pi (u-v)}    +O\left (\frac 1 n\right )
\end{equation}
by expanding
$\arccos(x+t)=\arccos x-t/\sqrt{1- x^2}+\cdots$ for fixed $x\in (-1, 1)$ and small $t$.  The large-$n$ limit \eqref{kernel-in-sine-final}  holds uniformly for  bounded real $u$ and $v$.  Thus we complete the proof of  Theorem \ref{thm-limit-kernel} (ii): The sine kernel limit.  We see that the   universality property
 is preserved in the bulk of the spectrum; cf. \cite{deift} and \cite{kv}.

\subsection{Proof of Theorem \ref{thm-limit-kernel} (i): The limiting eigenvalue density}
The  $O(1)$ term in  \eqref{kernel in sine} is uniform with respect to all  $x, y\in [-1+\delta, 1-\delta]$ for positive $\delta$.  Hence we can take the limit $y\rightarrow x$. As a result, we have
 \begin{equation}\label{limiting eigenvalue density}
K_n(x,x)=\frac{n}{ \pi  \sqrt{1-x^2}}+O(1),
\end{equation}
 for $x\in (-1, 1)$ fixed and $n\rightarrow\infty$. This proves Theorem \ref{thm-limit-kernel} (i), as stated in \eqref{limiting eigenvalue density introduction}.

\subsection{Proof of Theorem \ref{thm-limit-kernel} (iii): The Painlev\'{e} kernel limit}\label{sec:4.3}
Now we turn to the neighborhood $U(1, r)=\{z: ~|z-1|<r\}$, in which the parametrix $P^{(1)}(z)$ is constructed. A combination of
\eqref{Y to T}, \eqref{T to S}, \eqref{P-1-psi} and \eqref{S to R} gives
 $$Y_+(x)=2^{-n\sigma_3} R(x) E(x)\left\{ \left (\Psi_0\right )_+ (f_t(x), s)\right\} e^{-\frac {i\pi} 2 \left (\alpha+\beta\right )\sigma_3}\left(
                                                                                                                \begin{array}{cc}
                                                                                                                  1 & 0 \\
                                                                                                                  1 & 1 \\
                                                                                                                \end{array}
                                                                                                              \right)w(x)^{-\frac 1 2\sigma_3},~~1-r<x<1,$$where $s=2n\sqrt{\rho_t}=4n\ln\varphi(t)$; cf. \eqref{conformal mapping}.
Substituting it into \eqref{kernel in terms of Y},  we have
\begin{equation}\label{kernel-psi}
K_n(x,y)= \frac{
\left(
    - \psi_2\left (f_t(y)\right ), \psi_1\left (f_t(y)\right )\right )
E^{-1}(y)R^{-1}(y)R(x)E(x)
\left(
     \psi_1\left (f_t(x)\right ), \psi_2\left (f_t(x)\right)
\right)^T}{2\pi i(x-y)},
 \end{equation}
 where
 \begin{equation}\label{psi-Psi-0}
  \left(
                                \begin{array}{c}
                                  \psi_1(\zeta) \\
                                  \psi_2(\zeta) \\
                                \end{array}
                              \right) =
 \left(
                                \begin{array}{c}
                                  \psi_1(\zeta,s) \\
                                  \psi_2(\zeta,s) \\
                                \end{array}
                              \right)
 =\left (\Psi_0\right )_+ \left( \zeta,s\right )         \left(
                                                           \begin{array}{cc}
                                                           e^{- \frac{\pi i}{2} (\alpha+\beta)}    & e^{  \frac{\pi i}{2} (\alpha+\beta)}  \\
                                                           \end{array}
                                                         \right)^T\end{equation} for  $\zeta<0$
 and  $s=2n\sqrt{\rho_t}$.

Now specifying
\begin{equation}\label{u,v domain}
x=1- \frac {\rho_t} 2 u,~~y=1-\frac {\rho_t} 2v~~\mbox{with}~~u,v\in {D}, \end{equation} with $s^2{D}$ being a compact subset of $(0, +\infty)$,
where again  $s=2n\sqrt{\rho_t}=4n\ln\varphi(t)\approx 4\sqrt 2\; n\sqrt{ t-1 }$.
Then it follows from \eqref{conformal mapping} and \eqref{u,v domain} that
\begin{equation}\label{f-t(u)}f_t(x)=-u\left (1+O\left (n^{-2}\right )\right ),~~f_t(y)=-v\left (1+O\left (n^{-2}\right )\right ),\end{equation}
where the $O\left (n^{-2}\right )$ terms  are  uniform respectively for $u,v\in {D}$.

Since $E(z)$ is a matrix function analytic in $U(1, r)$, we have
\begin{equation*}\label{E difference}E(y)^{-1}E(x)=I+E(y)^{-1}(E(x)-E(y))=I+O(x-y)=I+O\left (n^{-2}\right ) \end{equation*}
for the above specified $x$ and $y$. The $O\left (n^{-2}\right )$ term is uniform $u,v\in {D}$.
Similarly, the analyticity  of $R(z)$ in $U(1, r)$  implies that
\begin{equation*}\label{R difference}R(y)^{-1}R(x)=I+O(x-y)=I+O\left (n^{-2}\right ),\end{equation*}
again with uniform    error terms. Hence, substituting all these into \eqref{kernel-psi} yields
\begin{equation}\label{kernel-psi-uniform}
K_n(x,y)= \frac{
\left(
    - \psi_2\left (f_t(y)\right ), \psi_1\left (f_t(y)\right )\right )
\left (I+O(x-y)\right )
\left(
     \psi_1\left (f_t(x)\right ), \psi_2\left (f_t(x)\right)
\right)^T}{2\pi i(x-y)},
 \end{equation}
the error term is actually uniform for $t\in (1, d]$ and for $1-r<x, y< 1$, with $d>1$ and $r>0$ being constants.

Now we consider the double scaling limit when $n^2 (t-1)$ approaches  a positive number as $n\to \infty$ and $t\to 1^+$. In such a case, we can regard $s$ as a positive constant.
The formula \eqref{f-t(u)} implies that
\begin{equation}\label{psi(u)} \psi_k(f_t(x),s)=\psi_k(-u,s)\left (1+O\left (n^{-2}\right )\right )~~\mbox{and}~~ \psi_k(f_t(y),s)=\psi_k(-v,s)\left (1+O\left (n^{-2}\right )\right )\end{equation}
 for $k=1,2$, where the error terms are uniform for  $u,v$ in compact subsets of $(0, \infty)$.

Thus, in view of  the fact that $\frac {\rho_t} 2=\frac {s^2} {8n^2}$,  a combination  of \eqref{kernel-psi-uniform} and  \eqref{psi(u)}  gives
\begin{equation}\label{psi-kernel}
\frac {s^2} {8n^2} K_n\left (1- \frac {s^2u} {8n^2}  , 1-\frac {s^2v} {8n^2} \right )=
K_{\Psi}(-u,-v;s)+O\left (\frac 1 {n^2}\right ),
 \end{equation}for large $n$,
where $$K_{\Psi}(-u,-v;s)=\frac{  \psi_1(-u,s)\psi_2(-v,s)- \psi_1(-v,s)\psi_2(-u,s)   }{2\pi i(u-v)}$$ is the Painlev\'{e} type kernel, and the error term $O\left (n^{-2}\right)$ is uniform for $u,v$ in compact subsets of $(0, \infty)$, and $t-1=O(1/n^2)$.  Thus completing the proof of   Theorem \ref{thm-limit-kernel}.

\section {Transition to the Bessel kernel $\mathbb{J}_\beta$ as $s\to \infty$}\label{sec:5}

When $t>1$ fixed, the weight in \eqref{p-jacobi weight} can  be written as $w(x)=(1-x^2)^{\beta} h_1(x)$, where $h_1(z)=(t^2-z^2)^\alpha h(z)$ is an analytic function for $z \in \Omega\backslash\{ (-\infty, -t]\cup [t, \infty)\}$, $\Omega$ being a neighborhood of $[-1, 1]$. This is a special case investigated in \cite{kmvv,kv}. The local behavior at $x=1$ is described via the kernel  $\mathbb{J}_{\beta}$ given  in \eqref{bessel-kernel}.

In the $\Psi$-kernel $K_{\Psi}(u,v;s)$,  we use the parameter $s=4n\ln\left ( t+\sqrt{t^2-1}\right )$ to describe the location of $t$. As $t$ varies to $d>1$ fixed, the parameter
$s\to \infty$ as $n\to\infty$. In the present  section, we begin with an asymptotic study   of the model RH problem for $\Psi_0(\zeta,s)$ with specified  parameters $\Theta=-\alpha$ and $\gamma=\beta-\frac 12$, and  as $s\to \infty$. Then, we apply the results to obtain a transition of the limit kernel from   $K_{\Psi}(u,v;s)$ to the classical Bessel kernel $\mathbb{J}_{\beta}$, as $s\to \infty$. As a by-product, we  obtain the asymptotics for the nonlinear equation $b(s)$, $u(s)$, and $y(s)$.

\subsection {Nonlinear steepest descent analysis of the RH problem for $\Psi_0(\zeta,s)$ as $s\to \infty$}\label{sec:5.1}

Taking the    normalization of  $\Psi_0(\zeta,s)$ at infinity  as
\begin{equation}\label{psi to U}
U(\zeta, s)=\Psi_0(\zeta,s) e^{-\frac {s\sqrt{\zeta}}2\sigma_3},~~\arg \zeta \in (-\pi, \pi),
 \end{equation}where $\Psi_0(\zeta,s)$ solves the model RH problem \eqref{psi-0 jump}-\eqref{psi-0 at 1/4},
we see that $U(\zeta, s)$ ($U(\zeta)$, for short) satisfies  the following RH problem:

\begin{description}
  \item(a)~~  $U(\zeta)$ is analytic in
  $\mathbb{C}\backslash\cup^4_{j=1}\Sigma_j$ (see Figure \ref{figure 1});

  \item(b)~~  $U(\zeta)$  satisfies the jump conditions,
   \begin{equation}\label{U-jump}
   U_+(\zeta)=U_-(\zeta)
   \left\{ \begin{array}{ll}
            e^{\pi i\alpha\sigma_3},   &  \mbox{for} \  \zeta \in \Sigma_1 , \\[.4cm]
            \left(
                               \begin{array}{cc}
                                 1 & 0 \\
                                 e^{-s\sqrt{\zeta}+\pi i(\alpha+\beta)} & 1 \\
                                 \end{array}
                             \right),   &   \mbox{for} \  \zeta \in \Sigma_2 ,\\[.4cm]
                       \left(
                               \begin{array}{cc}
                                 0 & 1 \\
                                 -1 & 0 \\
                                 \end{array}
                             \right), &    \mbox{for} \  \zeta \in \Sigma_3 ,  \\[.4cm]
                         \left(
                               \begin{array}{cc}
                                 1 & 0 \\
                                 e^{-s\sqrt{\zeta}-\pi i(\alpha+\beta)} & 1 \\
                                 \end{array}
                             \right),&    \mbox{for} \  \zeta \in \Sigma_4 ;
           \end{array}\right .
  \end{equation}

\item(c)~~  The asymptotic behavior of $U(\zeta)$  at infinity
  is
  \begin{equation}\label{U at infinity}U(\zeta)=\zeta^{\frac{1}{4}\sigma_3}\frac{I-i\sigma_1}{\sqrt{2}}
    \left (I+O\left (\frac 1{\sqrt{ \zeta}}\right )\right),~~\arg\zeta\in (-\pi, \pi)
  ;\end{equation}
\item(d)~~The behavior of $U(\zeta)$  at the origin is
\begin{equation}\label{U at 0}
U(\zeta) = O\left(1\right)\zeta^{ \frac 1 2\beta \sigma_3}
\left(
                                                                         \begin{array}{cc}
                                                                          O(1) &  O(1+c\ln \zeta) \\
                                                                           0 & O(1)\\
                                                                         \end{array}
                                                                       \right),
\end{equation}
for $\zeta\in \Omega_4$, $\zeta\to 0$, and the behavior in other sectors can be  determined by  the jump condition \eqref{U-jump}. Here  $c$ is given in \eqref{c-monodromy} such that $c=0$ for $\beta\not\in \mathbb{N}$;

\item(e)~~The behavior of $U(\zeta)$  at $\zeta=\frac 14$ is

\begin{equation}\label{U at 1/4}
U(\zeta)=  \widehat{\Psi}^{(0)}(\zeta)(\zeta-  1/4)^{-\frac 1 2 \Theta\sigma_3} e^{\frac {-s\sqrt{\zeta}}2\sigma_3},
\end{equation}where $\widehat{\Psi}^{(0)}(\zeta)$ is analytic at $\frac 14$.
\end{description}

We observe that the jumps along $\Sigma_2$ and $\Sigma_4$ in \eqref{U-jump} differ from the identical matrix by exponentially small errors, as $s\rightarrow +\infty$ and $\zeta$ being kept away from the origin. Hence, we may consider
  the following limiting RH problem for $G$:

\begin{description}
  \item(a)~~  $G(\zeta)$ is analytic in
  $\mathbb{C}\backslash\left\{\Sigma_1\cup\Sigma_3\right \}$ (see Figure \ref{figure 1});

  \item(b)~~  $G(\zeta)$  satisfies the jump conditions
   \begin{equation}\label{G-jump}
   G_+(\zeta)= G_-(\zeta)
   \left\{\begin{array}{ll}
           e^{\pi i\alpha\sigma_3}, & \mbox{for} \  \zeta \in \Sigma_1=(0, 1/4), \\[.4cm]
             \left(
                               \begin{array}{cc}
                                 0 & 1 \\
                                 -1 & 0 \\
                                 \end{array}
                             \right), &   \mbox{for} \  \zeta \in \Sigma_3=(-\infty, 0);
          \end{array}
   \right .
 \end{equation}

\item(c)~~  The asymptotic behavior of $G(\zeta)$  at infinity
  is
  \begin{equation}\label{G-infty}G(\zeta)=\zeta^{\frac{1}{4}\sigma_3}\frac{I-i\sigma_1}{\sqrt{2}}
    \left (I+O\left (\frac 1{\sqrt{ \zeta}}\right )\right)
  .\end{equation}
\end{description}
\vskip .2cm

For this
  RH problem with simple jump curve, a  solution  can be constructed as
\begin{equation}\label{G-large-s}G(\zeta)=\zeta^{\frac{1}{4}\sigma_3}\frac{I-i\sigma_1}{\sqrt{2}}
  \exp\left\{\left ( \frac {\alpha\sqrt\zeta} 2 \int^{1/4}_0 \frac 1 {\sqrt \tau} \frac {d\tau}{\tau-\zeta}\right )\sigma_3\right\}~~\mbox{for}~~\zeta\in \mathbb{C}\backslash (-\infty, 1/4],\end{equation}
where branches are chosen such that $\arg\zeta\in (-\pi, \pi)$.

At the origin, $G(\zeta)$ is no longer a good approximation of $U(\zeta)$; cf. the jumps \eqref{U-jump} and \eqref{G-jump}.
Hence, in the disk $|\zeta|<1/4$, we
 consider  a local parametrix $P^{(0)}(\zeta)$ , which obeys the same jump conditions \eqref{U-jump} and the same behavior \eqref{U at 0} at the origin as $U(\zeta)$, and fulfills the following matching condition at the boundary of the disk:
\begin{equation}\label{matching condition-2}P^{(0)}(\zeta)\sim G(\zeta)  \quad\mbox{as}~ |\zeta|=1/4.\end{equation}
We seek a solution, involving a re-scaling  of the variable,  of the form
\begin{equation}\label{P-0}P^{(0)}(\zeta)=E_1(\zeta)\Phi\left (\frac 1{16} {s^2\zeta} \right )\left\{\begin{array}{ll}
                                                     e^{-\frac {s\sqrt{\zeta}-\pi i\alpha}2\sigma_3}, \quad &\arg \zeta\in (0, \pi),\\
                                                     e^{-\frac {s\sqrt{\zeta}+\pi i\alpha}2\sigma_3},\quad &\arg \zeta\in (-\pi, 0),
                                                   \end{array}
\right. \end{equation}
with  analytic $E_1$  in the disk $|\zeta|<1/4$.  Here  $\Phi(\zeta)$ solves the following model RH problem:
\begin{figure}[t]
 \begin{center}
   \includegraphics[width=6cm]{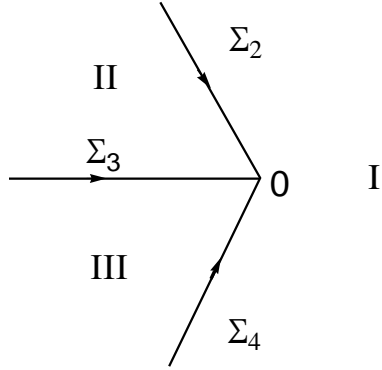} \end{center}
 \caption{\small{Contours and regions of the Bessel RH problem for $\Phi(\zeta)$.}}
 \label{figure 7}
 \end{figure}
\begin{description}
  \item(a)~~  $\Phi(\zeta)$ is analytic in
  $\mathbb{C}\backslash\cup^4_{j=2}\Sigma_j$ (see Figure \ref{figure 7});

  \item(b)~~  $\Phi(\zeta)$  satisfies the jump condition with parameter
   \begin{equation}\label{phi-jump}
  \Phi_+(\zeta)=\Phi_-(\zeta)
  \left\{ \begin{array}{ll}
            \left(
                               \begin{array}{cc}
                                 1 & 0 \\
                                 e^{\beta\pi i} & 1 \\
                                 \end{array}
                             \right), &   \mbox{for} \  \zeta \in \Sigma_2 ,\\[.4cm]
            \left(
                               \begin{array}{cc}
                                 0 & 1 \\
                                 -1 & 0 \\
                                 \end{array}
                             \right), &   \mbox{for} \  \zeta \in \Sigma_3,\\[.4cm]
                       \left(
                               \begin{array}{cc}
                                 1 & 0 \\
                                 e^{-\beta\pi i} & 1 \\
                                 \end{array}
                             \right), &     \mbox{for} \  \zeta \in \Sigma_4 ;
                         \end{array}
 \right .\end{equation}

\item(c)~~  The asymptotic behavior of $\Phi(\zeta)$  at infinity
  is
    \begin{equation}\label{phi at infinity}
  \begin{array}{rcl}
  \Phi(\zeta)&= &   \left (4\pi^2\zeta\right )^{-\frac 1 4 \sigma_3}\frac{I+i\sigma_1}{\sqrt{2}}\left (I+O\left (  \frac 1 {\sqrt{\zeta}}\right )\right )
    e^{2\sqrt \zeta\sigma_3 }\\[.2cm]
      &    =& i\sigma_1 \left (4\pi^2\zeta\right )^{\frac 1 4 \sigma_3}\frac{I-i\sigma_1}{\sqrt{2}}\left (I+O\left (  \frac 1 {\sqrt{\zeta}}\right )\right )
    e^{2\sqrt \zeta\sigma_3 }
  \end{array}
 \end{equation} for $\arg\zeta\in (-\pi, \pi)$, $\zeta\rightarrow\infty$.
\end{description}
\vskip .3cm

 A  solution to the RH problem for $\Phi$ can be constructed in terms of the modified Bessel functions as
 \begin{equation}\label{Bessel parametrix}
 \Phi(\zeta)=
 \left\{
 \begin{array}{ll}
   \left(
                               \begin{array}{cc}
                                 I_{\beta}(2\sqrt{\zeta}) &\frac{i}{\pi} K_{\beta}(2\sqrt{\zeta})  \\
                                 2\pi i\sqrt{\zeta}    I'_{\beta}(2\sqrt{\zeta})& -2\sqrt{\zeta}    K'_{\beta}(2\sqrt{\zeta}) \\
                                 \end{array}
                             \right),  &   \mbox{for} ~  \zeta \in I, \\[0.4cm]
  \left(
                               \begin{array}{cc}
                                 I_{\beta}(2\sqrt{\zeta}) &\frac{i}{\pi} K_{\beta}(2\sqrt{\zeta})  \\
                                 2\pi i\sqrt{\zeta}    I'_{\beta}(2\sqrt{\zeta})& -2\sqrt{\zeta}    K'_{\beta}(2\sqrt{\zeta}) \\
                                 \end{array}
                             \right)  \left(
                               \begin{array}{cc}
                                 1 & 0 \\
                                 -e^{\beta\pi i} & 1 \\
                                 \end{array}
                             \right),    & \mbox{for}~  \zeta\in II,  \\[0.4cm]
   \left(
                               \begin{array}{cc}
                                 I_{\beta}(2\sqrt{\zeta}) &\frac{i}{\pi} K_{\beta}(2\sqrt{\zeta})  \\
                                 2\pi i\sqrt{\zeta}    I'_{\beta}(2\sqrt{\zeta})& -2\sqrt{\zeta}    K'_{\beta}(2\sqrt{\zeta}) \\
                                 \end{array}
                             \right)\left(
                               \begin{array}{cc}
                                 1 & 0 \\
                                 e^{-\beta\pi i} & 1 \\
                                 \end{array}
                             \right), & \mbox{for}~\zeta\in III,
 \end{array}
 \right . \end{equation}where $\arg z\in (-\pi, \pi)$;
see Figure \ref{figure 7} for the regions, and see \cite{kmvv} for such a construction.
To verify the jump condition, one may use the analytic continuation   formulas  in \cite[(9.6.30)-(9.6.31)]{as}. The asymptotic at infinity \eqref{phi at infinity} can be obtained by expanding the Bessel functions asymptotically in sectors; see  \cite[(9.6.31) and Sec. 9.7]{as}, see also \cite{kmvv}.

Taking into consideration  the  matching condition
\eqref{matching condition-2},  and the asymptotic approximation  of $\Phi$ at infinity\eqref{phi at infinity}, we chose
 \begin{equation}\label{E-1}E_1(\zeta)=G(\zeta)\left\{\begin{array}{ll}
                                                     e^{ -\frac 1 2 \pi i\alpha \sigma_3}\frac{I-i\sigma_1}{\sqrt{2}} \left ( \frac {\pi^2} 4 s^2\zeta\right )^{\frac{1}{4}\sigma_3}, \quad& \arg \zeta\in (0, \pi),\\[.2cm]
                                                     e^{\frac 1 2 \pi i\alpha\sigma_3}\frac{I-i\sigma_1}{\sqrt{2}}\left ( \frac {\pi^2} 4 s^2\zeta\right )^{\frac{1}{4}\sigma_3}, \quad&   \arg \zeta\in (-\pi, 0),
                                                   \end{array}
\right.\end{equation}
which is analytic in  the disk $|\zeta|< 1/4$. Indeed, it is readily verified  that $E_1(\zeta)$ has no jump on $\Sigma_1\cup\Sigma_3$, and hence has at most an isolated singularity at the origin. Furthermore, from the fact that
 $\int^{\frac 1 4}_0\frac 1 {\sqrt\tau} \frac {d\tau}{\tau-\zeta} =\int^{\infty}_0\frac 1 {\sqrt\tau} \frac {d\tau}{\tau-\zeta}- \int^\infty_{\frac 1 4}\frac 1 {\sqrt\tau} \frac {d\tau}{\tau-\zeta}=\frac {\pi i}{\sqrt\zeta}+O(1)$ for $\zeta \sim 0$ and $\arg\zeta\in (0, 2\pi)$,   we see that the possible singularity of  $E_1(\zeta)$ at $\zeta=0$ is weak, and hence is removable.

We note that $P^{(0)}(\zeta)$ given in \eqref{P-0} behaves the same as $U(\zeta)$ in \eqref{U at 0} at the origin. Indeed, for $\zeta\in I$ ($|\arg \zeta|< 2\pi/3$), from \eqref{P-0} and \eqref{Bessel parametrix} we have
$P^{(0)}(\zeta)  =O(1) \zeta^{\frac 1 2 \beta \sigma_3}$ as $\zeta\to 0$.  Combining  it with    \eqref{U at 0}, we see that $U(\zeta) (P^{(0)})^{-1}(\zeta)$ has     isolated singularity at $\zeta=0$, such that
$U(\zeta) (P^{(0)})^{-1}(\zeta)=O\left (\zeta^{-  1/ 2}\ln \zeta\right )$ for $\zeta\to 0$ and $\zeta\in I$, so far as $\beta>-1$. The same order estimates can be obtained in other sectors. Therefore,    the singularity at the origin is weak, and is hence removable.

We are now in a position to introduce
\begin{equation}\label{R_1}R_1(\zeta)=\left\{\begin{array}{ll}
                                                     U(\zeta) (P^{(0)})^{-1}(\zeta), \quad &|\zeta|< 1/4,\\[.1cm]
                                                    U(\zeta)G^{-1}(\zeta),      \quad& |\zeta| > 1/4.
                                                   \end{array}
\right . \end{equation}So defined, $R_1(\zeta)$ is a piecewise analytic function in $\mathbb{C}\backslash \Sigma_{R_1}$, where the remaining contour $\Sigma_{R_1}$ consists of the circle $|\zeta|=1/4$ oriented clockwise,
and portions of $\Sigma_2$ and $\Sigma_4$, such that $|\zeta|>1/4$, cf. Figure \ref{figure 6}. From the matching condition
\eqref{matching condition-2}  and the jumps \eqref{U-jump},
we have the estimates
\begin{equation}\label{R-1 jump estimate}J_{R_1}(\zeta)=\left\{ \begin{array}{ll}
                     I+O(s^{-1}),& |\zeta|=1/4,\\
I+O(e^{-cs}), ~& \zeta\in \Sigma_j\cap \Sigma_{R_1}, ~j=2,4,
                 \end{array}\right.
\end{equation}
where $c$ is a positive constant.
Analysis similar to those  in Section 3.6 leads to
\begin{equation}\label{R-1 estimate}R_1(\zeta)=I+O( 1/ s),\end{equation}
with  uniform   error term    in $\mathbb{C}\backslash \Sigma_{R_1}$.

For later use, we need the following sharper estimate for $R_1(\zeta)-I$ for large $\zeta$, as can be derived from the jump estimate \eqref{R-1 jump estimate} and the Cauchy type resolvent operator of $R_1(\zeta)$:
\begin{equation}\label{R-1 estimate-large-zeta}R_1(\zeta)=I+O\left ( \frac 1{s\zeta}\right ),~~\mbox{as}~~\zeta\to\infty~~\mbox{and}~~s\to +\infty. \end{equation}

\subsection {Proof of Theorem \ref{thm-limit-kernel-big s}}
We prove  Theorem \ref{thm-limit-kernel-big s}  by
applying the  nonlinear steepest descendent analysis of $\Psi_0(\zeta,s)$,
as $s\to \infty$.

From \eqref{psi to U}, \eqref{R_1} and \eqref{R-1 estimate}, we get
\begin{equation}\label{psi-G}\Psi_0(\zeta,s) e^{\frac {-s\sqrt{\zeta}}2\sigma_3}=\left (I+O\left (\frac{1}{s\zeta }\right )\right )G(\zeta),\end{equation}
uniformly for $|\zeta|\geq r$ as $s\to +\infty$.

Expanding the integral representation for $G(\zeta)$  in \eqref{G-large-s}, we have
\begin{equation}\label{G-expand}\Psi_0(\zeta,s) e^{\frac {-s\sqrt{\zeta}}2\sigma_3}=\left (I+O\left (\frac{1}{s\zeta}\right )\right ) \zeta^{\frac{1}{4}\sigma_3}\frac{I-i\sigma_1}{\sqrt{2}}\left (I- \frac {\alpha \sigma_3}{2 \sqrt{\zeta}}+O\left (\frac{1}{\zeta}\right )\right )~~\mbox{as}~\zeta\to \infty.\end{equation}
Thus, from \eqref{psi-0 at infinity} and \eqref{G-expand},  we obtain
\begin{equation}\label{aymptotics of c-i} c_1(s)=\sigma(s)/s=-\alpha/2+O(1/s)~~\mbox{and}~~ c_2(s)=-iu(s)=O(1/s)~~\mbox{as}~s\to\infty.\end{equation}
Accordingly, we have
\begin{equation}\label{aymptotics of sigma u} u(s)=O(1/s)~~\mbox{and}~~ \sigma(s)=-s\alpha/2+O(1)~~\mbox{as}~s\to\infty.\end{equation}
Substituting
 \eqref{coefficient u} and \eqref{sigma-functon}  into \eqref{aymptotics of sigma u}, and recalling that $\Theta=-\alpha$ and $\gamma=\beta-\frac 1 2$ in the present case, we get
\begin{equation}\label{aymptotics of b beta big s} b(s)=O(1/s)~~\mbox{and}~~ y(s)=O(1)~~\mbox{as}~s\to\infty,\end{equation}
so long as  $\alpha\not=0$.

A combination  of \eqref{psi to U}, \eqref{P-0} and  \eqref{R_1} gives
\begin{equation}\label{Psi-0-Bessel}\Psi_0(\zeta)=R_1(\zeta)E_1(\zeta)\Phi\left (\frac 1 {16}s^2\zeta\right )
e^{\frac {\pi i\alpha}2\sigma_3}~~\mbox{for}~\arg\zeta\in (0,\pi)~\mbox{with}~|\zeta|<\frac 1 4, \end{equation}
where  $E_1$ is analytic in the disk and $\Phi$ is defined in  \eqref{Bessel parametrix}. Now further specifying  $\zeta\in II$; cf. Figure \ref{figure 6}, taking into account the   definitions in  \eqref{Bessel parametrix}  and \eqref{psi-Psi-0},    we have
$$  \left(
                                \begin{array}{c}
                                  \psi_1(\zeta,s) \\
                                  \psi_2(\zeta,s) \\
                                \end{array}
                              \right)
 = R_1(\zeta)E_1(\zeta)\left(
                               \begin{array}{cc}
                                 I_{\beta}(\frac 12s\sqrt{\zeta}) &\frac{i}{\pi} K_{\beta}(\frac 12s\sqrt{\zeta})  \\[.2cm]
                                 \frac {\pi i} 2 s\sqrt{\zeta}I'_{\beta}(\frac 1 2s\sqrt{\zeta})& -\frac 1 2s\sqrt{\zeta}    K'_{\beta}(\frac 12s\sqrt{\zeta}) \\
                                 \end{array}
                             \right)
                             \left(
                                \begin{array}{c}
                                  e^{-\frac\beta2\pi i } \\
                                0\\
                                \end{array}
                              \right) $$
Recalling that   $e^{-\frac 1 2 \beta \pi i}I_\beta(z)=   J_\beta (ze^{-\frac 1 2\pi i})$ for $\arg z\in (0, \pi/2]$ (corresponding to $\arg\zeta\in (0, \pi]$); cf. \cite[(9.6.3) and (9.6.30]{as}, we obtain
\begin{equation}\label{psi-J-v}
    \left(
                                \begin{array}{c}
                                  \psi_1(\zeta,s) \\
                                  \psi_2(\zeta,s) \\
                                \end{array}
                              \right)
=R_1(\zeta)E_1(\zeta)\left(
                               \begin{array}{c}
                                 J_{\beta}(\frac 1 2s\sqrt{\zeta} e^{-\frac \pi 2  i})  \\[0.2cm]
\frac \pi 2 s\sqrt{\zeta}J'_{\beta}(\frac 1 2s\sqrt{\zeta} e^{-\frac \pi 2  i}) \\
                                 \end{array}
                             \right).
\end{equation}

Thus, substituting \eqref{psi-J-v} into \eqref{psi-kernel},  by a similar argument as that in Section \ref{sec:4.3},  we obtain   the following reduction of the  $K_{\Psi}$  kernel, as $s\rightarrow +\infty$
  \begin{equation}\label{Bessel reduction big s}
  \frac 4 {s^2}K_{\Psi}\left (-\frac{4u}{s^2},-\frac{4v}{s^2}\right )=\frac{ \psi_1\left (-\frac{4u}{s^2},s\right )\psi_2\left (-\frac{4v}{ s^2},s\right ) -       \psi_1\left (-\frac{4v}{s^2},s\right )\psi_2\left (-\frac{4u}{ s^2},s\right )}{2\pi i(u-v)}=\mathbb{ J}_{\beta}(u,v)\left( I+O\left ( 1 /s\right )\right ),\end{equation}
where $\mathbb{ J}_{\beta}(u,v)$ is defined in \eqref{bessel-kernel}, with  $u$ and $v$ being in compact subsets of $(0, +\infty)$. In deriving the second equality, use has been made of the fact that $\det\{R_1(\zeta) E_1(\zeta)\}=1$, and that $X^T \left(
            \begin{array}{cc}
              0 & 1 \\
              -1 & 0 \\
            \end{array}
          \right)X=\left(
            \begin{array}{cc}
              0 & 1 \\
              -1 & 0 \\
            \end{array}
          \right )$ as $\det X=1$.  Here, as before, $X^T$ stands for the transpose of a matrix $X$.

It follows form \eqref{psi-kernel} and \eqref{Bessel reduction big s} that

\begin{equation}\label{Bessel kernel t big}
\begin{aligned}
 \frac{1}{2n^2}K_n\left (1-\frac{u}{2n^2},1-\frac{v}{2n^2}\right )&=\frac 4 {s^2} K_{\Psi}(-\frac{4u}{s^2},-\frac{4v}{s^2})\left (1+O\left (\frac 1{n^2}\right )\right )\\
 &=\mathbb{ J}_{\beta}(u,v) \left (1+O\left (\frac 1{n^2}\right )\right ),
   \end{aligned}
\end{equation}
where $\mathbb{ J}_{\beta}(u,v)$ is given in \eqref{bessel-kernel}, and the error terms are uniform in compact subsets of  $u,v\in(0,+\infty)$(that is  $ u/s^2, v/s^2\in {D} $, see \eqref{u,v domain} ).
Thus completes the proof of Theorem \ref{thm-limit-kernel-big s}.

\section {Reduction to Bessel kernel $\mathbb{J}_{\alpha+\beta}$ as $s\rightarrow 0$}\label{sec:6}

When $t=1$ fixed, the weight in \eqref{p-jacobi weight} can  be written as $w(x)=(1-x^2)^{\alpha+\beta}h(x)$.  The local behavior at $x=1$ is described via the kernel  $\mathbb{J}_{\alpha+\beta}$ given  in \eqref{bessel-kernel}  for $\alpha+\beta>-1$; cf. \cite{kv}.

In the $\Psi$-kernel $K_{\Psi}(-u,-v;s)$ in \eqref{psi-kernel}, the parameter $s=4n\ln\left ( t+\sqrt{t^2-1}\right )\to 0$ as $t$ varies to $1^+$. Similar to the derivation in   Section \ref{sec:5}, we study in the present section the asymptotics of the model RH problem for $\Psi_0(\zeta,s)$, with the parameters $\Theta=-\alpha$, $\gamma=\beta-\frac 12$ and $\alpha+\beta>-1$, as $s\to 0$. Then we apply the asymptotic results to reduce $K_{\Psi}(-u,-v;s)$ to the classical  Bessel kernel $\mathbb{J}_{\alpha+\beta}$. And we also obtain the asymptotics for the solution $b(s)$, $u(s)$ and $y(s)$ to the nonlinear equations given in Section \ref{sec:1.2}, as $s\to 0^+$.

\subsection {Nonlinear steepest descent analysis of the RH problem for $\Psi_0(\zeta,s)$ as $s\to 0$}
$\Psi_0(\zeta,s)$ solves the model RH problem  formulated in \eqref{psi-0 jump}-\eqref{psi-0 at 1/4}.  Accordingly, $\Psi_0  (  {\zeta}/{s^2},s)$
solves a re-scaled version  of the RH problem.    As $s\rightarrow 0$, the  jump contour $\Sigma_1$   for $\Psi_0({\zeta}/{s^2},s)$ becomes the  shrinking line segment $(0,s^2/4)$. Ignoring the constant jump on $\Sigma_1$,  the RH problem is then reduced to the  the Bessel model RH problem $\Phi$ formulated  in \eqref{phi-jump}-\eqref{phi at infinity}, with the parameter $\beta$ being replaced by $\alpha+\beta$.  Thus $\Psi_0({\zeta}/{s^2},s)$ is approximated by $\Phi$. However, the approximation is not true for $\zeta\in(0,s)$. So we need a local parametrix near the origin. A similar argument can be found in
\cite{cik, xz2013-2}.

First we recall the well-known    formulas for the modified Bessel functions
\begin{equation}\label{Bessel at 0}I_{\alpha+\beta}(2z)=z^{\alpha+\beta}\sum_{n=0}^{\infty}\frac{z^{2n}}{n!\Gamma(n+\alpha+\beta+1)},\quad \mbox{and}~K_{\alpha+\beta}(z)=\frac{\pi}{2}\frac{I_{-\alpha-\beta}(z)-I_{\alpha+\beta}(z)}{\sin ((\alpha+\beta)\pi)},
\end{equation}
where $\arg z\in(-\pi,\pi)$, and $\alpha+\beta\not\in \mathbb{Z}$; see \cite[(9.6.2) and (9.6.10)]{as}.
Applying  these formulas, we can rewrite   the    function   $\Phi$ in \eqref{Bessel parametrix}, using $\alpha+\beta$ instead of $\beta$,   as
 \begin{equation}\label{phi at 0}
 \Phi(\zeta)=E_{2}(\zeta)\zeta^{\frac{ (\alpha+\beta)\sigma_3}{2}}\left(
                                                          \begin{array}{cc}
                                                            1 &\frac{1}{2i\sin ((\alpha+\beta)\pi)}  \\
                                                            0 & 1 \\
                                                          \end{array}
                                                        \right)
\left \{
\begin{array}{ll}
J_{I}, &  \zeta\in I, \\
J_{II},&   \zeta\in II,\\
J_{III},  &  \zeta\in III;
\end{array}
\right .
\end{equation}cf. Figure \ref{figure 6} for the regions,
where
$$J_I=\left(
                                                          \begin{array}{cc}
                                                            1 &0 \\
                                                            0   & 1 \\
                                                          \end{array}
                                                        \right),~~
 J_{II}= \left(
                                                          \begin{array}{cc}
                                                            1 &0 \\
                                                            -e^{\pi i (\alpha+\beta)}  & 1 \\
                                                          \end{array}
                                                        \right),~~\mbox{and}~
J_{III}=  \left(
                                                          \begin{array}{cc}
                                                            1 &0 \\
                                                            e^{-\pi i (\alpha+\beta)}  & 1 \\
                                                          \end{array}
                                                        \right),$$
and $E_{2}(\zeta)$ is a matrix-valued entire function, explicitly given as
 \begin{equation}\label{E-2} E_{2}(\zeta)=
\left(
  \begin{array}{cc}
   \zeta^{-(\alpha+\beta)/2}I_{\alpha+\beta}(2\sqrt\zeta) & \frac {i}{2\sin((\alpha+\beta)\pi)} \zeta^{(\alpha+\beta)/2}I_{-(\alpha+\beta)}(2\sqrt\zeta) \\[.2cm]
  2\pi i   \zeta ^{(1-\alpha-\beta)/2}I'_{\alpha+\beta}(2\sqrt\zeta) & \frac {-\pi}{\sin((\alpha+\beta)\pi)} \zeta^{(1+\alpha+\beta)/2}I'_{-(\alpha+\beta)}(2\sqrt\zeta) \\
  \end{array}
\right).\end{equation}

Straightforward comparison shows that   $\Psi_0  (  {\zeta}/{s^2},s)$ and $\left ( \frac {\pi s} 2\right )^{-\frac 1 2\sigma_3}(-i\sigma_1) \Phi(\zeta/16)$ share the same jumps  and the same behavior at infinity, as long as $|\zeta|>s^2/4$; cf. \eqref{psi-0 jump}-\eqref{psi-0 at infinity} and \eqref{phi-jump}-\eqref{phi at infinity}, in which $\beta$ being replaced with $\alpha+\beta$. For  $|\zeta|<s^2/4$, $\Phi(\zeta)$ fails to approximate  $\Psi_0  (  {\zeta}/{s^2},s)$ due to the appearance of the  extra contour $\Sigma_1$ for $\Psi_0$. Then it is natural to consider a local parametrix, say,  $M(\zeta)$,   in a small neighborhood  $U_\epsilon:|\zeta|< \epsilon$, $0<\epsilon<1$. For small $s$, we see that the re-scaled $\Sigma_1$ lies in  $U_\epsilon$.

We state the RH problem for $M(\zeta)$  as follows:
\begin{description}
  \item(a)~~  $M(\zeta)$ is analytic in {$U_\epsilon\backslash\cup_{j=1}^4 \Sigma_j$}, where $\Sigma_j$ are re-scaled version of those depicted  in Figure \ref{figure 1}, such that $\Sigma_1=(0, s^2/4)$;
  \item(b)~~   $M(\zeta)$  shares  the same constant jump conditions \eqref{psi-0 jump}, with $\Psi_0$  on { $U_\epsilon\cap\Sigma_j$}, $j=1,2,3,4$, specifying  $\Theta=-\alpha$ and  $\gamma=\beta-\frac 1 2$;

\item(c)~~  The matching condition on the boundary  {$\partial U_\epsilon$}, as the parameter $s\rightarrow 0$, is
{\begin{equation}\label{matching condition M-phi}
 M(\zeta)=(I+O(s^{l}))\Phi(\zeta/16),~~   |\zeta|=\epsilon,
\end{equation}
where $l=2\min\{1,\alpha+\beta+1\}$, and $\alpha+\beta>-1.$}
\end{description}

We seek a solution of the form
\begin{equation}\label{M-def}
  M(\zeta)= \tilde E_2(\zeta)
\left(
                                                          \begin{array}{cc}
                                                            1 & m(\zeta/s^2)  \\
                                                            0 & 1 \\
                                                          \end{array}
                                                        \right)
\zeta^{\frac{\beta\sigma_3}{2}}\left (\zeta-\frac{s^2}{4}\right )^{\frac{\alpha\sigma_3}{2}}  \left(
                                                          \begin{array}{cc}
                                                            1 &\frac{1}{2i \sin(\alpha+\beta)\pi }  \\
                                                            0 & 1 \\
                                                          \end{array}
                                                        \right)
\left \{
\begin{array}{ll}
J_{I}, &  \zeta\in I, \\
J_{II},&   \zeta\in II,\\
J_{III},  &  \zeta\in III
\end{array}
\right .
 \end{equation}
 for $\arg \zeta\in(-\pi, \pi)$ and $\arg (\zeta-\frac{s^2}{4})\in(-\pi, \pi)$,
where the constant matrices  $J_I$-$J_{III}$    are given in  \eqref{phi at 0}, the sectors $I$-$III$ are illustrated in Figure \ref{figure 7}, and  $\tilde E_2(\zeta)=E_2(  \zeta/{16})\; 4^{-(\alpha+\beta)\sigma_3}$ is entire,  in which   $E_{2}(\zeta)$   is explicitly defined in
\eqref{E-2}.

 Assuming that  $m(\zeta)$ is an analytic scalar function in $\mathbb{C}\backslash[0,\frac{1}{4}]$, it is easily shown that the jump conditions \eqref{psi-0 jump}  on  $\Sigma_2$-$\Sigma_4$ are satisfied automatically by $M(\zeta)$.
The remaining jump condition  for  $M(\zeta)$ on the re-scaled contour $\Sigma_1=(0, s^2/4)$ is equivalent to the jump for $m(\zeta)$ in  \eqref{m-jump} below. Hence it suffices to solve the scalar RH problem:
\begin{description}
  \item(a)~~  $m(\zeta)$ is analytic in $\mathbb{C}\backslash[0,\frac{1}{4}]$;
  \item(b)~~   $m(\zeta)$  satisfies the  jump condition
  \begin{equation}\label{m-jump}m_{+}(\zeta)-m_{-}(\zeta)= -\frac {\sin(\alpha\pi)} {\sin(\alpha+\beta)\pi} s^{2(\alpha+\beta)} \zeta^{\beta} \left (\frac{1}{4}-\zeta\right )^{\alpha}~~\mbox{for}~\zeta\in (0, 1/4);
\end{equation}

\item(c)~~  The behavior of $m(\zeta)$   at infinity is
 \begin{equation}\label{m-infty}m(\zeta)=O( 1/{\zeta}).
\end{equation}
\end{description}
\vskip .5cm

The RH problem can be solved  by using the Sokhotski-Plemelj formula. We have
 \begin{equation}\label{m}m(\zeta)=-\frac{\sin(\alpha\pi)\;  s^{2(\alpha+\beta)}}{2\pi i\sin(\alpha+\beta)\pi}
 \int_0^{\frac 1 4}\frac{\tau^{\beta }\left (\frac 1 4-\tau\right )^{\alpha}d\tau}{\tau-\zeta},~~ \zeta\in \mathbb{C}\backslash [0, 1/4]    .\end{equation}

In view of  \eqref{M-def} and \eqref{m},  we see that the matching condition \eqref{matching condition M-phi} is fulfilled.

\noindent
{\rmk{
The function $m(\zeta)$ is related to a hypergeometric function. Indeed, recalling the integral representation
 \begin{equation}\label{hypergeometric function }F(a,b;c;z)=\frac{\Gamma(c)}{\Gamma(b)\Gamma(c-b)}\int_0^{1}x^{b-1}(1-x)^{c-b-1}(1-zx)^{-a }dx,\end{equation}
 where $\Re c>\Re b>0$, $z\not \in[1,+\infty)$; cf. \cite[(15.3.1)]{as}, we have
 \begin{equation}\label{m-hypergeometric}m(\zeta)= \frac {\sin(\alpha\pi)} {8\pi i \zeta \sin(\alpha+\beta)\pi} \frac {\Gamma(\alpha+1)\Gamma(\beta+1)}{\Gamma(\alpha+\beta+2)}\left (    \frac s 2\right ) ^{2(\alpha+\beta)} F\left (1,\beta+1; \alpha+\beta+2;\frac{1}{4\zeta}\right )  \end{equation}
for $\zeta  \not \in[0,  {1}/4]$.
}}\vskip .3cm

 We note that $m(\zeta)$ can also be defined for $\alpha<-1$ by applying  Gauss' relations for contiguous functions to \eqref{m-hypergeometric}; see, e.g., \cite[(15.2.25)]{as}.
\vskip .3cm

\noindent
{\rmk{   Also, for integer $\alpha+\beta$, a special treatment should be brought in from \eqref{Bessel at 0} on.
  The relations of Bessel functions in  \eqref{Bessel at 0} should be modified, and
   a logarithmic singularity  may appear in the off-diagonal  entry in \eqref{phi at 0}.  Instead of \eqref{phi at 0}, we have
$$  \Phi(\zeta)=\hat{E}_{2}(\zeta)\zeta^{\frac{ (\alpha+\beta)\sigma_3}{2}}\left(
                                                          \begin{array}{cc}
                                                            1 &\frac{(-1)^{\alpha+\beta}}{2\pi i}\ln \zeta  \\
                                                            0 & 1 \\
                                                          \end{array}
                                                        \right),$$
where $\hat{E}_{2}(\zeta)$ is an analytic matrix function and $\zeta\in I$; cf. Figure \ref{figure 7} for the region,  see also  \cite[Sec.\;5.1]{xdz2014} for similar formulas.
Then, by the same argument as in the non-integer $\alpha+\beta$ case, we construct a local parametrix  $M(\zeta)$ in the form  of  \eqref{M-def}, with $m(\zeta)$ defined as
$$m(\zeta)=\frac{\sin(\alpha\pi)(-1)^{\alpha+\beta} s^{2(\alpha+\beta)}}{2\pi i}
 \int_0^{\frac 1 4}\frac{\tau^{\beta }\left (\frac 1 4-\tau\right )^{\alpha}\ln(s^2\tau)d\tau}{\tau-\zeta},~~ \zeta\in \mathbb{C}\backslash [0, 1/4]. $$
And, the matching condition \eqref{matching condition M-phi} is now slightly modified as
$$ M(\zeta)=(I+O(s^{2}\ln s))\Phi(\zeta/16),~~   |\zeta|=\epsilon, ~~\mbox{for}~~\alpha+\beta=0,$$
  and
$$ M(\zeta)=(I+O(s^2))\Phi(\zeta/16),~~   |\zeta|=\epsilon,$$
for $\alpha+\beta$ being a positive integer, where use is made of the condition that $\alpha+\beta>-1$.
}}

\vskip .5cm

Now we proceed to consider
\begin{equation}\label{R-2}R_2(\zeta)=\left\{\begin{array}{ll}
                                                   i\sigma_1 \left (\frac{\pi s} 2\right )^{\frac 1 2\sigma_3}\Psi_0\left (\frac{\zeta}{s^2},s\right ) M^{-1}(\zeta),  \quad & |\zeta|< \epsilon,\\[.2cm]
                                                   i\sigma_1 \left (\frac {\pi s} 2\right )^{\frac 1 2\sigma_3}\Psi_0\left (\frac{\zeta}{s^2},s\right )\Phi^{-1}\left (\frac \zeta {16}\right ),      \quad &|\zeta| > \epsilon,
                                                   \end{array}
\right. \end{equation}
The matrix function $R_2$ is analytic in $|\zeta|\neq \epsilon$.  Indeed, we need only to verify,   in a straightforward manner,   that the isolated singularities at $\zeta=0, s^2/4$ are removable. For example, a combination of \eqref{psi-0 at 1/4} (with $\Theta=-\alpha$),  \eqref{M-def}, and \eqref{R-2} gives
$$R_2(\zeta)=O(1) \left(
                    \begin{array}{cc}
                      O(1) & O(1)\\
                      0 & O(1) \\
                    \end{array}
                  \right) O(1)~~\mbox{as}~\zeta\to s^2/4. $$ Thus $\zeta=s^2/4$    is a  weak singularity, and hence is removable. Similar argument applies to $\zeta=0$. Here use has been made of the fact that the scalar function defined in \eqref{m} has the boundary behavior
                  $$m(\zeta)=O(1) + O\left (\zeta^\beta\right ),~~\zeta\to 0~~\mbox{and}~~ m(\zeta)=s^{2(\alpha+\beta)} \left [O(1)- \frac {\zeta^\beta (\zeta- 1/4)^\alpha}{2i\sin(\alpha+\beta)\pi}\right ]
,~~\zeta\to 1/4$$
 for $\zeta\not\in [0, 1/4]$, where $\arg\zeta\in (-\pi, \pi)$  and  $\arg(\zeta-1/4) \in (-\pi, \pi)$ in the  approximation at $\zeta= 1/4$.
Also, it follows from the matching condition that  the jump
\begin{equation}\label{R-2 jump estimate}J_{R_2}(\zeta)=
                     I+O(s^{l}),\quad  |\zeta|=\epsilon,
\end{equation}
where  $l=2\min\{1,\alpha+\beta+1\}$, with $\alpha+\beta>-1$.
So, by an argument   similar to   Section \ref{sec:3.6}, we have
 \begin{equation}\label{R-2 estimate}R_2(\zeta)=I+O(s^{l}),~~s\rightarrow 0^+,\end{equation}
where the error term $O(s)$ is uniform in $\zeta$. Sharper estimate is available for large $\zeta$, namely,
 \begin{equation}\label{R-2 estimate-large-zeta}R_2(\zeta)=I+O\left (\frac {s^{l}}\zeta \right ),~~\mbox{as}~~s\rightarrow 0^+~~\mbox{and}~~\zeta\to \infty,\end{equation}
where  $l=2\min\{1,\alpha+\beta+1\}$, and $\alpha+\beta>-1$.

\subsection{Proof of Theorem \ref{thm-limit-kernel-small s}}
We apply the asymptotic formulas to obtain the asymptotic properties   of several  functions introduced in Section \ref{sec:2.1}, as $s\to 0^+$. To begin with,
we see from
 \eqref{phi at infinity}, \eqref{R-2} and \eqref{R-2 estimate-large-zeta} that
\begin{equation}\label{psi-0 at circle small s}\Psi_0\left (\frac{\zeta}{s^2},s\right ) e^{\frac  {-\sqrt{\zeta}}2\sigma_3}=s^{-\frac 12\sigma_3}
\left(I+O\left (\frac {s^l}\zeta \right )\right )
\zeta^{\frac{1}{4}\sigma_3}\frac{I-i\sigma_1}{\sqrt{2}}\left (I+O\left (\frac 1 {\sqrt{\zeta}}\right )\right) \end{equation}
  for $\zeta\to \infty$ and  $s\to 0^+$,  where $l=2\min\{1,\alpha+\beta+1\}$, and $\alpha+\beta>-1$.
Refinement is available by using \eqref{psi-phi-small s} below and expanding $\Phi(\zeta)$ in  \eqref{Bessel parametrix} for large $\zeta$,  with  $\beta$ being replaced by $\alpha+\beta$.  As a result, we have
  \begin{equation}\label{psi-0 at circle small s-2}
 \begin{aligned}
  \Psi_0\left (\frac{\zeta}{s^2},s\right ) e^{\frac  {-\sqrt{\zeta}}2\sigma_3}&=  s^{-\frac 12\sigma_3}\left(I+O\left (\frac {s^l}\zeta \right )\right )
\zeta^{\frac{1}{4}\sigma_3}
\frac{I-i\sigma_1}{\sqrt{2}}\left (I+\frac {C_{R,1}}{\sqrt{\zeta}}+\frac {C_{R,2}} \zeta+O\left (\frac 1{\zeta^{3/2}}\right )\right) \\
 &=   \left ( \frac \zeta {s^2}\right )^{\frac{1}{4}\sigma_3}\frac{I-i\sigma_1}{\sqrt{2}}
\left (I+\frac {C_{R,1}+O(s^{l}) }{\sqrt{\zeta}}+\frac {C_{R,2}+O(s^{l}) }{\zeta}+O\left (\frac 1 {\zeta^{3/2}}\right ) \right)
\end{aligned}
\end{equation} as $\zeta\to \infty$ and  $s\to 0$,
 where the first two coefficients of the large-$\zeta$ expansion for  $\Phi(\zeta)$ are
  $$C_{R,1}=-\frac{i}{2}\sigma_1-\left\{(\alpha+\beta)^2+\frac  14\right \}\sigma_3~~\mbox{and}~~
  C_{R,2}= \frac {4(\alpha+\beta)^2-1}{8}\left \{\left ((\alpha+\beta)^2+\frac 34\right )I+3\sigma_2\right \}.$$

Thus, comparing  \eqref{psi-0 at infinity} with \eqref{psi-0 at circle small s-2},  we have the  behavior for  $\sigma(s)$, $u(s)$ and $\hat{c}_2(s)$, such that
 \begin{equation}\label{asymptotics of b and u amall s} \sigma =-\left ((\alpha+\beta)^2+\frac 14\right )+O\left (s^{l}\right ), ~   u =-\frac 1 {2s}+O\left (s^{l-1} \right )~\mbox{and}~  \hat{c}_2= \frac {3(4(\alpha+\beta)^2-1)}{8s^2}+O\left (s^{l-2}\right )\end{equation}
as $s\to 0^+$, where $\sigma$, $u$ and $\hat{c}_2= \frac 1s\left(u+u\sigma-\frac 12 \left (\frac by+(b-\alpha)y\right )\right)$ appear in the coefficient of the asymptotic behavior at infinity of $\Psi$ and  $\Psi_0$; cf.     \eqref{Psi at infinity} and \eqref{psi-0 at infinity}.
Then, a combination of \eqref{coefficient u},  \eqref{sigma-functon} and \eqref{asymptotics of b and u amall s}, with $\Theta=-\alpha$ and $\gamma=\beta-\frac 1 2$,    yields
 \begin{equation}\label{asymptotics of beta amall s}
 \left\{ \begin{aligned}
  b(s) &= - \frac {(\alpha+\beta)^2} s+\frac \alpha 2+O\left (s^{l-1}\right ), \\
  y(s) &=  1+ O\left (s^l\right )
  \end{aligned}\right .
 \end{equation} as $s\to 0^+$, where $l=2\min\{1,\alpha+\beta+1\}$ and  use has been made of the fact that
 $$y=\frac {\gamma+2u+2u\sigma-us-2\hat{c}_2s}{2(b-\alpha)}.$$

Now we are in a position to prove Theorem \ref{thm-limit-kernel-small s}.
For $|\zeta|>\epsilon$, it follows directly form \eqref{R-2} that
\begin{equation}\label{psi-phi-small s} \Psi_0\left (\frac{\zeta}{s^2},s\right )=-i\left ( \frac {\pi s} 2\right )^{-\frac{\sigma_3}{2}}\sigma_1R_2(\zeta)\Phi\left(\frac \zeta {16}\right ).\end{equation}

Thus, for $\zeta\in II$ and $|\zeta|>\epsilon$, a combination of \eqref{psi-Psi-0}, \eqref{psi-phi-small s}, and \eqref{Bessel parametrix}, again  with $\alpha+\beta$   taking the place of $\beta$,  gives
\begin{equation}\label{psi-bessel small s}
\begin{aligned}
\left(
                                \begin{array}{c}
                                  \psi_1\left (\frac{\zeta}{s^2},s\right ) \\[.2cm]
                                  \psi_2\left (\frac{\zeta}{s^2},s\right ) \\
                                \end{array}
                              \right)
 &=     \left (\frac {\pi s} 2\right )^{-\frac 1 2 \sigma_3}(-i\sigma_1) R_2(\zeta)e^{-\frac{(\alpha+\beta)\pi i}2 }\left(
                               \begin{array}{c}
                                 I_{\alpha+\beta}\left (\frac {\sqrt{\zeta}} 2\right )  \\[.2cm]
                                 \frac {\pi i\sqrt{\zeta}} 2 I'_{\alpha+\beta}\left (\frac {\sqrt{\zeta}} 2\right ) \\
                                 \end{array}
                             \right)_{+} \\
  &=
                             \left (\frac {\pi s} 2\right )^{-\frac 1 2 \sigma_3}(-i\sigma_1)R_2(\zeta)\left(
                               \begin{array}{c}
                                 J_{\alpha+\beta}\left ( \frac {\sqrt{ |\zeta| }} 2 \right )  \\[.2cm]
                                 \frac {\pi i\sqrt{|\zeta|}} 2 J'_{\alpha+\beta}\left ( \frac {\sqrt{ |\zeta |}} 2 \right ) \\
                                 \end{array}
                             \right).
  \end{aligned}
 \end{equation}
Here
use has been made of the fact that  $e^{-\frac 1 2 \nu \pi }I_\nu(z)=   J_\nu (ze^{-\frac 1 2\pi i})$ for $\arg z\in (0, \pi/2]$.

Thus by a similar argument leading to \eqref{psi-kernel}, or to \eqref{Bessel reduction big s}, we get from \eqref{psi-bessel small s}
the approximation of $K_{\Psi}$ by the Bessel kernel as follows,
\begin{equation}\label{K-psi-bessl-small s}  \frac{\psi_1(-\frac{4u}{ s^2},s)\psi_2(-\frac{4v}{ s^2},s)-\psi_1(-\frac{4v}{ s^2},s)\psi_2(-\frac{4u}{s^2},s)}{2\pi i(u-v)}=\mathbb{J}_{\alpha+\beta}(u,v) \left(1+O\left(s^l\right)\right),\end{equation}
where   $l=2\min\{1,\alpha+\beta+1\}$, $\alpha+\beta>-1$,   and  the Bessel kernel is  defined in \eqref{bessel-kernel}, and the error term is uniform in compact subsets of  $u,v\in(0,\infty)$.

Thus by \eqref{psi-kernel} and \eqref{K-psi-bessl-small s}, we obtain
\begin{equation}\label{Bessel kernel reduction small s}
\begin{aligned}
 \frac{1}{2n^2}K_n\left (1-\frac{u}{2n^2},1-\frac{v}{2n^2}\right )&=  \frac{\psi_1\left (-\frac {4u} {s^2},s\right )\psi_2\left (-\frac {4v} {s^2},s\right )-\psi_1\left (-\frac{4v} {s^2},s\right )\psi_2\left (-\frac {4u} {s^2},s\right )}{2\pi i(u-v)}
 \left (1+O\left (\frac {1}{n^2} \right ) \right) \\
 &=    \mathbb{ J}_{\alpha+\beta}(u,v) (1+O(s^l)+O(1/n^2))  ,
  \end{aligned}
 \end{equation}
where  $l=2\min\{1,\alpha+\beta+1\}$, $\alpha+\beta>-1$ , $\mathbb{ J}_{\alpha+\beta}(u,v)$ is given in \eqref{bessel-kernel},  and the error terms are uniform in compact subsets of  $u,v\in(0,+\infty)$.
And we complete the proof of Theorem \ref{thm-limit-kernel-small s}.

\section*{Acknowledgements}
 The authors would like to thank the editor and the referees for their   valuable suggestions and comments. The authors are also grateful to Zhao-Yun Zeng for a careful reading of the previous version of the manuscript. The work of Shuai-Xia Xu  was supported in part by the National
Natural Science Foundation of China under grant number
11201493, GuangDong Natural Science Foundation 
 under grant numbers S2012040007824 and 2014A030313176,
 Postdoctoral Science Foundation of China under grant number 2012M521638 and the Fundamental Research Funds for the Central Universities under grant number 13lgpy41.
 Yu-Qiu Zhao  was supported in part by the National
Natural Science Foundation of China under grant numbers 10471154 and
10871212.

\end{document}